\documentclass[11pt,oneside,letterpaper]{article}
\usepackage{amssymb}
\usepackage{amsmath}
\usepackage{amsthm}
\usepackage[dvips]{graphicx}
\usepackage{setspace}
\usepackage{amsfonts}
\usepackage{fancyhdr}
\usepackage{xcolor}
\usepackage{graphicx}
\usepackage{rotating}
\usepackage{comment}
\usepackage{color}
\usepackage{cite}
\usepackage{braket}
\usepackage{caption}
\usepackage[utf8]{inputenc}
\usepackage{mathtools}
\usepackage{tikz}
\usepackage{overpic}
\usepackage{stackengine}
\usepackage[normalem]{ulem}
\usepackage{bbold}

\definecolor{darkgreen}{rgb}{0,0.5,0}
\definecolor{darkblue}{rgb}{0,0,0.6}
\definecolor{purple}{rgb}{0.4,.2,0.7}
\newcommand{\p}{\partial}

\newcommand{\be}{\begin{equation}}
\newcommand{\ee}{\end{equation}}

\usepackage[colorlinks=true,citecolor=darkgreen,linkcolor=black,urlcolor=purple]{hyperref}

\usepackage{pdfsync}
\makeatletter
\newcommand*{\defeq}{\mathrel{\rlap{%
                     \raisebox{0.3ex}{$\m@th\cdot$}}%
                     \raisebox{-0.3ex}{$\m@th\cdot$}}%
                     =} 
\makeatother

\def\be{\begin{eqnarray}}
\def\ee{\end{eqnarray}}

\newcommand{\tr}{\textrm{Tr}\,}

\newcommand{\bea}{\begin{eqnarray}}
\newcommand{\eea}{\end{eqnarray}}
\def\ben{\begin{equation}}
\def\een{\end{equation}}

    \let\p=\phi \let\r=v

  \let\D=\Delta  \let\L=\Lambda
   
\let\C=\Chi 

\def\be{\begin{equation}}
\def\ee{\end{equation}}
\def\ba{\begin{array}}
\def\ea{\end{array}}

\def\ba#1\ea{\begin{align}#1\end{align}}
\def\bs#1\es{\begin{split}#1\end{split}}

\renewcommand{\p}{\partial}

\newcommand{\Pvc}{\tilde{P}_V}

\renewcommand{\L}{{\cal L}}

\renewcommand{\O}{{\cal O}}

\definecolor{vert}{rgb}{0.1367 0.543 0.1367}

\newcommand{\id}{\mathbb{1}}

\allowdisplaybreaks  % allow page breaks in displayed eqs

\numberwithin{equation}{section}
\def \be {\begin{equation}}
\def \ee {\end{equation}}
	
\def \JM#1 {{\color{blue}  JM: #1 }}
\def \AAl#1 {{\color{red}  AA: #1 }}

\begin{document}
\onehalfspacing

\begin{center}

{\LARGE  {
Coarse graining pure states in AdS/CFT
}}

\vskip1cm

Jeevan Chandra and Thomas Hartman

\vskip5mm
Department of Physics, Cornell University, Ithaca, New York, USA

\vskip5mm

{\tt jn539@cornell.edu, hartman@cornell.edu }

\end{center}

\vspace{4mm}

\begin{abstract}

\noindent We construct new Euclidean wormhole solutions in AdS$_{d+1}$ and discuss their role in UV-complete theories, without ensemble averaging. The geometries are interpreted as overlaps of GHZ-like entangled states, which arise naturally from coarse graining the density matrix of a pure state in the dual CFT. In several examples, including thin-shell collapsing black holes and pure black holes with an end-of-the-world brane behind the horizon, the coarse-graining map is found explicitly in CFT terms, and used to define a coarse-grained entropy that is equal to one quarter the area of a time-symmetric apparent horizon. Wormholes are used to derive the coarse-graining map and to study statistical properties of the quantum state. This reproduces aspects of the West Coast model of 2D gravity and the large-$c$ ensemble of 3D gravity, including a Page curve, in a higher-dimensional context with generic matter fields.

 \end{abstract}
%\vspace{.2in}
%\vspace{.3in}

\pagebreak
\pagestyle{plain}

\setcounter{tocdepth}{2}
{}
\vfill

\ \vspace{-2cm}
\renewcommand{\baselinestretch}{1}\small
\tableofcontents
\renewcommand{\baselinestretch}{1.15}\normalsize

\newcommand{\brho}{\bar{\rho}}
\newcommand{\Svn}{S_{\rm vN}}
\newcommand{\N}{{\cal N}}
\newcommand{\C}{{\cal C}}
\renewcommand{\H}{{\cal H}}

\section{Introduction}

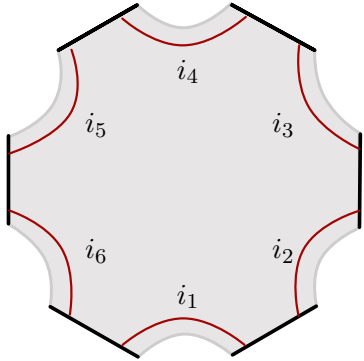
\begin{figure}
\begin{center}
\begin{overpic}[width=2in,grid=false]{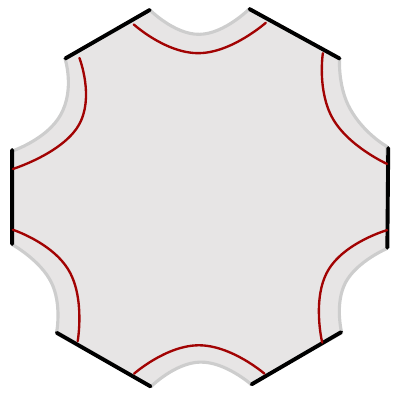}
\put (47,17) {$i_1$}
\put (72,29) {$i_2$}
\put (72,62) {$i_3$}
\put (47,76) {$i_4$}
\put (23,62) {$i_5$}
\put (23, 29) {$i_6$}
\end{overpic}
\end{center}
\caption{An on-shell Euclidean AdS$_{d+1}$ wormhole with six disconnected boundaries. Each boundary (shown in black) is $\mathbb{R}^d$ with the flat metric or $S^d$ with the round metric. The red lines are matter sources with planar or spherical symmetry. We consider examples where these matter sources are thin shells of pressureless fluid. There are similar wormholes sourced by end-of-world branes; in that case the spacetime ends at the red lines, and only replica-symmetric `fractional' wormholes  with $k < 2$ are on shell. The  $i_m$'s label the particular matter configuration, or the flavor of the EOW brane.
\label{fig:introWormhole}
}
\end{figure}

In this paper we construct asymptotically-AdS$_{d+1}$ Euclidean wormholes with $k$ disconnected conformal boundaries, sourced by matter going through the wormhole. The structure is illustrated in figure \ref{fig:introWormhole}. These are multiboundary solutions of Einstein gravity with negative cosmological constant.
They appear generically in any theory of gravity, although the allowed range of $k$ depends on the details, and in some cases only `fractional' wormholes with non-integer $k$ exist as gravitational saddles. We will focus on two cases: Thin shell wormholes, where the matter is a shell of pressureless perfect fluid, and $B$-states, where the matter is an end-of-the-world (EOW) brane on which the spacetime terminates.  Thin shell wormholes exist for integer $k$, while $B$-states admit only fractional wormholes.

Euclidean solutions with disconnected boundaries are puzzling from the point of view of the AdS/CFT correspondence, because the duality predicts that they ultimately cannot contribute to the gravitational path integral with the standard boundary conditions \cite{Maldacena:2004rf,Arkani-Hamed:2007cpn}. On the other hand, these solutions should not just be ignored, because in some cases, Euclidean wormholes are known to calculate ensemble-averaged quantities in the dual quantum theory \cite{Cotler:2016fpe,Saad:2019lba} (see also \cite{Penington:2019kki,Afkhami-Jeddi:2020ezh,Maloney:2020nni,Chandra:2022bqq}). This is known as the factorization puzzle. Similar Euclidean wormholes also play a role in calculating the entropy of Hawking radiation \cite{Penington:2019kki,Almheiri:2019qdq}.

There is a CFT interpretation for the wormholes described above which does not require ensemble averaging: These  wormholes compute the inner products of states with a GHZ-like pattern of entanglement in the UV. 
Consider a Euclidean wormhole in $(d+1)$ bulk dimensions with $k$ disjoint boundaries, like that in figure \ref{fig:introWormhole}. %Red lines are matter sources, and $i_m$ labels the particular matter configuration joining boundaries $(m-1,m)$. 
This wormhole calculates
\begin{align}\label{zoverlap}
Z_{\rm wormhole} &\approx \langle \Psi_k| \Psi_k\rangle  \ , 
\end{align}
where
\begin{align}\label{psik}
|\Psi_k\rangle &= \sum_{n} V^{i_1}\psi_n^{i_1}|n\rangle  V^{i_2}\psi_n^{i_2} |n\rangle \cdots V^{i_k}\psi_n^{i_k} |n\rangle \ . 
\end{align}
In this relation,  $Z_{\rm wormhole} \approx e^{-I_{\rm wormhole}}$ is the gravitational path integral on this topology, which we treat to leading order in the semiclassical approximation. The relation \eqref{zoverlap} is approximate because there may be other topologies that contribute to this overlap, but we will study examples in which the wormhole is the leading term. The state $|\Psi_k\rangle$ lives in $k$ copies of the CFT Hilbert space, and comes with a specific normalization so that its norm is a meaningful physical quantity. 
The sum is over black hole microstates $|n\rangle$ in a single copy of the CFT, in the energy basis. The numerical coefficients $\psi^i_n$ are UV data that can be studied statistically, but not exactly, in the low-energy theory. The $V^i$ are special CFT operators, built from single traces, that we call semiclassical isometries; a \textit{semiclassical isometry} is  an invertible map from the CFT Hilbert space around energy $E$ to the CFT Hilbert space around energy $E'$, with $E' > E$, and in \eqref{psik} it corresponds to adding massive particles that do not go through the wormhole.

\begin{figure}
\begin{center}
\begin{overpic}[width=2in,grid=false]{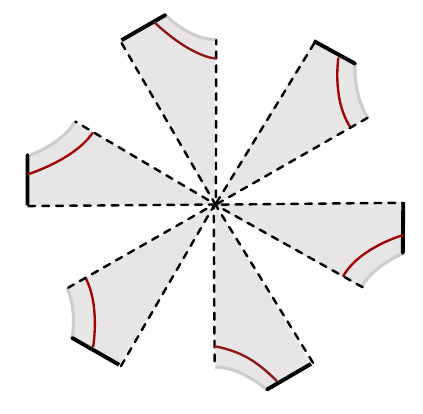}
\put (-30, 50) {{\Large $|\Psi_k\rangle   = $}}
\end{overpic}
\end{center}
\caption{
The Euclidean spacetime associated to the GHZ-like quantum state $|\Psi_k\rangle$ is a `windmill' geometry. In the CFT, $|\Psi_k\rangle$ is defined on the $t=0$ slice of $k$ copies of the CFT, which is an $(S^{d-1})^k$ that cannot be extended smoothly into a bulk Cauchy slice.  States of this type are not required to satisfy holographic entanglement inequalities. Any pair of boundaries can be connected by a spatial slice, so the quantum states on all $k$ boundaries are correlated by the gravitational constraints.  Gluing this to another windmill for $\langle \Psi_k|$ produces the wormhole geometry in figure \ref{fig:introWormhole}. \label{fig:pinwheel}}
\end{figure}

A GHZ state in a finite-dimensional quantum system is a state with diagonal, $k$-party entanglement, $|\mbox{GHZ}\rangle = \sum_n |n\rangle^{\otimes k} $ \cite{greenberger1989going}. The state in \eqref{psik} has GHZ-like entanglement among the UV microstates $|n\rangle$, but because of the dressing by the operators $V^i$, the IR entanglement can be more general. At this point the reader may object that it is a well known fact that holographic states cannot have GHZ entanglement, because GHZ states violate the entropy inequalities demanded by the Ryu-Takayanagi formula \cite{Hayden:2011ag,Balasubramanian:2014hda,Susskind:2014yaa,Bao:2015bfa}.  However, this logic only applies to CFT states defined on spatial slices in the boundary that can be extended into smooth Cauchy slices in the bulk. In fact, it is possible to study GHZ-like states holographically by exploiting this loophole. The Euclidean bulk region associated to $|\Psi_k\rangle$ is the non-smooth `windmill' geometry illustrated in figure \ref{fig:pinwheel}. 
The intuition for the ansatz \eqref{psik} is that when the boundaries are glued together at the central vertex in this diagram, the Hamiltonian constraint in the bulk enforces a projection onto states that are correlated on all $k$ boundaries.

This interpretation does not require that the wormholes contribute to the gravitational path integral for ordinary $k$-copy observables, $(Z_{\rm cft}[J])^k$, so there is no tension with factorization.

Although there is no ensemble average, there is a close connection to coarse graining. As a matter of terminology, the distinction is that a `coarse-grained' observable is defined in a particular CFT, whereas an `ensemble average' is defined by averaging over CFTs (or CFT data, such as matrix elements) with respect to some measure. The notion of coarse graining in CFT is neither unique nor well understood, so one goal of the paper will be to propose a definition of coarse graining and to develop a replica formalism to calculate the coarse-grained entropy.  The conclusion is that the wormholes above, when taken to be $\mathbb{Z}_k$ symmetric, are the replica wormholes for coarse-grained entropy, and furthermore can be used to find the coarse-grained density matrix associated to a bulk region outside a time-symmetric apparent horizon. 
%EDIT67: Added following
On the CFT side, the replica method for coarse-grained entropy involves a projection onto states with GHZ-like entanglement. The projection is implemented in the bulk by a topological constraint that glues the manifold together into a wormhole.

Similar wormholes have been discussed in 2D gravity \cite{Penington:2019kki,Stanford:2020wkf} (the `West Coast' model) and in 3D gravity \cite{Chandra:2022bqq} (the large-$c$ ensemble), where they were interpreted in terms of ensemble averaging. The current paper provides an alternative interpretation in terms of coarse graining in a single CFT. These two different points of view are entirely compatible --- the same wormholes can compute coarse-grained observables in a single theory as well as ensemble averages. We will show that the higher-dimensional wormholes can also be ascribed an ensemble interpretation, \textit{if} we make the (perhaps implausible) assumption that they are the dominant contributions to the gravitational path integral. Under this assumption the results are parallel to those in \cite{Penington:2019kki}, including Page-like behavior when the black hole is entangled with an external reservoir. 
%EDIT68 removed
%One new ingredient compared to 2D is a curious effect where the island rule can be violated at leading order due to a phase transition.

Our conclusions including \eqref{zoverlap} apply to a variety of different wormholes, including those involving thin shells, EOW branes, and massive probe matter in AdS$_{d+1}$, as well as wormholes in 3D gravity created by heavy local operators. All of these cases are special in that the metric is locally identical to an eternal black hole, or an eternal black with a small perturbation, away from localized matter sources. For this special class of black holes we will also find the explicit coarse-graining map in the dual CFT.

The matter content we consider is generic --- massive particles coupled to Einstein gravity --- so the results appear to be readily embedded into top-down theories, with one important caveat: We will not study the question of stability. Instabilities can affect the interpretation, as in the case of the Schwarzschild black hole in flat spacetime \cite{Gross:1982cv}, or even remove the wormhole contributions entirely, depending on the UV completion. See \cite{Maldacena:2004rf, Arkani-Hamed:2007cpn, Marolf:2021kjc} for other types of Euclidean wormholes in AdS and related discussion of instabilities.

The remainder of this introduction is a summary of how wormholes are used to find the coarse-grained density matrix of a pure-state black hole.

\subsection{Coarse graining and apparent horizons}

\begin{figure}
\begin{center}
\begin{overpic}[grid=false]{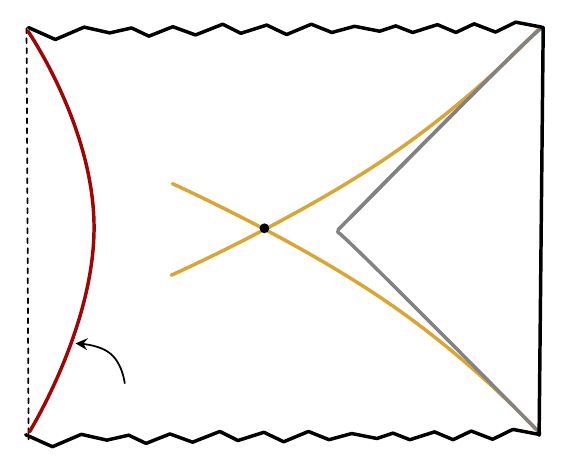}
\put(22,10) {\footnotesize matter}
\put(45,35) {\footnotesize $\gamma$}
\put(55,56) {\footnotesize $\theta_+=0$}
\put(55,22) {\footnotesize $\theta_-=0$}
\end{overpic}
\end{center}
\caption{Penrose diagram of a black hole with a time-symmetric apparent horizon, $\gamma$, at $t=0$ (which is also extremal). 
The corresponding Euclidean geometry is shown in eq. \eqref{psi1overlap} and the $t=0$ spatial slice is in eq. \eqref{spatialslice}. For thin shells and $B$-states, the apparent horizon coincides with the event horizon (gray), but sending in additional matter from the boundary leads to the figure shown. \label{fig:penrosediagram}}
\end{figure}
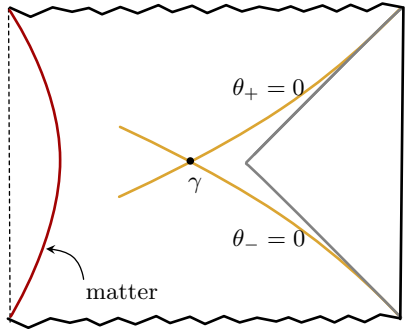

Consider a black hole pure-state $|\Psi_1\rangle$, prepared by a Euclidean path integral, with a time-symmetric apparent horizon. The Euclidean geometry of $|\Psi_1\rangle$ is the saddlepoint that computes the norm of this state, schematically
\begin{align}\label{psi1overlap}
\langle \Psi_1 | \Psi_1 \rangle  = \vcenter{ \hbox{\includegraphics[width=0.9in]{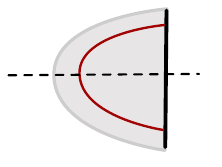} }} \ .
\end{align}
The diagram represents a Euclidean gravitational path integral that prepares the state $|\Psi_1\rangle$ on the dashed line, $t=0$. The black line is the $S^d$ boundary and the red line is a matter source. Specifically, we consider EOW-brane solutions found in \cite{Cooper:2018cmb}, thin shells of pressureless fluid similar to the solutions studied in \cite{Keranen:2015fqa,Anous:2016kss}, and perturbations of these. Assuming the existence of a time-symmetric apparent horizon, the geometry on the $t=0$ slice  looks like this:
\begin{align}\label{spatialslice}
\vcenter{\hbox{
\begin{overpic}[width=1.5in,grid=false]{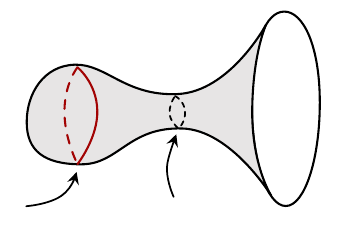}
\put (-20,5) {{\footnotesize matter}}
\put (50,5) {{\footnotesize $\gamma$}}
\end{overpic}
}} \ ,
\end{align}
possibly with additional matter outside the black hole.
The minimal surface $\gamma$ is the apparent horizon, which has vanishing inward and outward-pointing null expansions and is therefore also extremal. The corresponding Lorentzian geometry is shown in the Penrose diagram in fig.~\ref{fig:penrosediagram}. The extremal surface is homotopically trivial, so it is subdominant in the sense of the Ryu-Takayanagi-formula \cite{Ryu:2006bv,Hubeny:2007xt}, and indeed, the state $\rho = |\Psi_1 \rangle \langle \Psi_1|$ is pure so it has vanishing von Neumann entropy.

%%%%%%%%%%%%%%%%%%%%%%%
%%%%%%%%%%%%%%%%%%%%%%%

Although it has zero fine-grained entropy, from the bulk point of view, $|\Psi_1\rangle$ is naturally assigned a coarse-grained entropy equal to one quarter the area of $\gamma$ \cite{Engelhardt:2017aux}. We will show that the wormholes described above, taken to have a $\mathbb{Z}_k$ symmetry permuting the boundaries, are the gravitational replica manifolds that compute this coarse-grained entropy. They are found by inserting a $\frac{2\pi}{k}$ conical defect at the apparent horizon and allowing it to backreact. We can therefore use \eqref{zoverlap} to find the coarse-grained density matrix of $|\Psi_1\rangle$ corresponding to the region outside the apparent horizon, as follows. The $k \to 1$ limit of \eqref{psik} gives a decomposition of the microstate in the form
\begin{align}\label{decomp}
|\Psi_1 \rangle = V \sum_n \psi_n |n\rangle \ ,
\end{align}
where $V$ is a semiclassical isometry. This decomposition of the quantum state is interpreted geometrically as
\begin{align}\label{cutslice}
|\Psi_1\rangle \quad \Leftrightarrow \quad \vcenter{\hbox{
\begin{overpic}[grid=false]{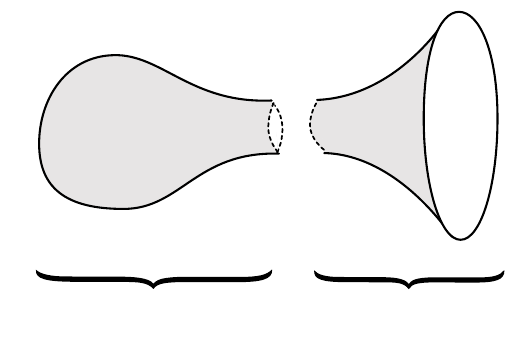}
\put (14, 5) {$\sum_n \psi_n |n\rangle$}
\put (76, 5) {$V$}
\end{overpic}
}}  \ ,
\end{align}
in the spirit of a holographic tensor network \cite{Swingle:2009bg,Yang:2015uoa,Hayden:2016cfa} or quantum error-correcting code \cite{Almheiri:2014lwa,Pastawski:2015qua}.
The expansion coefficients $\psi_n$ define a particular black hole microstate, and the isometry $V$ is associated to the region outside the apparent horizon. 
This is a CFT realization of a holographic code \cite{note:toappear}. Interestingly, the operator $V$ ---  which will be found explicitly as a CFT operator in some examples --- is approximately isometric, up to normalization, if the CFT satisfies eigenstate thermalization (ETH). This draws a direct connection between ETH in the dual CFT and the success of random tensor networks \cite{Hayden:2016cfa} in reproducing aspects of holographic duality. 

Using the decomposition \eqref{decomp} we can now define the coarse-graining map. Given the pure state
\begin{align}
\rho = |\Psi_1 \rangle \langle \Psi_1| = \sum_{m,n} \psi_m \psi^*_n V|m\rangle\langle n|V^\dagger \ ,
\end{align}
define the coarse-grained density matrix
\begin{align}\label{defbrho}
\brho &:= \sum_n |\psi_n|^2 V |n \rangle \langle n| V^\dagger \ .
\end{align}
The general result \eqref{zoverlap}, specialized to have $\mathbb{Z}_k$ symmetry, can be restated as
\begin{align}
Z_{\rm wormhole} = \tr \brho^k \ .
\end{align}
This takes the form of a replica partition function.
Therefore, $\brho$ is interpreted as the coarse-grained density matrix, and applying the gravitational replica method \cite{Lewkowycz:2013nqa}, we have derived a CFT formula for the area of the apparent horizon,
\begin{align}\label{introSAH}
S(\brho)  =  \frac{1}{4G} \mbox{Area($\gamma$)} \ ,
\end{align}
where $S(\brho)$ is the von Neumann entropy (for a normalized state, $S(\sigma) := -\tr \sigma \log \sigma$). 

In other words, wormholes provide the answer to the question: \textit{What CFT quantity does the area of a subdominant extremal surface calculate?} Given a particular black hole, to answer this question, the recipe is to construct the corresponding $k$-boundary wormholes, calculate their action, read off $V$ and $|\psi_n|^2$ from \eqref{zoverlap}-\eqref{psik}, and then define $\brho$ by \eqref{defbrho}. We will do this in several examples and find explicit formulae for $\brho$.

From the CFT point of view, it is very natural to define the coarse-grained density matrix of a pure state by \eqref{defbrho}. Consider the case in which $V = \mathbb{1}$; this holds for pure states in which the geometry outside the horizon is exactly that of the eternal black hole, including spherically symmetric $B$-states and thin shell geometries with all the matter behind the horizon. Then $\brho$ defined in \eqref{defbrho} is the diagonal projection of $\rho$ in the energy basis, and the coarse-grained entropy reduces to
\begin{align}
S(\brho) &=  S_{\rm diag}(\rho) := - \sum_n (\rho)_{nn} \log (\rho)_{nn} \ .
\end{align}
The right-hand side is a quantity known as the diagonal entropy, introduced by Barankov and Polkovnikov \cite{Barankov:2008qq}, and it is a common definition of coarse-grained entropy used to study pure states in chaotic quantum many-body systems out of equilibrium. In simple quantum mechanical models, the diagonal entropy has several nice properties: It agrees with the von Neumann entropy in stationary states, it increases under stochastic evolution, and in chaotic systems obeying the eigenstate thermalization hypothesis, it obeys a thermodynamic first law \cite{Barankov:2008qq,DAlessio:2015qtq}.\footnote{That there are similarities between gravity and a `diagonal approximation' is already well known; see especially \cite{Cotler:2016fpe}, as well as \cite{Roy:2015pga,Sonner:2017hxc,Hunter-Jones:2017raw,Saad:2019lba,Penington:2019kki,Pollack:2020gfa,Marolf:2020vsi,Altland:2021rqn,Saad:2021uzi,Freivogel:2021ivu,Chandra:2022bqq,Cotler:2022rud}.}

More generally, when $V$ is a nontrivial operator, the coarse-grained density matrix defined by \eqref{defbrho} is not diagonal. The coarse graining map decoheres the black hole microstates while retaining the infrared quantum correlations associated to certain single-trace excitations, or in bulk language, matter outside the apparent horizon. This is a natural notion of coarse graining for a low-energy observer at the boundary.

We have described a \textit{gravity} procedure to find the coarse-grained density matrix $\brho$, but we have not defined it purely in CFT terms. Without using wormholes, or any other input from the gravity side, what is the definition of $\brho$? We suggest, tentatively, that large-$N$ holographic CFTs are naturally equipped with a quantum channel $\C$ such that $\brho = \C(\rho)$. In the examples that we will discuss, the coarse-graining map does act like a quantum channel, but there is no general argument for this; see the discussion section.

\subsection{Comments on the literature}

Engelhardt and Wall proposed \cite{Engelhardt:2017aux,Engelhardt:2018kcs} that the coarse-grained entropy associated to an apparent horizon is the `simple entropy', defined by maximizing over density matrices subject to holding fixed a class of simple observables (\textit{i.e.}, one-point functions with time-ordered sources). There is convincing evidence for this proposal under the assumption that the optimal density matrix is achieved by a classical Lorentzian geometry \cite{Engelhardt:2017aux,Engelhardt:2018kcs,Engelhardt:2021mue}. Our definition of coarse graining is different, but there is no conflict. We will find the explicit coarse-grained density matrix $\brho$ in the CFT, as opposed to defining it implicitly by a maximization procedure. On the other hand, our method in its current form is limited to certain black holes.

This also connects to recent discussions of complexity and the `python's lunch' \cite{Brown:2019rox,Engelhardt:2021mue}. The spatial slice depicted in \eqref{spatialslice} is an example of a python, and the region enclosed by the extremal surface $\gamma$ is the `lunch'. Pythons are related to regions of the bulk that are very complex to reconstruct from the boundary, and the area of the extremal surface quantifies this complexity \cite{Brown:2019rox,Engelhardt:2021mue}.  Therefore our results can also be interpreted as a way to define and calculate the complexity of reconstruction. Other than this qualitative similarity, we will not make a direct connection to the information-theoretic results in \cite{Brown:2019rox,Engelhardt:2021mue,Engelhardt:2021qjs}, but it would be very interesting to explore this further.

Other perspectives on the relation between black hole interiors and coarse graining appear in the recent papers \cite{Renner:2021qbe,Almheiri:2021jwq,Qi:2021oni}. Our approach has some similarities, in particular to the idea in \cite{Almheiri:2021jwq,Qi:2021oni} that black hole interiors can be removed by classical measurements on a pointer system that cause the microstates to decohere. In the simplest examples, with $V = \id$, the coarse-graining map discussed in section \ref{s:replicaformalism} is the completely dephasing channel, i.e., total decoherence in the energy basis.

\subsection{Plan}

In section \ref{s:wormholes} we describe the wormhole solutions on the gravity side. The construction starts with single-boundary, pure-state black holes, and proceeds by adding conical defects and backreacting the geometry to produce $\mathbb{Z}_k$-symmetric multiboundary wormholes. We also discuss how the $k \to 1$ limit matches onto the replica geometries of Lewkowycz and Maldacena \cite{Lewkowycz:2013nqa} so that the thermodynamics of the wormholes is related to the apparent horizon entropy of the single-boundary black holes. 

In section \ref{s:replicaformalism} we develop a replica formalism for coarse-grained density matrices and the calculation of coarse-grained entropy. This is phrased in CFT language but for the most part can be applied to any quantum mechanical system.

In section \ref{s:cftdual} we find the CFT dual of the wormholes. This is used to infer the coarse-grained density matrix, $\brho$, associated to the region outside the horizon of a single-boundary pure-state black hole.

Up to to this point we have assumed spherically symmetric, unperturbed black holes, which have $V=\id$. In section \ref{s:addmatter} we add massive particles to linear order. This introduces a new conceptual ingredient: the coarse-graining map is no longer a diagonal projection, but a more intricate operation, with $V \neq \id$. Again there is a match between gravity and CFT.

In section \ref{s:ensemble}, we discuss the ensemble interpretation of the wormholes. This is the only section that makes use of an ensemble average. The analysis is quite similar to the ensemble interpretation of EOW branes in JT gravity in \cite{Penington:2019kki}, except that we discuss only the classical solutions rather than doing an off-shell path integral.

In the discussion section, we summarize the 3-step process used to infer the coarse-grained density matrix of a region outside an extremal surface, and discuss open questions and limitations of this approach.

The four appendices have details of the gravity calculations and a toy model for holographic coarse graining with a quantum channel.

\subsubsection*{3D Gravity}
The coarse graining procedure can also be applied to a wide class of states in AdS$_3$/CFT$_2$, including many of the solutions studied in \cite{Chandra:2022bqq} which motivated this work. The 3D thin shells that we will discuss can be viewed as $n$-point correlation functions, like those studied in \cite{Chandra:2022bqq}, in the regime where $n$ is of order the central charge. The CFT techniques in $d=2$ are more powerful due to Virasoro symmetry, but also rather technical, so this application will be described in a separate paper.

\subsubsection*{Readers' guide}
For a first read through this paper we recommend the following path: See figure \ref{fig:quotientwormholes} to understand how the wormhole solutions are constructed; peruse section \ref{s:replicaformalism} for the information-theoretic basis for coarse graining; browse the introductory parts of sections \ref{s:cftdual}, \ref{s:addmatter}, and \ref{s:ensemble} for a summary of the CFT calculations; and, lastly, read the summary in the discussion section.

\section{Wormhole solutions}\label{s:wormholes}

We will consider two examples of $k$-boundary wormholes:  $B$-states, which are supported by EOW branes, and thin shells. We work in AdS$_{d+1}$ with the canonical metric (flat or round) on the boundaries. 
We assume spherical symmetry, so outside the thin shell, or away from the EOW brane, these solutions are locally identical to eternal black holes. (Extra matter in the bulk will be added in section \ref{s:addmatter}.) The metric of the eternal black hole is AdS-Schwarzschild,
\begin{align}\label{ds2bh}
ds^2_{BH} &= f(r) d\tau^2 + \frac{dr^2}{f(r)} + r^2 d\Omega_{d-1}^2 \\
f(r) &= 1 + r^2  - \left( \frac{r_H}{r} \right)^{d-2}(1+r_H^2) \ ,
\end{align}
where $\tau \sim \tau + \beta$ with the inverse temperature
\begin{align}
\beta = \frac{4\pi r_H}{dr_H^2+d-2} \ .
\end{align}

\subsection{Black hole pure states}
Let us start with the single-boundary solutions, $k=1$. These are black hole pure states with matter behind the horizon. The matter, which is either a thin shell of pressureless fluid or an EOW brane, is spherically symmetric and follows a trajectory $\tau = u(r)$. The main difference between $B$-state black holes and thin shell black holes is the equation of motion for $u(r)$. We will consider each of these in turn.

\subsubsection{$B$-state black holes}\label{ss:bstate}

For our purposes, an EOW brane is just a particular type of matter. Quantum states in the boundary CFT will be defined on slices that avoid the branes, so despite the boundary in Euclidean signature, the Lorentzian theory is not a BCFT --- the brane defines an excited state in an ordinary CFT. 

\begin{figure}
\begin{center}
\begin{overpic}[grid=false]{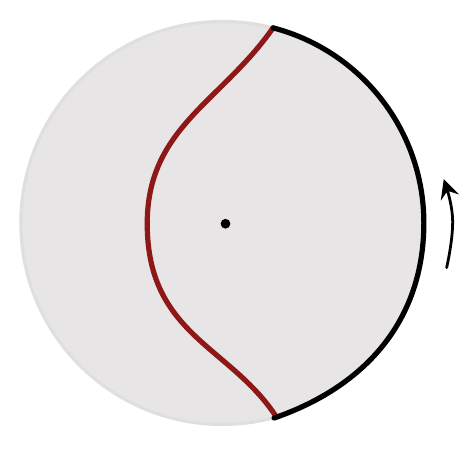}
\put (97,45) {$\tau$}
\put (55,92) {$\tau_0$}
\put (52,0) {$-\tau_0$}
\end{overpic}
\end{center}
\caption{Euclidean $B$-state black hole. The physical region, to the right in the figure, is a portion of the eternal black hole bounded by the asymptotic AdS boundary (black) and the EOW brane (red). Positive tension branes lead to solutions that cover at most half the thermal circle  at the boundary. \label{fig:bstate}}
\end{figure}

Bulk solutions with an EOW brane have been studied extensively \cite{Karch:2000ct,Takayanagi:2011zk,Fujita:2011fp,Hartman:2013qma,Kourkoulou:2017zaj,Cooper:2018cmb,Chen:2020uac,Chen:2020hmv,Miyaji:2021ktr,Akal:2021foz,Suzuki:2022yru,Izumi:2022opi,Rozali:2019day}. 
We will consider the black hole solution found in \cite{Cooper:2018cmb} (see also \cite{Hartman:2013qma} for a discussion of the tensionless case). 
The geometry, shown in figure \ref{fig:bstate}, is a portion of the eternal AdS black hole that terminates on the EOW brane. The gravitational action, including the brane, is
{\small
\begin{align}\label{eowaction}
    S=-\frac{1}{16\pi G}\int_{\text{bulk}}d^{d+1}x \sqrt{g}(R-2\Lambda)-\frac{1}{8\pi G}\int_{\text{brane}}d^dy \sqrt{h}(K-(d-1)T)-\frac{1}{8\pi G}\int_{\text{bdry}}d^dy \sqrt{h}K
\end{align}
}(plus counterterms). We set $\Lambda=-\frac{d(d-1)}{2}$. The brane tension $T>0$ is a free parameter that controls how far the brane is behind the horizon; positive tension branes are never outside the horizon at $t=0$. Assuming the geometry away from the brane is AdS-Schwarzschild, this action leads to an equation of motion that describes how the brane is embedded into the eternal black hole. Denote the brane trajectory by $\tau = \pm u_B(r)$;  we take $\tau \in (-\frac{\beta}{2}, \frac{\beta}{2})$, and the physical region is $|\tau| \leq \tau_B(r)$. The solution to the equation of motion, reviewed in appendix \ref{app:bstatedetails} along with various other details of the $B$-state geometries, is \cite{Cooper:2018cmb}
\begin{align}\label{uBsolution}
u_B(r') &= \frac{\beta}{2} - \int_{r_0}^{r'} \frac{dr}{f} \frac{Tr}{\sqrt{f-T^2r^2}} \ ,
\end{align}
where $r_0$ is the brane turning point, which is determined by $f(r_0) = T^2 r_0^2$ and has been placed at $\tau = \frac{\beta}{2}$. The brane meets the AdS boundary at $\tau = \pm \tau_0$, with
\begin{align}\label{tau0b}
\tau_0 &=u_B(\infty) \ .
\end{align}
$\tau_0$ is a complicated function of the temperature and brane tension, but it is easily plotted numerically. The allowed range for the tension is $T \in [0, T_{max})$ where the upper bound depends on $d$, and comes from requiring the brane to hit the boundary before intersecting with itself.
With the tension in the allowed range, the EOW brane endpoint is found to satisfy
\begin{align}
0 < \tau_0 \leq \frac{\beta}{4} \ ,
\end{align}
with the upper bound saturated when $T=0$. 
Thus the asymptotic boundary of the $B$-state geometry, $\tau \in (-\tau_0, \tau_0)$, covers at most half of the thermal circle of the eternal black hole, as in figure \ref{fig:bstate}. 

Although we have phrased the calculation as fixing $\beta$ and computing $\tau_0(\beta)$, it is really $\tau_0$ that should be viewed as the independent parameter that defines the quantum state, since $\tau_0$ labels the boundary condition in the Euclidean path integral. The relation $\tau_0 = u_B(\infty)$ is then an equation for the temperature $\beta(\tau_0)$ of the black hole produced by the backreaction of a brane inserted at Euclidean time $\tau_0$.

\subsubsection{Thin shell black holes}\label{ss:thinshell}
Thin shell black holes are constructed by gluing a patch of vacuum AdS to a patch of the eternal black hole across a codimension-1 shell of matter. For simplicity we take the matter to be a pressureless perfect fluid, but we expect similar wormholes to exist for other types of matter. Thin shell solutions in Lorentzian signature, with matter outside the horizon at $t=0$, were studied in \cite{Keranen:2015fqa}. We are interested in Euclidean geometries where the matter is behind the horizon at $t=0$ so the solutions are a bit different, but we will follow the method in \cite{Keranen:2015fqa} closely. Details of the gravity calculation are in appendix \ref{app:thinshelldetails}.

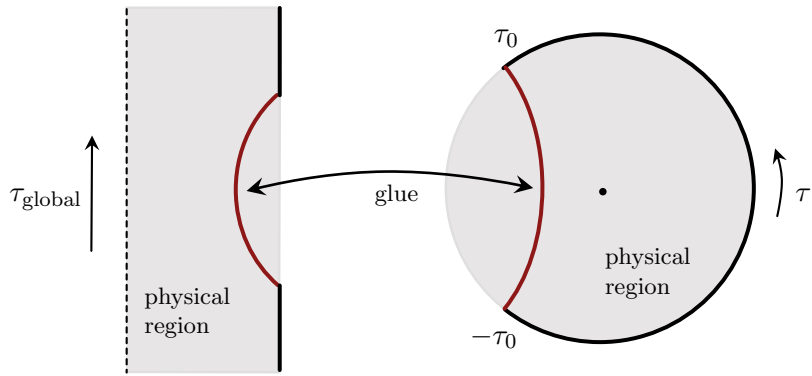
\begin{figure}
\begin{center}
\begin{overpic}[grid=false]{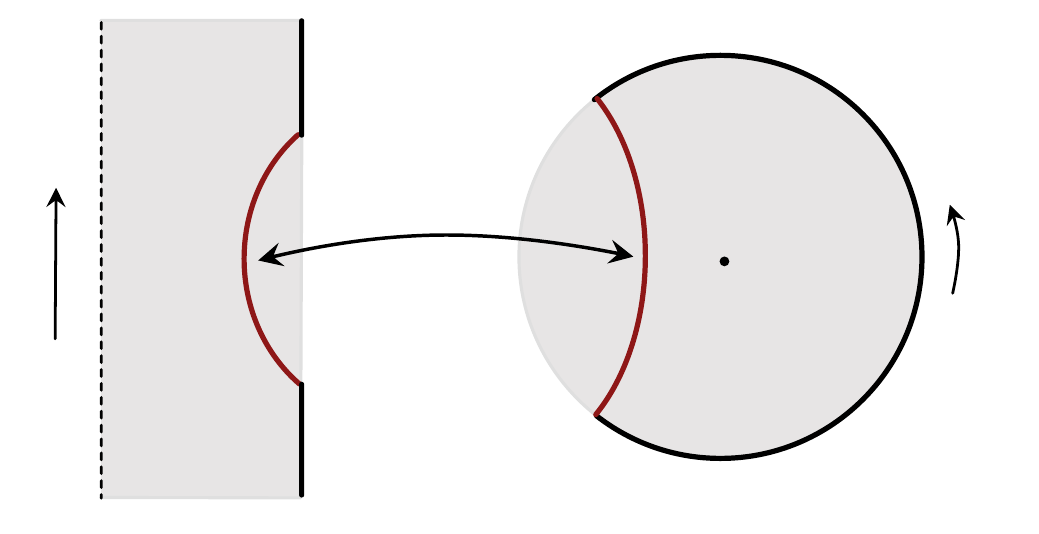}
\put (94, 25) {$\tau$}
\put (-5,25) {$\tau_{\rm \tiny global}$}
\put (41,25) {\footnotesize glue}
\put (56,45) {$\tau_0$}
\put (53,7) {$-\tau_0$}
\put (12,10) {\parbox{1in}{\footnotesize{ physical \\ region}}}
\put (70,15) {\parbox{1in}{\footnotesize{physical \\ region}}}
\end{overpic}
\end{center}
\caption{Euclidean thin-shell black hole, obtained by gluing a piece of global AdS (left) to a piece of the eternal black hole (right) across a thin shell (red). Each point on the diagram is $S^{d-1}$, except for the dashed line, which is the center of Euclidean global AdS. \label{fig:thinshell}}
\end{figure}

To construct the spherical thin shell black hole, we glue the region $|\tau|< u_S(r)$ of the Schwarzschild-AdS solution \eqref{ds2bh} to a portion of global AdS. The Israel junction conditions lead to an equation of motion for the shell trajectory, $u_S(r)$. The 
solution to this equation of motion is given in \eqref{uSsolution}.

From this description it is clear that thin shells and $B$-states are quite similar, but with different trajectories for the matter. There is one important difference. For $B$-states, we found that the brane hits the boundary at $\tau_0 \leq \frac{\beta}{4}$. For thin shells, there is no such restriction --- the shell endpoint,
\begin{align}\label{tau0s}
\tau_0 = u_S(\infty) \ ,
\end{align}
can land anywhere on the thermal circle for some choice of the parameters $\beta, r_0$ that define the shell.

\subsection{$k$-boundary wormholes}\label{ss:quotientwormholes}

We now seek $k$-boundary wormholes with cyclic symmetry, treating $B$-states and thin shells simultaneously. An example of a thin shell solution with $k=6$ is shown in the introduction in figure \ref{fig:introWormhole}.
The geometry is $(d+1)$-dimensional and spherically symmetric; the figure shows the $(r,\tau)$ directions. The boundary condition is that on each asymptotic boundary, the matter (thin shell or EOW brane) is offset into Euclidean time by $\pm \tau_0$, where $\tau_0$ is the landing point found in \eqref{tau0b} for branes and \eqref{tau0s} for thin shells. This boundary condition is chosen to match $k$ copies of the boundary condition of the $k=1$ solution. 

\begin{figure}
\begin{center}
\begin{overpic}[grid=false]{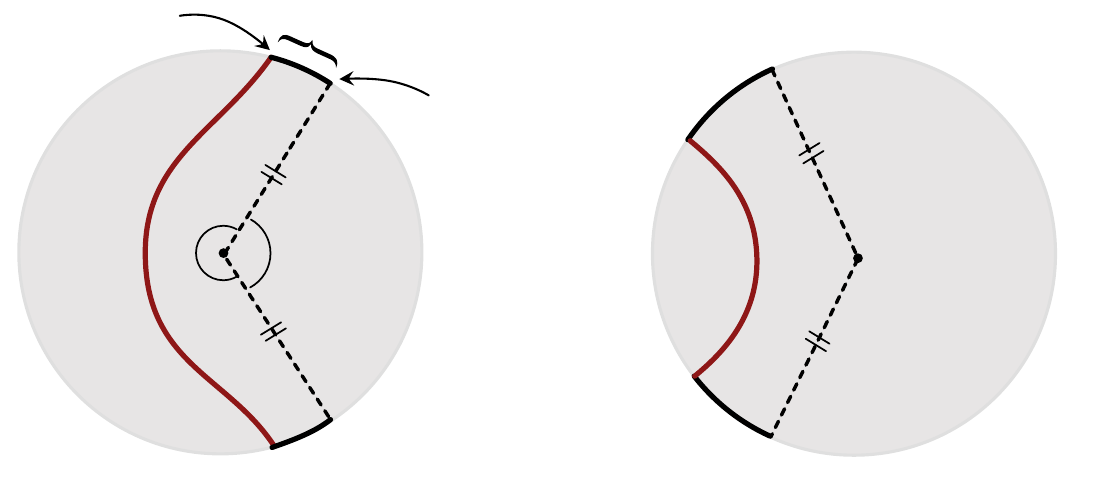}
\put (38,36) { $\frac{\theta_k \beta_k}{4\pi}$}
\put (29,43) {$\tau_0$}
\put (11.5,43) {$\tau_{0k}$}
\put (26,21) {$\theta_k$}
\put (14.4,21) {$\frac{2\pi}{k}$}
\put (19,-2) {$(a)$}
\put (78,-2) {$(b)$}
\end{overpic}
\end{center}
\caption{Wormholes constructed from $(a)$ $B$-states, and $(b)$ thin shells. Each diagram shows a single sheet of the wormhole, i.e., a quotient geometry ${\cal M}_k/\mathbb{Z}_k$ where ${\cal M}_k$ is the smooth wormhole. The metric is that of the eternal black hole at temperature $\beta_k$. In $(a)$ the red curve is an EOW brane and this is a fractional wormhole, with $k = 1.5$ boundaries. In $(b)$ the red curve is a thin shell where the spacetime is glued to global AdS, and this is a wormhole with $k=3$ boundaries. The full 3-boundary solution for $(b)$ is obtained by cutting the diagram along the horizontal line from the center to the shell and gluing together 3 copies across this cut.\label{fig:quotientwormholes}}
\end{figure}
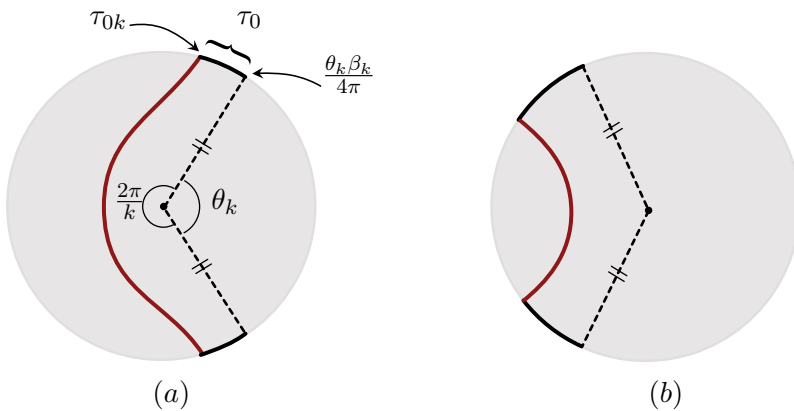

These wormholes are constructed by assuming a $\mathbb{Z}_k$ cyclic symmetry, taking the $\mathbb{Z}_k$ quotient, and then allowing $k$ to be non-integer. This means looking for solutions with a single boundary, with matter sources separated by Euclidean time $2\tau_0$, and a spherically symmetric conical defect in the interior with angle $\frac{2\pi}{k}$. The solutions with these properties are shown in figure \ref{fig:quotientwormholes}. They are obtained by starting with a single-boundary black hole (a $B$-state or thin shell geometry described in the previous section) at a different temperature, $\beta_k$, with matter endpoint $\tau_{0k}$. Then excise the wedge $\tau \in (-\beta_k \frac{\theta_k}{4\pi}, \beta_k \frac{\theta_k}{4\pi})$ with $\theta_k = 2\pi(1 - \frac{1}{k})$ by gluing its edges together. The remaining boundary has length $2(\tau_{0k}-\beta_k \frac{\theta_k}{4\pi})$, and setting this equal to $2\tau_0$ requires
\begin{align}\label{bbc1}
\tau_{0k} = \tau_0 + \frac{\beta_k}{2}(1 - \frac{1}{k}) \ .
\end{align}
Let us write the matter endpoint as a function of temperature $\tau_0 = G(\beta)$. $G$ also depends implicitly on whether the matter is an EOW brane or thin shell, and its tension or mass. Then the boundary condition \eqref{bbc1} is
\begin{align}\label{bbc2}
G(\beta_k) = G(\beta) + \frac{\beta_k}{2}(1 - \frac{1}{k}) \ .
\end{align}
This should be read as an implicit equation for $\beta_k$ --- it tells us the temperature $\beta_k$ of the black hole that can be used to construct a $k$-fold replica of the pure-state black hole at temperature $\beta$.

The construction only makes sense for $\frac{\theta_k}{4\pi} \beta_k < \tau_{0k}$, because otherwise the excised wedge would hit the matter. This places an upper bound on $k$ for a given $\tau_0$. For example, consider the tensionless EOW brane, which has $\tau_0 = \frac{\beta}{4}$ and $\tau_{0k} = \frac{\beta_k}{4}$. Then $\frac{\theta_k}{2\pi} < \frac{1}{4}$ requires
\begin{align}
k < 2 \qquad (\mbox{tensionless $B$-state}) \ .
\end{align}
Turning on a positive tension leads to $\frac{\tau_{0k}}{\beta_k}< \frac{1}{4}$, since the $B$-state never covers more than half of the boundary of the eternal black hole. Therefore, in general, $B$-state wormholes have
\begin{align}
k < k_{max}(d,T) , \quad k_{max}(d,T) \leq 2 \qquad (\mbox{$B$-state with tension}) \ .
\end{align}
When $k$ is non-integer, we refer to these solutions as `fractional wormholes.' Conical excesses with $k \in (0,1)$ also make sense as fractional wormholes.

Thin shells, unlike $B$-states, can have $\tau_0 > \frac{\beta}{4}$. This allows for on-shell wormholes with integer $k \geq 2$. For any given matter shell, there is still an upper bound $k_{max}$ on the number of boundaries, but shells exist with arbitrarily large $k_{max}$. In particular, the $k=6$ wormhole in figure \ref{fig:introWormhole} is an actual solution to the equations of motion.

Since there are on-shell geometries with multiple boundaries, it is also possible to study thin shell wormholes without a cyclic symmetry, for example by choosing different $\tau_0$'s on each boundary, or several different shell masses.

It is an important question whether the wormholes are stable, as this can affect their interpretation. We will not answer this question here, but remark that similar multiboundary wormholes in AdS are often unstable perturbatively and/or under brane nucleation \cite{Maldacena:2004rf,Marolf:2021kjc}. These instabilities tend to be more severe for larger $k$ and could influence the value of $k_{max}$. For $k$ near 1, our wormholes are very similar to those in \cite{Lewkowycz:2013nqa} which are usually assumed to be legitimate contributions to the path integral.

\subsection{Action}\label{ss:action}
Denote the on-shell action of the eternal black hole by $I_0(\beta)$, so that by the usual AdS/CFT dictionary we have
\begin{align}
I_0(\beta) = -\log Z(\beta) 
\end{align}
with $Z$ the thermal partition function (if the temperature is above the Hawking-Page transition). Let $\tau = u(r)$ be the matter trajectory, found in \eqref{uBsolution} for the EOW brane and \eqref{uSsolution} for the thin shell, and $\tau_0 = u(\infty)$. Denote the action of the single-boundary, pure-state black hole by $I(\beta)$ and let us separate this into two parts,
\begin{align}\label{singleI}
I(\beta) &= \frac{2\tau_0}{\beta}I_0(\beta) + I_{L}(\beta) \ .
\end{align}
The first term is the contribution from $|\tau| < \tau_0$, and $I_{L}$ is the action of the remaining portion $|\tau|\geq \tau_0$, including the matter. (An explicit formula for $I_L$ in $B$-states is derived in appendix \ref{app:bstatedetails}.) Then it follows immediately from the quotient construction in section \ref{ss:quotientwormholes} that the action of a $k$-boundary wormhole is
\begin{align}\label{Ik}
I_k(\beta) &= \frac{2\tau_0}{\beta_k}k I_0(\beta_k) + k I_{L}(\beta_k) \ .
\end{align}
In this equation, $\beta_k$ is defined implicitly as the solution to the boundary condition \eqref{bbc2} with $\tau_0 = G(\beta)$. Using \eqref{bbc2} we can also write the action of a single sheet as
\begin{align}
I_k(\beta)/k &=  I_1(\beta_k)  + ( \frac{1}{k}-1) I_0(\beta_k) \ .
\end{align}
Another useful relation is 
\begin{align}\label{energyrel}
 I_0'(\beta_k) = E(\beta_k)  = \frac{1}{2 k} \p_{\tau_0} I_k(\beta) \ .
\end{align}
Here $E$ is the ADM mass of a single boundary, so the first equality is the usual thermodynamic relation for the eternal black hole. The second equality is the Hamilton-Jacobi equation for the wormhole, i.e., the statement that the Hamiltonian is the generator of time translations; moving the matter insertion point is equivalent to evolving in Euclidean time, and the factor of $\frac{1}{2}$ is because $\p_{\tau_0}$ affects both endpoints of the matter. For $B$-states, we have also checked \eqref{energyrel} by explicit calculation of the regulated on-shell action as described in appendix \ref{app:bstatedetails}, but this is quite involved.

\subsection{Apparent horizon entropy}\label{s:bulkentropy}
Let 
\begin{align}\label{defS0}
S_0(\beta) &= (-1+\beta \p_\beta)I_0(\beta)
\end{align}
be the entropy of the eternal black hole. This is clearly equal to one quarter the area of the apparent horizon of the $B$-state or thin shell black hole, 
\begin{align}
S_0(\beta) = \frac{1}{4}\mbox{Area}(\gamma) \ , 
\end{align}
where $\gamma$ is the time-symmetric apparent horizon at $t=0$.\footnote{For the unperturbed black holes discussed in this section, the $t=0$ apparent horizon also happens to be the bifurcation point of the event horizon. However it is the apparent horizon, not the event horizon, that is important --- this is clear from the application of the gravitational replica method, which requires an extremal surface, and when we add a massive particle outside the black hole in section \ref{s:addmatter} the apparent horizon and event horizon no longer coincide.} This follows from the fact that these solutions are locally identical to the eternal black hole and the matter is behind the horizon. From the quotient construction of the wormhole, shown in figure \ref{fig:quotientwormholes}, we see that the apparent horizon of the single-boundary black hole is the fixed point of the $\mathbb{Z}_k$ symmetry of the wormhole as $k \to 1$. Therefore, by the gravitational replica method of Lewkowycz and Maldacena \cite{Lewkowycz:2013nqa}, the apparent horizon entropy can also be calculated by the formula
\begin{align}\label{LMrel}
S_0(\beta) =  \left. \p_k \left( \frac{I_k(\beta)}{k} \right) \right|_{k=1}\ .
\end{align}
In the context of the Ryu-Takayanagi formula and its usual (fine-grained) generalizations, this equation would be interpreted microscopically as a von Neumann entropy, $S = -\tr \hat{\rho} \log \hat{\rho}$ with  $\tr (\rho^k) = e^{-I_k}$ and $\hat{\rho} = \rho  / \tr \rho$. In the present context, this cannot be the correct interpretation, because the $B$-state and thin shell black holes are pure states --- the von Neumann entropy vanishes, and indeed, there is a trivial extremal surface (the empty set) that satisfies the homology condition and gives zero von Neumann entropy. We will return to the interpretation below, but for now, we will make use of \eqref{LMrel} as a purely gravitational statement to derive a useful identity. By plugging the wormhole action \eqref{Ik} into \eqref{LMrel} and comparing to \eqref{defS0}, we find 
\begin{align}
I_{L}'(\beta) &=  \frac{1}{\p_k \beta_k|_{k=1}}S_0(\beta) - \frac{2\tau_0}{\beta^2} S_0(\beta) \ . 
\end{align}
The boundary condition \eqref{bbc2} implies $\p_k \beta_k|_{k=1} = \frac{\beta}{2\tau_0'(\beta)}$ and this gives the relation
\begin{align}\label{Iinprime}
I_{L}'(\beta) &= 2S_0(\beta) \p_\beta( \frac{\tau_0}{\beta} ) \ .
\end{align}
This identity, which applies to both $B$-state black holes and thin shells, is in fact equivalent to \eqref{energyrel}. This is straightforward to check by acting with the $\tau_0$-derivative and using the boundary condition. \eqref{Iinprime} is useful for the explicit calculation of the action, $I_L$, as discussed in appendix \ref{app:bstatedetails}. In the planar limit, $\tau_0 \propto \beta$ and therefore \eqref{Iinprime} implies $I_L' = 0$. 

In the introduction, we claimed that $k$-boundary wormholes of the type studied in this paper are generic in any theory of gravity. We can now explain this remark. For $k = 1 + \epsilon$, with $\epsilon \ll 1$, our wormholes are simply the gravitational replicas of Lewkowycz and Maldacena, for the case of a time-symmetric apparent horizon. These are always solutions to the equations of motion, because we can add a conical deficit at the apparent horizon and follow the logic of \cite{Lewkowycz:2013nqa}. As $k$ is increased above one, we may reach a point $k_{max}$ where the solution no longer exists, depending on the details of the black hole under consideration. This is exactly what we have seen in the examples above.\footnote{It is an interesting question whether the breakdown of the replica method at large enough $k$ is generic, and whether it has physical consequences. Strictly speaking, the analytic continuation in $k$ is ambiguous if we cannot control the behavior of $I_k$ at large $k$. Similar comments apply to the Ryu-Takayanagi formula. We will ignore this issue and define $I_k$ by the `obvious' analytic continuation. 
}

\subsection{Special cases}\label{ss:specialcases}

The general expressions above are sufficient for the comparison to CFT, but it is worth mentioning some special cases where the action can be written in closed form.

\subsubsection*{The planar limit}
In the planar limit, $r_H \to \infty$, $\beta \to 0$, the black hole becomes a black brane, with a flat metric on the boundary $\mathbb{R}^{d}$. This leads to two nice simplifications. First, the temperature scales out of the metric, so that the brane or shell endpoint is proportional to $\beta$, and we may write it as 
\begin{align}
\tau_0 = \frac{\beta}{2}(1-\alpha) 
\end{align}
where $\alpha$ depends only on the brane tension or shell energy density. An explicit formula for $\alpha$ in $B$-states is given in \eqref{FT}. The boundary condition \eqref{bbc2} may now be solved explicitly to find
\begin{align}\label{betakplanar}
\beta_k = \frac{2k \tau_0}{1-k \alpha } \ .
\end{align}
The upper bound on the number of boundaries is $k_{max} = \frac{1}{\alpha}$. For tensionless EOW branes, $\alpha(T) = \frac{1}{2}$, so $\beta_k = \frac{\beta}{\frac{2}{k}-1} = \frac{4k \tau_0}{2-k}$ and $k_{max}=2$.

The second simplification is that $B$-states in the planar limit have $I_L = 0$; see appendix \ref{app:bstatedetails}. The contribution from the EOW brane exactly cancels the bulk action from the left wedge, $|\tau|> \tau_0$. Therefore the action of the $k$-boundary planar $B$-state wormhole is simply
\begin{align} \label{BBaction}
I_k  = \frac{2\tau_0}{\beta_k} k I_0(\beta_k)
\end{align}
with $I_0$ the action of the eternal black hole and $\beta_k$ given by \eqref{betakplanar}. 

For planar thin shells, the identity \eqref{Iinprime} implies $I_L'(\beta) = 0$, so the final result is similar:
\begin{align}
I_k = k(\frac{2\tau_0}{\beta_k}  I_0(\beta_k)+\mbox{const.})
\end{align}
We have not calculated the constant for a thin shell, but in any case it can be set to zero by a choice of normalization for the dual operator. 

\subsubsection*{$B$-states in 3D gravity}
For spherical or planar $B$-states in $d=2$, the action simplifies, since the endpoint is always $\tau_0 = \frac{\beta}{4}$, the boundary condition \eqref{bbc2} implies $\beta_k=\frac{\beta}{\frac{2}{k}-1}$, and as shown in appendix \ref{app:bstatedetails}, $I_L = 0$. Therefore, using the eternal BTZ action quoted in \eqref{d2action}, the action of the $d=2$ $B$-state wormhole is
\begin{equation}
    I_k =-k\left(\frac{2}{k}-1\right)^2\frac{c\pi^2}{24\tau_0} \ .
\end{equation}

\section{Replica formalism for coarse-grained entropy}\label{s:replicaformalism}

In this section we discuss the general formalism for the replica calculation of coarse-grained entropy.  This leads to a definition of coarse graining for holographic CFTs that we will later match to wormholes.

\renewcommand{\D}{\mathcal{D}}

\subsection{Warm-up: Diagonal and block-diagonal entropy}\label{ss:warmup}

We begin by working through a quantum-mechanical toy model where the coarse graining map is a projection onto diagonal or block-diagonal density matrices in the energy eigenbasis. The holographic coarse graining map is generally not of this form, but it is for states where the metric is that of the eternal black hole outside the apparent horizon, including $B$-states and thin shells.

\subsubsection*{Diagonal projection}
Let us first consider the diagonal projection. Define a coarse graining map $\D$ by
\begin{align}
\brho = \D(\rho) &:= \sum_n P_n \rho P_n \ , 
\end{align}
where $P_n = |n\rangle \langle n|$ is the projector onto energy eigenstate $|n\rangle$. In quantum information theory, $\D$ is referred to as the completely dephasing channel. The associated coarse-grained entropy is the diagonal entropy \cite{Barankov:2008qq}, 
\begin{align}
S_{\rm diag}(\rho) = S(\brho)  = -\sum_n (\rho)_{nn} \log (\rho)_{nn} \ , 
\end{align}
where $S(\brho) = - \tr \brho \log \brho$ is the von Neumann entropy. The coarse-grained entropy is bounded below by the fine-grained entropy, $S(\rho)$. There is a standard proof of this fact using the positivity of relative entropy:
\begin{align}\label{entropyIncrease}
0 \leq S(\rho | \D(\rho) )  &= \tr \rho \log \rho - \tr \rho \log \D(\rho) \\
&= \tr \rho \log \rho - \tr \D(\rho) \log \D(\rho) \notag \\
&= -S(\rho) + S(\D(\rho)) \ . \notag
\end{align}
Physically this makes sense because the diagonal projection can be implemented by performing a projective measurement and discarding the result; this cannot decrease the entropy.

To apply the replica method, define the replica partition function
\begin{align}
Z(k) &:= \tr (\brho^k) = \sum_n \langle n| \rho | n \rangle^k  \ . 
\end{align}
The coarse-grained entropy is 
\begin{align}\label{cgderiv}
S(\brho) &= -\left. \p_k \log \left(\frac{Z(k)}{Z(1)^k}\right)  \right|_{k=1} \ .
\end{align}
Now suppose the original state is pure, 
\begin{align}
\rho_\psi = |\psi\rangle \langle \psi| , \quad
|\psi\rangle = \sum_n \psi_n |n\rangle , \quad
\sum_n |\psi_n|^2 = 1 \ .
\end{align}
Then we define the replica state
\begin{align}
|\psi_k\rangle &= \sum_n (\psi_n)^k |n\rangle^{\otimes k}  \ ,
\end{align}
which lives in $k$ copies of the Hilbert space. We refer to such states as `GHZ-like' because they have a diagonal pattern of entanglement. The replica partition function for $\rho_\psi$ is the norm of this state,
\begin{align}
 Z(k) &= \langle \psi_k | \psi_k\rangle  = \sum_n |\psi_n|^{2k}  \ ,
\end{align}
and the diagonal entropy is
\begin{align}
S(\brho) &= - \sum_n |\psi_n|^2 \log |\psi_n|^2 \ .
\end{align}

\subsubsection*{Block-diagonal projection}
These results are easily generalized to a coarse-graining map defined by a projection onto block-diagonal matrices in some basis. Given a collection of projectors $P_i$, satisfying $P_i^2=P_i$ and $\sum_i P_i=\id$, define a coarse-graining map $\N$ by 
\begin{align}
\brho = \N(\rho) &:= \sum_i P_i \rho P_i   \ . 
\end{align}
This takes the form of a Choi-Kraus decomposition and is therefore a quantum channel --- a completely positive trace-preserving linear map. It is sometimes called a pinching channel. We assume the $P_i$ commute with the Hamiltonian so that $\brho$ is block-diagonal in the energy basis, with blocks labeled by $i$.

Once again, the coarse-grained entropy $S(\brho)$ is greater than or equal to the fine-grained entropy, $S(\rho)$. The proof is identical to \eqref{entropyIncrease}, using the fact that $\log \N(\rho)$ is also block-diagonal, and the same physical intuition applies here: a block-diagonal projection can be implemented by performing a measurement and throwing away the result, so it cannot decrease the entropy.

The replica partition functions are
\begin{align}
Z(k) &= \sum_i \tr (P_i \rho P_i)^k \ .
\end{align}
In the block of states corresponding to projector $P_i$, call the lowest-energy eigenstate $|i\rangle$. For a pure state, we can decompose the density matrix $|\psi\rangle\langle \psi|$ into blocks as
\begin{align}
\rho_\psi &= \sum_{i,j} A_i |i\rangle \langle j| A^\dagger_j \ , 
\end{align}
where $A_i$ is an operator that acts within block $i$ and $\sum_i \tr (A^\dagger_i A_i) = 1$. The action of the coarse-graining map is
\begin{align}
\brho_\psi &= \N(\rho_\psi) =  \sum_i A_i |i\rangle \langle i| A^\dagger_i \ .
\end{align}
Therefore we can rewrite the replica partition functions as
\begin{align}\label{bdreplica}
Z(k) &= \sum_i \langle i| A_i^\dagger A_i |i\rangle^k \ .
\end{align}
We now  apply \eqref{cgderiv} to calculate the entropy. The conclusion is that under a block-diagonal coarse graining channel, the coarse-grained entropy of a pure state $|\psi\rangle = \sum_i A_i|i\rangle$ is
\begin{align}
S(\brho_\psi)  =  - \sum_i \langle i|A_i^\dagger A_i|i\rangle \log \langle i |A_i^\dagger A_i|i\rangle  \ .
\end{align}

\subsection{Holographic coarse graining}
As described in the introduction, we will see that the $k$-boundary wormholes described in section \ref{s:wormholes} calculate a CFT quantity of the form
\begin{align}\label{zworm}
Z_{\rm wormhole} &= \sum_n |\psi_n|^{2k} \langle n| V^\dagger V |n\rangle^{k} \ ,
\end{align}
where $V$ is built from single-trace operators.\footnote{We use the symbol $V$ to evoke an isometric map, which is standard notation in quantum information. $V$ does turn out to be approximately isometric, however it is not normalized, so $V^\dagger V \neq \id$.} (Actually,  the unperturbed black holes in section \ref{s:wormholes} have $V = \id$. But when we add matter outside the horizon in section \ref{s:addmatter} below, $V$ becomes nontrivial, so we will include it in the present discussion.)

Our goal now is to reintepret the wormhole result \eqref{zworm} as a coarse-grained replica partition function, i.e. to write it as
\begin{align}\label{wantzworm}
Z_{\rm wormhole} = \tr \brho^k \ . 
\end{align}
There is already a very close resemblance to the block-diagonal coarse graining map; compare \eqref{zworm} to  \eqref{bdreplica}. However, the similarity is imperfect, because we do not know how to decompose the CFT Hilbert space into blocks to make these two expressions identical. We will therefore define a new, holographic coarse-graining map to reproduce the wormhole answer. Given a state expressed in the form
\begin{align}\label{rhoform}
\rho = V \sum_{m,n} a_{mn} |m\rangle \langle n| \tilde{V}^\dagger\ , 
\end{align}
where $m,n$ label energy eigenstates, 
we define the coarse-graining map $\C$ by projecting onto the energy-basis diagonal inside the sum,
\begin{align}
\brho = \C(\rho) &:=  V\sum_{n} a_{nn} |n\rangle \langle n| \tilde{V}^{\dagger} \ .
\end{align}
Suppose we have a pure state decomposed into the form
\begin{align}
\rho_\psi = |\psi\rangle\langle \psi| , \quad |\psi\rangle = V \sum_n \psi_n |n\rangle \ .
\end{align}
Then the coarse-grained density matrix is
\begin{align}\label{defC}
\brho_\psi =  \C(\rho) = V \sum_n |\psi_n|^2 |n\rangle \langle n| V^\dagger \ , 
\end{align}
and the wormhole partition function agrees with \eqref{wantzworm}. 

It is now guaranteed that the von Neumann entropy of $\brho_\psi$ agrees with the area of the apparent horizon in the semiclassical limit:
\begin{align}
S(\brho_\psi) &= \frac{1}{4}\mbox{Area}(\gamma) \ . 
\end{align}
We therefore interpret $\brho_\psi$ as the coarse grained density matrix for the region outside the apparent horizon.

An important point is that unlike the diagonal projection $\D$ and the block-diagonal projection $\N$, it is not clear that the holographic coarse-graining map $\C$ is a quantum channel. A quantum channel is defined on density matrices, whereas the definition of $\C$ involves a decomposition of the state $\rho$ into the form \eqref{rhoform}, and it is not clear that every density matrix in a CFT can be unambiguously expressed in this form. For our purposes, the map defined as in \eqref{defC} is sufficient, whether or not there is a corresponding quantum channel. A toy model for $\C$ that is formulated explicitly as a quantum channel is discussed in appendix \ref{app:toychannel}.

The coarse graining map $\C$ has the following physical interpretation. The expansion coefficients $\psi_n$ encode the UV details of a particular black hole microstate. These coefficients can be studied statistically using the low-energy theory --- for example, we will calculate $|\psi_n|^2$ averaged over a small energy window --- but the precise values of individual coefficients are UV-sensitive. The coarse-graining procedure eliminates all of the phase information in $\psi_n$ by decohering the UV microstates, like the diagonal coarse-graining discussed above. However, unlike the diagonal coarse-graining, $\C$ retains the quantum correlations in the IR corresponding to the operators $V$. This is very natural, since these are low-energy operators, obtained by acting with single traces to create matter outside the apparent horizon. Thus an experimenter with access to $\brho_\psi$ can study superpositions of matter outside the horizon, but cannot form delicate superpositions of UV microstates that are indistinguishable from the outside.

\subsection{Coarse graining in 2D CFT}
\newcommand{\V}{\mathcal{V}}
In 2D CFT, the holographic coarse-graining map can be found explicitly for a much wider class of states by taking advantage of the infinite-dimensional conformal symmetry. The details will be reported elsewhere but here is a brief summary. In a 2D CFT, energy eigenstates are organized into lowest-weight representations of two copies of the Virasoro algebra. The lowest weight vectors are primary states, $|p\rangle$. Define a quantum channel $\V$ that projects onto matrix elements connecting two states in the same representation,
\begin{align}
\V(\rho) = \sum_p P_p \rho P_p \ , 
\end{align}
where the sum is over primaries, and $P_p$ projects onto the representation with lowest weight $|p\rangle$. This is an example of a block-diagonal projection as discussed above.

Now consider a theory of 3D gravity in which all nontrivial primaries have $\Delta \gg 1$. Let $|\psi\rangle$
be a state created by operator insertions in Euclidean time which has a nontrivial apparent horizon at $t=0$, with no matter outside the horizon (other than boundary gravitons). Such states were considered recently in \cite{Chandra:2022bqq} and used to construct multiboundary wormholes --- an example is the state $\O_H(-\tau_2) \O_H(-\tau_1)|0\rangle$ where $\O_H$ is a scalar primary of scaling dimension $\Delta_H > \frac{c}{16}$, with $c$ the central charge. Let us act on $|\psi\rangle$ with local operators $\O$ having $1 \ll \Delta \ll c$, which add particles outside the horizon,
\begin{align}
|\Psi_1\rangle &= \O(x_1) \O(x_2) \cdots|\psi\rangle \ .
\end{align}
Then the holographic coarse-graining map is
\begin{align}
\C( |\Psi_1 \rangle \langle \Psi_1| ) &= \O(x_1) \O(x_2) \cdots \V\left(|\psi\rangle\langle \psi|\right) \cdots \O(x_2)^{\dagger} \O(x_1)^{\dagger} \ .
\end{align}
Unlike in higher dimensions, we have \textit{not} assumed that the metric outside the horizon is spherically symmetric --- in particular, this map can be applied to black holes created by heavy local operator insertions as in \cite{Chandra:2022bqq}. The spherically symmetric thin shell black hole can be obtained by a  limit of the black holes studied in \cite{Chandra:2022bqq} by inserting a large number $n$ of operators with scaling dimensions $\Delta \sim \frac{c}{n}$ and taking $n \to \infty$ \cite{Anous:2016kss}.

\section{CFT dual of EOW branes and thin shells}\label{s:cftdual}

EOW brane geometries are dual to CFT states created by Euclidean evolution on a strip of Euclidean time, with a boundary condition at $\tau = -\tau_0$. Denote the CFT state on $S^{d-1}$ exactly at the EOW boundary by $|B\rangle$. This state is non-normalizable. Evolving by Euclidean time $\tau_0$ prepares a normalizable state,
\begin{align} \label{Psi1B}
|\Psi_1\rangle = e^{-\tau_0 H} |B\rangle \ .
\end{align}
Pictorially, the path integral preparation of the CFT state is
\begin{align}
\vcenter{\hbox{
\begin{overpic}[grid=false]{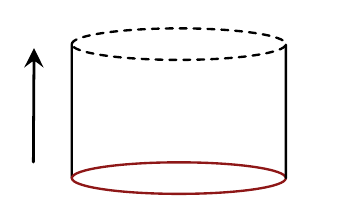}
\put (-1,35) {$\tau$}
\put (95,10) {$|B\rangle$}
\put (95,52) {$e^{-\tau_0 H}|B\rangle$}
\end{overpic} }}
\end{align}
The state $|\Psi_1\rangle$ is dual to an EOW brane geometry in which the brane hits the boundary at $\tau = \pm \tau_0$. The bulk saddle computes the norm
\begin{align}
\langle \Psi_1 | \Psi_1 \rangle \quad=  \quad
\vcenter{\hbox{\includegraphics{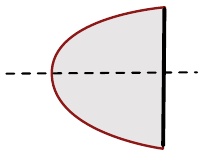}}}
\end{align}
with this diagram representing the geometry found in section \ref{ss:bstate}. 

The CFT description of thin shell black holes was explored in \cite{Anous:2016kss} and it can be phrased in a way that is similar to the $B$-state. Denote the state at the shell by $|S\rangle$. This state, assuming the matter is a pressureless perfect fluid, is created either by adding a source $J$ to the action or by acting on the vacuum with a large number of single-trace operators $\O$, with $1 \ll \Delta_{\O} \ll N^2$, uniformly over the transverse space. In the latter case, each operator insertion creates a `dust' particle in the bulk, and the state can be obtained by taking a continuum limit of the expression  \cite{Anous:2016kss}
\begin{align}
|S\rangle &\sim \left(\prod_j \O(\tau=0, x_\perp^j)\right) |0\rangle \ .
\end{align}
The details of this limit will not be important. The state defined at the shell is not normalizable, but once again we can prepare a normalizable state by evolving in Euclidean time,
\begin{align}
|\Psi_1\rangle = e^{-\tau_0 H} |S\rangle \ .
\end{align}
The bulk saddle found in section \ref{ss:thinshell} computes the norm of this state,
\begin{align}
\langle \Psi_1 | \Psi_1 \rangle \quad=  \quad
\vcenter{\hbox{\includegraphics{figures/psi1overlap.pdf}}} \ .
\end{align}

Given the similarities, it is convenient to discuss $B$-states and thin shells simultaneously. Denote the state at the matter insertion by $|A\rangle$, so $|A\rangle = |B\rangle$ or $|A\rangle = |S\rangle$ for $B$-states and thin shells, respectively. The state at $t=0$ is
\begin{align}
|\Psi_1\rangle = e^{-\tau_0 H} |A\rangle \ .
\end{align}
Expand in energy eigenstates, 
\begin{align}\label{psi1expansion}
|\Psi_1\rangle = \sum_n e^{-\tau_0 E_n} a_n |n\rangle \ .
\end{align}
The density matrix for this pure state is
\begin{align}
\rho &= |\Psi_1\rangle\langle \Psi_1|  = \sum_{m,n} e^{-\tau_0 (E_n+E_m)} a_m a_n^* |m\rangle\langle n| \ .
\end{align}
We now define a coarse-grained density matrix by projecting onto the diagonal in the energy basis,
\begin{align}
\brho &= \sum_n e^{-2\tau_0 E_n} |a_n|^2 |n\rangle\langle n| \ .
\end{align}
In the terminology of section \ref{s:replicaformalism}, we have chosen the coarse-graining map to be the completely dephasing channel $\D$, with $\brho = \D(\rho)$. This is equivalent to the holographic coarse-graining map $\C$ defined in \eqref{defC} in the special case $V = \id$.

In this section we will demonstrate the following by a CFT calculation. Choose
\begin{align}\label{anf}
a_n = e^{w(E_n)} c_n \ , 
\end{align}
where $w(E)$ is a smooth function of energy determined by matching to the $k=1$ black hole, and $c_n$ is a UV-sensitive coefficient that will not be determined but satisfies $|c_n|^2 \approx 1$ when averaged over a small energy window. Then
\begin{align}\label{zweq}
Z_{\rm wormhole} \approx e^{-I_k} \approx \tr \brho^k \ ,
\end{align}
where $I_k$ is the classical action of the $k$-boundary wormhole. The conclusion is that the bulk theory coarse-grains these black holes by the action of the completely dephasing channel. 

Equivalently, define a state in $k$ copies of the CFT with a GHZ-like, diagonal pattern of entanglement,
\begin{align}\label{apsik}
|\Psi_k\rangle = \sum_n \left( e^{-\tau_0 E_n}a_n \right)^k |n\rangle^{\otimes k} \ .
\end{align}
Then
\begin{align}
Z_{\rm wormhole} \approx \langle \Psi_k | \Psi_k \rangle \ .
\end{align}
It follows from the general discussion in section \ref{s:replicaformalism} that the entropy of the apparent horizon in the geometry dual to $|\Psi_1\rangle$ is the von Neumann entropy of $\brho$,
\begin{align}
 \frac{1}{4G} \mbox{Area}(\gamma) &= S(\brho) \\
 &= -  \sum_n \frac{  e^{-2\tau_0 E_n + 2 w (E_n)} }{\langle \Psi_1 | \Psi_1\rangle} \log  \frac{  e^{-2\tau_0 E_n+2w(E_n)} }{\langle \Psi_1 | \Psi_1\rangle} \ . 
 \end{align}
The logic that we will follow is to use the single-boundary black hole to determine the weighting function $w(E)$ (which depends on whether this is a $B$-state or thin shell, and on the brane tension or shell mass), and then match to the $k$-boundary wormholes, without any further input from the gravity side. The match \eqref{zweq} is nontrivial because it is not obvious that this CFT procedure should lead to the correct $k$ dependence.

There is an added complication for $B$-states: there is a second gravitational saddle, with a single boundary and two disconnected EOW branes, that contributes to $\langle \Psi_1 |\Psi_1\rangle$. We will ignore this for now and add in the effects of the disconnected phase in subsection \ref{ss:cylinder}.\footnote{$B$-states are also special in that they have anomalously large one-point functions for light operators, which decay in time as the coefficients randomize \cite{Hartman:2013qma}. This has consequences for the $c_n$'s that we will not explore here; see \cite{Kourkoulou:2017zaj} for a 0+1D analogue.}

\subsection{Planar black holes}
We shall first illustrate the strategy outlined above for the planar black holes. As discussed in section \ref{ss:specialcases}, this case is simpler than spherical black holes because $\tau_0 \propto \beta$.  The norm of the black hole state defined in \eqref{psi1expansion}, \eqref{anf} can be expressed as an integral over energy,
\begin{align}\label{psi1o}
\langle \Psi_1 |\Psi_1\rangle &\approx \int dE \, \exp\left( S(E) - 2 \tau_0 E + 2 w(E) \right) \ ,
\end{align}
with $S(E)$ the thermodynamic entropy. To determine the function $w(E)$, we set this equal to the gravitational action of the black hole, $\langle \Psi_1 | \Psi_1\rangle = e^{-I_1}$. The solution of the resulting saddlepoint equations is 
\begin{align}
w(E) &= -\frac{1}{2}(I_1(\tau_0(E))+S(E)) +  \tau_0(E) E \label{planarsaddlef}\\
E &= \frac{1}{2}\p_{\tau_0} I_1(\tau_0) \ , \label{planarsaddleE}
\end{align}
where the second equation defines $\tau_0(E)$.
The planar wormhole action \eqref{BBaction} is 
\begin{align}
I_1(\tau_0) = \frac{2\tau_0}{\beta_1}I_0(\beta_1) , \qquad \beta_1 = \frac{2\tau_0}{1-\alpha} \ ,
\end{align}
where $I_0$ is the action of the eternal black hole. This can be used to show that \eqref{planarsaddleE} agrees with the ordinary thermodynamic energy-temperature relation, $E = I_0'(\beta_1)$. Therefore $S(E) = -I_0(\beta_1) + \beta_1 E$, and \eqref{planarsaddlef} simplifies to
\begin{align}\label{planarfans}
w(E) = -\frac{\alpha}{2} S(E) \ . 
\end{align}
Thus we have shown that in the planar limit, $B$-states and thin shells have the following expansion in the energy basis:
\begin{align}\label{Aexp}
|A\rangle &= \sum_n e^{-\frac{\alpha}{2} S(E_n)}c_n |n\rangle \ ,
\end{align}
with $\alpha$ determined by the relation $\tau_0 = \frac{\beta}{2}(1-\alpha)$ and $|c_n|^2 \approx 1$ (when averaged over a small energy window).\footnote{The norm of the state in \eqref{Aexp} is $\langle A|A\rangle \sim \int dE e^{S(E)(1-\alpha)} = \int dE e^{2\tau_0 S(E) /\beta}$, which diverges, so the state is indeed non-normalizable at the matter insertion. It becomes normalizable upon evolution by $e^{-\tau_0 H}$.} For EOW branes in $d=2$, we have $\alpha = \frac{1}{2}$ independent of tension, so the weighting factor in \eqref{Aexp} is $e^{-S/4}$.

Now we turn to the wormholes. The coarse-grained replica partition function in the CFT is
\begin{align}
\tr \brho^k = \langle \Psi_k |\Psi_k\rangle &\approx \int dE \, \exp\left( S(E) - 2k \tau_0 E + 2k w(E) \right) \\
&\approx \int dE \exp\left( S(E) (1- k \alpha)  - 2 k \tau_0 E \right)
\end{align}
Evaluating this integral at the saddlepoint gives
\begin{align}
 \langle \Psi_k |  \Psi_k \rangle &\approx e^{-(1-k\alpha) I_0(\frac{2k\tau_0}{1-k\alpha}) }\ .
\end{align}
Comparing to the gravity calculation \eqref{BBaction} there is a perfect match to leading order,
\begin{align}
\langle \Psi_k|\Psi_k\rangle \approx e^{-I_k} \ .
\end{align}
In particular, the CFT saddle disappears for $k \geq k_{max}$, with the same value of $k_{max}$ found in the bulk. Therefore the fact that $B$-states only admit fractional wormhole saddles is in agreement with the dual CFT.

\subsection{Spherical black holes}\label{ss:cftsphere}
For spherical black holes, the logic is the same: We use the $k=1$ black hole to determine $w(E)$, then check that the CFT reproduces the $k$-boundary wormhole, $\tr \brho^k \approx e^{-I_k}$. It is a bit simpler to do both steps at once, as follows. Set the CFT and gravity answers equal:
\begin{align}\label{wantsph}
 \int dE \, \exp\left( S(E) - 2k \tau_0 E + 2k w(E) \right) = e^{-I_k}
\end{align}
with $I_k$ given by \eqref{Ik}. This is viewed as an equation for $w(E)$. The saddlepoint analysis gives a result for $w(E)$ that \textit{a priori} depends on $k$; but if $w(E)$ is actually independent of $k$, then we have a successful match, because in that case $w(E)$ is entirely determined by the $k=1$ black hole.

The solution to \eqref{wantsph} in the saddlepoint approximation is
\begin{align}\label{fe1}
w(E) &= -\frac{1}{2k}(I_k(\tau_0(E)) + S(E)) + \tau_0(E) E  \ , 
\end{align}
where $\tau_0(E)$ is defined by
\begin{align}
E &= \frac{1}{2 k} \p_{\tau_0} I_k(\tau_0) \ .
\end{align}
We need to show that the function of $E$ on the right-hand side of \eqref{fe1} has no $k$ dependence.
Using \eqref{Ik} and \eqref{bbc2} it can be rewritten as
\begin{align}
w(E) &= -\frac{1}{2}\left[ \frac{2\tau_0}{\beta_k}I_0(\beta_k) + I_L(\beta_k) \right] - \frac{1}{2k}S(E) + \tau_0(E) E \\
&= \left(\frac{1}{2}-\frac{\tau_{0k}}{\beta_k}\right) \left( I_0(\beta_k)-\beta_k E \right) - \frac{1}{2}I_L(\beta_k) - \frac{1}{2k}\left[ I_0(\beta_k)+S(E)-\beta_k  E\right]\label{fe3}
\end{align}
where $\beta_k = \beta_k(\tau_0(E))$ is the inverse temperature of the wormhole determined implicitly by the boundary condition \eqref{bbc2}, and $\tau_{0k}$ is the corresponding endpoint, $\tau_0(\beta_k)$.  
As discussed around \eqref{energyrel},  $\frac{1}{2 k} \p_{\tau_0} I_k(\tau_0) = I_0'(\beta_k)$, so $E$ is in fact the ordinary thermodynamic energy at inverse temperature $\beta_k$. Therefore we can invoke the relation $I_0(\beta_k) = \beta_k E - S(E)$ and find
\begin{align}\label{fe4}
w(E) &= \left(\frac{\tau_{0k}}{\beta_k} -   \frac{1}{2}\right)S(E) - \frac{1}{2}I_L(\beta_k) \ , \qquad E = I_0'(\beta_k) \ .
\end{align}
The explicitly $k$-dependent term in \eqref{fe3}, proportional to $1/k$, has dropped out. The variable $\beta_k$ in \eqref{fe4} is now just a dummy variable so it can be renamed $\beta_k \to \beta$. Thus
\begin{align}\label{fsfinal}
w(E) = -\frac{\alpha(E)}{2}S(E) - \frac{1}{2}I_L(\beta(E)) 
\end{align}
where $E=I_0'(\beta)$ and we have defined $\alpha(E)$ by the relation
\begin{align}
\alpha = 1 - \frac{2 \tau_0}{\beta} \ .
\end{align}
The final answer \eqref{fsfinal} is manifestly independent of $k$. In the planar limit, $I_L = 0$ and $\alpha$ is a constant, so we recover the results of the previous subsection. 

In this form, it is clear that the only gravity input used to determine $w(E)$, and therefore the CFT state, is the action of the $k=1$ black hole. The action of the $k$-boundary wormhole, as a function of $k$, is the nontrivial agreement between gravity and CFT.

\subsection{The cylinder phase}\label{ss:cylinder}
For $B$-states, there is a second bulk geometry that contributes to the overlap $\langle \Psi_1 |\Psi_1\rangle$, in addition to the black hole \cite{Cooper:2018cmb}. It is a portion of global AdS bounded by a pair of disconnected EOW branes:
\begin{align}\label{cyl}
\begin{overpic}[grid=false,width=0.7in]{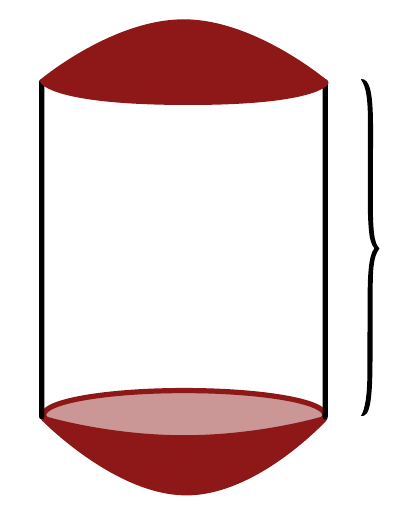}
\put (80,50) {$2\tau_0$}
\end{overpic}
\end{align}
The cylinder is filled in, and the caps are the EOW branes.
Thus the total overlap for a $B$-state is schematically
\begin{align}
\langle \Psi_1 |\Psi_1\rangle \approx 
\vcenter{\hbox{\includegraphics{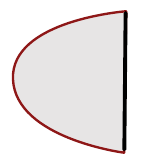}}}
+
\vcenter{\hbox{\includegraphics{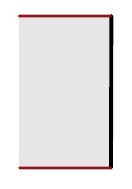}}}
 \ ,
\end{align}
where the first diagram represents the black hole and the second diagram represents the cylinder solution.
There is a phase transition analogous to the Hawking-Page transition as we tune $\tau_0$, with the tension $T$ held fixed \cite{Cooper:2018cmb}. The black hole phase dominates at small $\tau_0$. The cylinder phase is accounted for in the dual CFT by writing the state as
\begin{align}\label{psi1vacterm}
|\Psi_1\rangle = e^{-I_{\rm cyl}/2}|0\rangle + \sum_{n\neq 0}e^{-\tau_0 E_n + w(E_n)}c_n|n\rangle \ ,
\end{align}
where $I_{cyl}$ is the gravitational action of the saddle in \eqref{cyl}, and the second term is the contribution from the black hole. There are also other, non-vacuum contributions from the cylinder phase, but the semiclassical saddlepoint never lands on them (unless additional operators are inserted), so they are invisible to this leading-order analysis.
The cylinder action, calculated in appendix \ref{app:cylinderdetails}, takes the form
\begin{align}
I_{\rm cyl} = 2\tau_0 E_{\rm vac}+ I_G  , 
\end{align}
where $E_{vac}$ is the vacuum energy, and $I_G$ is a contribution from the branes that is independent of $\tau_0$. Thus the first term in \eqref{psi1vacterm} is $e^{-I_G/2  - \tau_0 H}|0\rangle$. In the tensionless limit, for any $d$, $I_G$ vanishes, and in AdS$_3$, as a function of the tension it is $I_{G} = -\frac{c}{3}\tanh^{-1}(T)$ \cite{Cooper:2018cmb,Miyaji:2021ktr}.

For thin shells, whether there is a similar, disconnected phase depends on the details of how the shell is constructed. Let us assume that the shell consists of a large number of dust particles, carrying a flavor charge; then there is no way for the dust worldlines to terminate without hitting a conjugate shell insertion, so there is no disconnected phase.

\section{Adding a massive particle}\label{s:addmatter}
We will now add a massive particle to the EOW-brane or thin-shell black hole, and consider its effect on the coarse-grained density matrix.  The coarse-graining map relies on the decomposition of a quantum state into the form
\begin{align}
|\Psi\rangle = V \sum_n \psi_n |n\rangle \ .
\end{align}
We will show that particles added behind the horizon change the microstate coefficients $\psi_n$ while particles added outside the horizon become part of $V$. Thus there is a sharp distinction between particles added inside or outside the extremal surface --- particles  behind the apparent horizon are effectively hidden by the coarse-graining map, while particles  outside are not.  This is very natural from the bulk, but nontrivial in CFT. 

\subsection{Setup on the gravity side}

The mass of the particle is taken to satisfy $\ell_{\rm AdS}^{-1} \ll m \ll M_{\rm planck}$, so that it travels on a geodesic but its backreaction is small. We will work to first order in the backreaction (it cannot be neglected). 

In the Euclidean path integral, the particle is added to the black hole by inserting the dual operator $\O$ at $\tau = \pm \tau_1$. The resulting saddlepoint has a black hole, plus a particle on a geodesic.  In the case of a thin shell black hole, the particle travels on a connected geodesic, so the two options, corresponding to whether the particle is inside or outside the apparent horizon at $t=0$, are:
\begin{align}\label{probeinout}
\mbox{inside:} \quad \vcenter{\hbox{
\begin{overpic}[grid=false,width=1.3in]{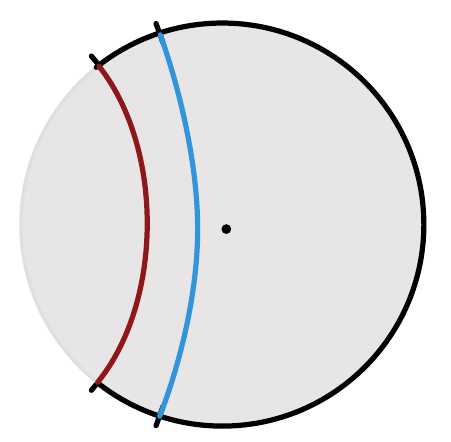}
\put (12,90) {$\tau_0$}
\put (28,97) {$\tau_1$}
\end{overpic}
}} 
\qquad
\mbox{outside:} \quad \vcenter{\hbox{
\begin{overpic}[grid=false,width=1.3in]{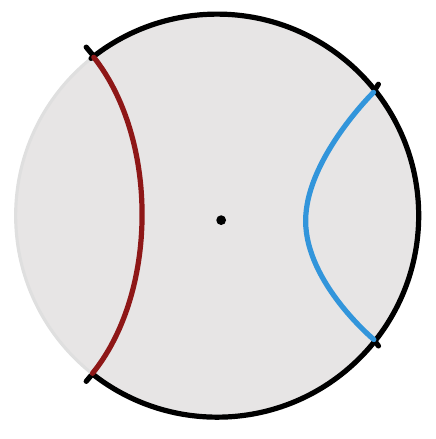}
\put (11,93) {$\tau_0$}
\put (87,82) {$\tau_1$}
\end{overpic}
}}
\end{align}
The blue line is the massive particle, created by CFT operator insertions $\O(- \tau_1)\O(\tau_1)$.\footnote{In the `inside' case, the worldline of the particle can also cross the shell into the vacuum region. This doesn't affect the ensuing calculation.} These are local operators inserted at a point in the transverse space, which may be different for the two operators. For $B$-states, since $\tau_0 < \frac{\beta}{4}$, there are no connected geodesics that go behind the horizon. However, the geodesic can end on the EOW brane, so the dominant contribution may look like this:
\begin{align}
\vcenter{\hbox{
\begin{overpic}[grid=false,width=1.3in]{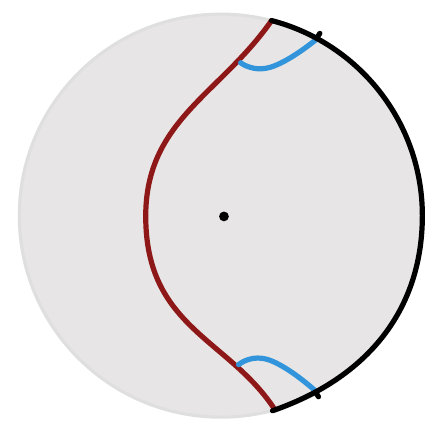}
\put (55,98) {$\tau_0$}
\put (71,93) {$\tau_1$}
\end{overpic}
}}
\end{align}
and in this case we consider the particle to be `behind the horizon' at $t=0$.

\subsection{Particle inside the apparent horizon}

Consider a (possibly fractional) wormhole with $k$ boundaries corresponding to a replica of the $B$-state or thin-shell geometry with a massive particle behind the horizon. By adding a massive particle to the solution in section \ref{ss:quotientwormholes}, we see that if the particle is behind the horizon at $t=0$ then it goes through the wormhole.\footnote{We have assumed that after turning on the conical defect, the particle still goes behind the horizon; this is certainly true for $k-1 \ll 1$, where the backreaction of the defect is small, but at large enough $k$ there may be a transition where the particle crosses to the outside. We restrict to the range of $k$ where this does not occur.} This is clearest for a thin shell wormhole, say for $k=6$, where the geodesics are connected:
\begin{align}
\vcenter{\hbox{
\begin{overpic}[grid=false]{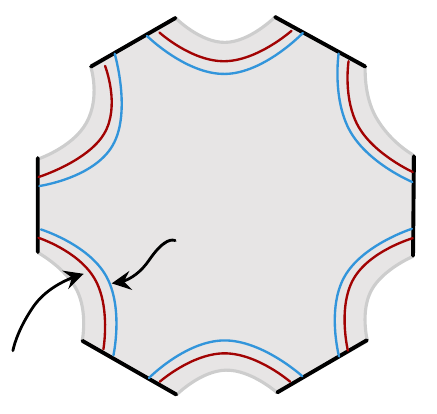}
\put (-3,8) {{\footnotesize shell}}
\put (43,39) {{\footnotesize particle}}
\end{overpic}
}}
\end{align}
To analyze this situation from the dual CFT, we will simply treat the massive particle as part of the initial matter insertion; then the entire calculation carries through almost unchanged, so we will be brief. Insert the operator $\O$ at $\pm \tau_1$ and place the EOW brane or thin shell endpoint at $\pm \tau_0$, with
\begin{align}
\tau_0 = \tau_1 + \Delta \tau \ .
\end{align}
Fix $\Delta \tau$ to a constant, and consider the state to be parameterized by $\tau_1$. Now $\tau_1$ will play the role that was played by $\tau_0$ in the calculations in sections \ref{s:wormholes} and \ref{s:cftdual}. Divide the bulk action into
\begin{align}
I_k  = k \left( \frac{2\tau_1}{\beta_k}I_0(\beta_k) + \widehat{I}_L(\beta_k) \right) \ , 
\end{align}
where the first term is the contribution from $|\tau| < \tau_1$, and $\widehat{I}_L$ is the contribution from the region $|\tau| \geq \tau_1$, including the massive particle --- so $\widehat{I}_L$ includes a term $m \ell(\beta_k)$ with $\ell(\beta_k)$ the length of the probe geodesic in the wormhole background. Now we repeat exactly the steps of section \ref{ss:cftsphere}, with the replacement $\tau_0 \to \tau_1$ and $I_L \to \widehat{I}_L$.  Every step is identical, after these replacements, so we reach the following conclusion. The quantum state in the CFT at $\tau = 0$ is
\begin{align}
|\widehat{\Psi}_1\rangle &= \O(-\tau_1) |\Psi_1\rangle \ , \
\end{align}
where $|\Psi_1\rangle$ is the state analyzed in section \ref{s:cftdual}, i.e. without the extra particle. This state
has an expansion in the energy basis of the form
\begin{align}
|\widehat{\Psi}_1\rangle &=
 \sum_n e^{-\tau_1 E_n + \hat{w}(E_n)} \hat{c}_n |n \rangle \ , 
\end{align}
with $|\hat{c}_n|^2\approx 1$ when averaged over a small energy window. The weight function is given by \eqref{fsfinal} but with $\tau_0 \to \tau_1$ and $I_L \to \widehat{I}_L$. The coarse-grained density matrix is the diagonal projection in the energy basis,
\begin{align}
\brho = \D(\rho) &= \sum_{n} e^{-2\tau_1 E_n + 2 \hat{w}(E_n)} |\hat{c}_n|^2 |n\rangle\langle n| \ .
\end{align}
The replica partition function reproduces the wormhole action,
 \begin{align}
Z_{\rm wormhole} = \tr \brho^k \ ,
\end{align}
and the von Neumann entropy $S(\brho)$ matches the area of the apparent horizon, including the $O(m)$ contribution from  backreaction after adding the extra particle.

It is clear from this calculation that it works for any linearized deformation of the thin shell or $B$-state geometry in which the extra matter is entirely behind the horizon at $t=0$. The step in the calculation where we assumed the matter was behind the horizon was in using the relation $\frac{1}{2}\p_{\tau_1} I_k = I_0'(\beta_k)$ (see below \eqref{fe3}). It is always true that the ADM mass of the solution including the contribution of the particle is $E_{\rm ADM} =\frac{1}{2}\p_{\tau_1} I_k$, since this is the statement that $E_{\rm ADM}$ is the Hamiltonian (and this is the Hamilton-Jacobi equation), but the relation $E_{\rm ADM} = I_0'(\beta_k)$ is only applicable when there is no matter outside the horizon.

\subsection{Particle outside the apparent horizon}
\newcommand{\tPsi}{\widehat{\Psi}}
Let us now suppose the particle is outside the horizon at $t=0$. The $k=1$ solution is in the right-hand diagram of \eqref{probeinout}. In the corresponding $k$-boundary solution (for $k$ sufficiently close to 1 to avoid a phase transition) the massive particle does not go through the wormhole. For example:
\begin{align}\label{particleout}
\vcenter{\hbox{
\begin{overpic}[grid=false]{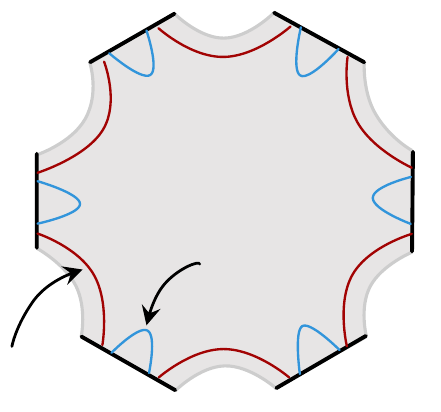}
\put (-3,8) {\footnotesize{shell}}
\put (48,33) {\footnotesize{particle}}
\end{overpic}
}}
\end{align}
From a bulk point of view, we now expect the coarse graining procedure to preserve the quantum correlations associated to the particle. Let $|\Psi_1\rangle$ be the undeformed black hole state studied in section \ref{s:cftdual}, and 
\begin{align}
|\widehat{\Psi}_1\rangle = \O(-\tau_1) |\Psi_1\rangle \ .
\end{align}
Expand in the energy basis,
\begin{align}
|\widehat{\Psi}_1\rangle = \O(-\tau_1) \sum_n c_n e^{-\tau_0 E_n  + w(E_n)} |n\rangle \ ,
\end{align}
with $w(E)$ derived in section \ref{s:cftdual}. 
The pure-state density matrix $\rho = |\tPsi_1\rangle \langle \tPsi_1|$ is
\begin{align}
\rho = \O(-\tau_1) \sum_{m,n} c_m c_n^* e^{-\tau_0(E_m+E_n)+w(E_n)+w(E_m)}|m\rangle\langle n| \O(\tau_1) \ .
\end{align}
This takes the form of \eqref{rhoform} with $V = \tilde{V} = \O(-\tau_1)$. Note that in this step, we are \textit{choosing} $V = \O(-\tau_1)$ as part of the definition of the coarse-graining map; this is an ansatz that will be matched to the bulk. Thus we define the coarse-grained density matrix
\begin{align}
\brho = \C(\rho) = \O(-\tau_1) \sum_n |c_n|^2 e^{-2\tau_0 E_n+2w(E_n)}|n\rangle\langle n|  \O(\tau_1) \ ,
\end{align}
 and the replica partition function in the CFT is
 \begin{align}\label{rhokO}
\tr \brho^k &= \sum_n |c_n|^{2k} e^{-2\tau_0 k E_n+2kw(E_n)} \big(\langle n| \O(\tau_1) \O(-\tau_1) |n\rangle\big)^k \ . 
\end{align}
This can also be expressed as $\langle \tPsi_k | \tPsi_k\rangle$, with the $k$-copy entangled state defined by
\begin{align}
|\tPsi_k\rangle &= \sum_n \left( c_n e^{-\tau_0 E_n + w(E_n)}\O(-\tau_1)|n\rangle\right)^{\otimes k} \ .
\end{align}
To simplify \eqref{rhokO} we need to make a further assumption about the matrix elements of the CFT operator $\O$. We will assume that it satisfies the eigenstate thermalization hypothesis \cite{deutsch1991quantum,Srednicki:1994mfb,DAlessio:2015qtq}, so that the 2-point function in the eigenstate $|n\rangle$ is well approximated by the thermal 2-point function at the temperature corresponding to energy $E_n$. Denote the thermal 2-point function by
\begin{align}
G(2\tau_1; \beta) &= \frac{1}{Z(\beta)} \tr e^{-\beta H} \O(\tau_1) \O(-\tau_1) \ .
\end{align}
Applying this to \eqref{rhokO} and converting the sum to an integral using $|c_n|^2 \approx 1$, we have
\begin{align}
\tr \brho^k &\approx \int dE e^{S(E)-2k\tau_0 E+2kw(E)} G(2\tau_1; \beta(E) ) ^k \ .
\end{align}
Without the extra particle, we saw in section \ref{s:cftdual} that the saddlepoint reproduces the unperturbed wormhole action $e^{-I_k}$, and that the saddlepoint lands at same energy as the wormhole, $E = \frac{1}{2k} \p_{\tau_0}I_k(\beta) = I_0'(\beta_k)$. The particle contribution can now be evaluated at the saddle to obtain
\begin{align}\label{outsiderhok}
\tr \brho^k &\approx e^{-I_k} G(2\tau_1; \beta_k)^k \ .
\end{align}
This CFT result manifestly agrees with the bulk calculation from a wormhole with a non-traversing geodesic on each boundary, like the one shown for $k=6$ in \eqref{particleout}. Of course, we need the bulk to calculate the thermal 2-point function $G$, but this is done at $k=1$ and then the CFT reproduces the $k$-boundary wormholes with no further input from the gravity side.

The entropy calculated from \eqref{outsiderhok} is
\begin{align}
S(\brho) &= \p_k\left( \frac{I_k(\beta)}{k} - \log G(2\tau_1, \beta_k) \right)_{k=1} \ .
\end{align}
The first term is one quarter the area of the \textit{unperturbed} apparent horizon. The second term is the correction $\delta (\mbox{Area})/4$ from the first order backreaction. To see this, use the fact that the perturbed black hole mass is the ADM energy evaluated on the Euclidean radial slice $\tau = -\tau_0$; therefore $E_{\rm BH} = \frac{1}{2}\p_{\tau_0} I_{\rm total}$, with $I_{\rm total} = I_1   - \log G(2\tau_1; \beta)$ and this $\p_{\tau_0}$ derivative is taken at fixed $\tau_1$. Thus $\delta E_{\rm BH} = -\frac{1}{2}\p_{\tau_0}\log G$ and using the gravitational first law, 
\begin{align}
\delta \frac{\mbox{Area}}{4}  = \beta \delta E_{\rm BH}= - \frac{\beta}{2}\p_{\tau_0}\log G(2\tau_1; \beta_k) = 
-\p_k \left. \log G(2\tau_1; \beta_k)\right|_{k=1}
\end{align} 
where the last equality uses \eqref{bbc2}. Thus $S(\brho) = \frac{1}{4}$Area, including the first order backreaction.

\subsection{Semiclassical isometries and random tensor networks}
A natural question is how the CFT distinguishes between operators that add particles inside the horizon vs.~outside the horizon. This will be explored in detail in a separate paper \cite{note:toappear}. The key point is that an operator $\O$ that adds matter outside the horizon acts around the semiclassical saddle as a random map from the CFT Hilbert space near energy $E$ to the CFT Hilbert space near energy $E'$, with $E'>E$. This is a random map from a large Hilbert space to a much larger Hilbert space, and it is a general property of such maps that $\O^\dagger \O$ is proportional to the identity matrix, as an operator \cite{Hayden:2016cfa}. We therefore refer to $\O$ as a `semiclassical isometry'. By contrast, adding matter behind the apparent horizon acts around the semiclassical saddle as a random map \textit{downward} in energy in the CFT Hilbert space. It is therefore conjugate to an approximate isometry, up to normalization. These features of the CFT operator algebra agree with the expectation from information-theoretic arguments \cite{Hayden:2016cfa,Brown:2019rox,Engelhardt:2021mue,Akers:2021fut} and can be used to construct a holographic tensor network directly in the dual CFT that is exactly dual to Einstein gravity with the standard boundary conditions.

\section{Ensemble interpretation}\label{s:ensemble}

So far, all of the results of this paper have been about individual CFTs such as ${\cal N}=4$ Super Yang-Mills (with the caveat of possible instabilities). We will now step slightly outside these bounds and ask whether the wormholes found in section \ref{s:wormholes} have an ensemble interpretation, along the lines of the West Coast EOW-brane model in \cite{Penington:2019kki}. The answer is yes, in a limited sense. The sum over known gravitational saddles --- i.e., the black holes, wormholes, and cylinder phase ---  is consistent with a Gaussian average over the coefficients $c_n$ that were left undetermined in the CFT analysis of section \ref{s:cftdual}.
This does not mean gravity is an ensemble average; this is an effective field theory calculation in the bulk and should be interpreted as such in the dual CFT. It ignores the (very likely) possibility of additional, UV-sensitive contributions to the path integral of the same magnitude, so it cannot be used it to make any firm conclusions about string theory examples like ${\cal N}=4$ SYM. 

We expect there to be additional saddles that lead to non-gaussian statistics \cite{Belin:2021ryy}.\footnote{Furthermore, $B$-states have large 1-point functions for probe operators, and this requires the $c_n$'s to be correlated with the matrix elements of local operators. For this reason we will not consider matter probes in this section.}
The observation that a Gaussian average is sufficient to account for the wormhole saddles is therefore only a starting point for a more complete analysis.

Consider $N_f$ flavors of $B$-states or thin shells. The non-normalizable state at the matter insertion, $|A^i\rangle$, now has an additional flavour index $i = 1 \dots N_f$. Let us assume the tension parameter for $B$-states or mass parameter for thin shells is the same for all flavours, and expand the flavoured matter states in the energy basis,
\begin{align}\label{flavorexpA}
|A^i\rangle  = \sum_n e^{w(E_n)} c_n^i |n\rangle \ ,
\end{align}
where $w(E)$ is the smooth function found in section \ref{s:cftdual} and the UV-sensitive coefficients $c_n^i$ satisfy $|c_n^i|^2\approx 1$ when averaged over a small energy window of nearby eigenstates. The flavoured black hole microstates $|\Psi^i\rangle$ are defined by a Euclidean evolution over time $\tau_0$ of the flavoured matter states,
\begin{align}\label{flavorexp}
|\Psi^i\rangle = e^{-\tau_0 H} |A^i\rangle \ .
\end{align}

\subsection{Overlaps}
The overlap of two flavoured black hole microstates $\langle \Psi^i |\Psi^j\rangle$ is calculated, in principle, by a gravitational path integral with the boundary condition that matter with flavour $j$ hits the boundary at $\tau = -\tau_0$, and matter with flavour $i$ hits the boundary at $\tau = \tau_0$. 
For thin shells, the only gravitational solution satisfying these boundary conditions is the black hole if the flavours at the two end-points match, and there is no solution for different flavours at the end-points. For the EOW brane, the dominant phase is the black hole when the flavours match (as we are always assuming $\tau_0$ is chosen so that this phase dominates). However, there is a classical solution even when the $B$-states have two different flavours: The cylinder phase bounded by disconnected EOW branes at the two endpoints; see section \ref{ss:cylinder}. Thus in the effective gravitational theory we find
\begin{equation} \label{overlap}
\begin{split}
\mbox{shell:}  \qquad & \langle \Psi^i\ket{\Psi^j}_{\rm saddles} = 
\vcenter{\hbox{\begin{overpic}[grid=false]{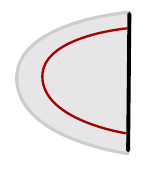}
\put (83,20) {$j$}
\put (83,80) {$i$}
\end{overpic}
}}\ 
 = e^{-I_{BH}}\delta_{ij}\\
\mbox{brane:}  \qquad & \langle \Psi^i\ket{\Psi^j}_{\rm saddles} = 
\vcenter{\hbox{
\begin{overpic}[grid=false]{figures/BphaseNoDash.pdf}
\put (87,5) {$j$}
\put (87,87) {$i$}
\end{overpic}
}}
\quad +
\vcenter{\hbox{
\begin{overpic}[grid=false]{figures/cylinderschematic.pdf}
\put (70,7) {$j$}
\put (70,85) {$i$}
\end{overpic}
}}
\quad = e^{-I_{BH}}\delta^{ij}+e^{-I_{cyl}}
\end{split}
\end{equation}
where $I_{\rm BH} := I_1$ is the action of the black hole \eqref{singleI}.
The subscript `saddles' is to remind us that this is the sum over known saddles but there may be other important contributions.
Let us compare to the Gaussian average in CFT. Using \eqref{flavorexp}, and separating out a vacuum state contribution (which is zero for thin shells), 
\begin{align}\label{cftij}
\langle \Psi^i | \Psi^j\rangle &= \langle \Psi^i|0\rangle \langle 0|\Psi^j\rangle  + \sum_{n\neq 0} e^{-2\tau_0 E_n + 2 w(E_n)} c_n^{i*} c_n^j 
\end{align}
Define the ensemble average over $c_n^i$ by treating them as Gaussian random variables with zero mean and unit variance,
\begin{align}
\rule[1em]{2em}{0.03em}\hspace{-2em}c_n^{i*} c_m^j := \delta_{mn} \delta^{ij} \ ,
\end{align}
with higher moments calculated by Wick contractions.
Then the average of \eqref{cftij} is
\begin{align}
\overline{ \langle \Psi^i | \Psi^j\rangle } &\approx \langle \Psi^i |0\rangle \langle 0 |\Psi^j\rangle + \sum_{n \neq 0}e^{-2\tau_0 E_n + 2 w(E_n)} \ . 
\end{align}
This agrees with the bulk results in \eqref{overlap}. 

It is clear that this agreement continues for higher powers of the overlap, which are calculated by multiboundary wormholes. Gaussian contractions simply pair the boundaries and project onto the terms with matching energy eigenstates, and these are exactly the terms calculated by wormholes. 

As a multiboundary example consider the modulus squared of the overlap, $|\langle \Psi^i\ket{\Psi^j}|^2$. The boundary condition for computing  $|\langle \Psi^i\ket{\Psi^j}|^2$ is that we have two boundaries, each with a matter insertion $i,j$ at $\tau = \pm \tau_0$. 
For EOW branes, there is no wormhole saddle with $k=2$ boundaries, because for any value of the tension, $k_{max} \leq 2$. Therefore in this case
\begin{align}\label{sqoverlapBrane}
|\langle \Psi^i\ket{\Psi^j}|^2_{\rm saddles} = \left( \langle \Psi^i | \Psi^j\rangle_{\rm saddles} \right)^2
= \left( e^{-I_{BH}}\delta^{ij}+e^{-I_{cyl}}\right)^2 \ .
\end{align}
For thin shells, assuming $k_{max} > 2$, the wormhole is on shell, so we find --- summing over known saddles --- the overlap
\begin{align} \label{sqoverlapShell}
  |\langle \Psi^i\ket{\Psi^j}|^2_{\rm saddles}= 
  \vcenter{\hbox{\begin{overpic}[grid=false]{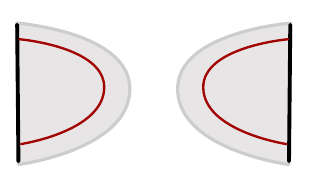}
  \put (-2,10) {$j$}
  \put (-2,43) {$i$}
  \put (96,10) {$j$}
  \put (96,43) {$i$}
  \end{overpic}}}
  +
    \vcenter{\hbox{\begin{overpic}[grid=false]{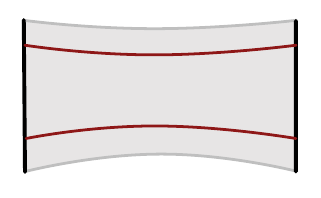}
  \put (0,11) {$j$}
  \put (0,43) {$i$}
  \put (95,11) {$j$}
  \put (95,43) {$i$}    
  \end{overpic}}}
   = e^{-2I_{BH}}\delta^{ij}+e^{-I_2}
\end{align}
 It is straightforward to check that these equations agree with the Gaussian averages calculated in CFT. 

For $B$-states, the $k=2$ results \eqref{overlap} and \eqref{sqoverlapBrane} were obtained earlier in \cite{Miyaji:2021ktr}, and a notion of coarse-grained states similar to our Gaussian average is discussed in \cite{Freivogel:2021ivu}. However the overlaps that we obtain for non-integer $k$, $|\langle \Psi^i|\Psi^j\rangle|^k$, have an extra term from on-shell fractional wormholes when $k<k_{max}$, and the wormhole contributions will be important for the calculation of the Page curve.

\subsection{Page-like behavior}
We now consider a toy model for an evaporating black hole, analogous to that described in \cite{Penington:2019kki}. To this end, we introduce an auxiliary system $R$ which plays the role of a non-gravitational reservoir to collect the outgoing Hawking radiation, and interpret the flavoured matter states $\ket{A^i}$ as representing the interior partners of the Hawking modes. Let the Hilbert space of $R$ denoted $\mathcal{H}_R$  be spanned by an orthonormal basis $\{\ket{i}_R\}$ of radiation states with $\text{dim}(\mathcal{H}_R)=N_f$ equal to the number of flavours. 
Consider a state $\ket{\psi}$ (unnormalised) of the black hole entangled with the radiation system $R$,
\begin{equation}
    \ket{\psi}=\sum_{i=1}^{N_f}\ket{\Psi^i}\ket{i}_R
\end{equation}
where $\ket{i}_R$ represents the state of the radiation corresponding to the flavoured matter state, $\ket{A^i}$, of the EOW brane or the thin shell. The reduced density matrix $\rho_R$ (unnormalised) for the radiation is obtained by tracing out the black hole degrees of freedom, 
\begin{equation} \label{Raddensitymatrix}
    \rho_R=\text{Tr}_{BH}(\ket{\psi}\bra{\psi})=\sum_{i,j=1}^{N_f}\langle \Psi^j\ket{\Psi^i}\ket{i}\bra{j} \ ,
\end{equation}
so the matrix elements of the radiation state are given by the overlaps between black hole microstates,
\begin{equation}
   (\rho_R)_{ij}=\langle \Psi^j\ket{\Psi^i} \ . 
\end{equation}
We use the gravitational path integral to compute the Renyi entropies of the radiation state, which are given by 
\begin{equation} \label{Renyi}
    S_k=\frac{1}{1-k}\log \text{Tr}(\hat\rho_R^k)
\end{equation}
where $\hat \rho_R$ is the normalised density matrix, $\hat \rho_R=\frac{\rho_R}{\text{Tr}\rho_R}$ associated with the radiation state. 

\subsubsection{Purity of the radiation state}

We start by computing the purity, $S_2=\text{Tr}(\hat\rho_R^2)$ of the state of radiation obtained by the evaporation of a thin-shell black hole with $k_{max}>2$. Using (\ref{Raddensitymatrix}), we have
\begin{equation}
   \text{Tr}(\rho_R^2)=\sum_{i,j=1}^{N_f}|\langle \Psi^j\ket{\Psi^i}|^2
\end{equation}
Thus the purity of the radiation state is closely related to the square of the overlap between black hole microstates in (\ref{sqoverlapBrane},\ref{sqoverlapShell}). Denoting the contraction of flavour indices by a dotted line, the gravitational saddles contributing to the purity are
\begin{align}
   \text{Tr}(\rho_R^2)_{\rm saddles} &= 
\vcenter{\hbox{\includegraphics{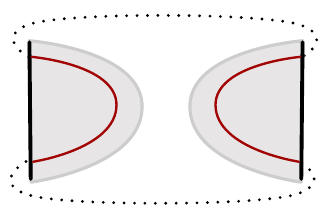}}}   
+
\vcenter{\hbox{\includegraphics{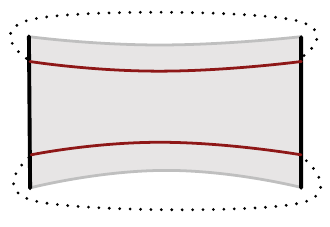}}}   \notag
   \\
   &= N_f e^{-2I_{BH}}+N_f^2 e^{-I_2}
\end{align}
Each flavour loop comes with a factor of $N_f$. As shown in the figure, the disconnected phase has one index loop while the connected wormhole phase has two index loops. Using $\text{Tr}(\rho_R)=Z_1=N_f e^{-I_{BH}}$ to normalise the density matrix $\rho_R$, the purity of the radiation state is given by
\begin{equation}
   \text{Tr}(\hat{\rho}_R^2)_{\rm saddles}=\frac{1}{N_f}+e^{-(I_2-2I_{BH})} \ .
\end{equation}
The first term is the contribution from the disconnected phase and the second term is the contribution from the connected phase. As we vary $N_f$, the saddles exchange dominance, with the connected phase dominating at large $N_f$. At $\log N_f= I_2-2I_{BH}$ , there is a phase transition and the connected phase begins to dominate and the purity assumes a constant value independent of $N_f$.

Let us now turn to the purity of the radiation state for the $B$-state black hole. In this case the saddles are
\begin{align}
     \text{Tr}(\rho_R^2)_{\rm saddles} &=
     \vcenter{\hbox{\includegraphics{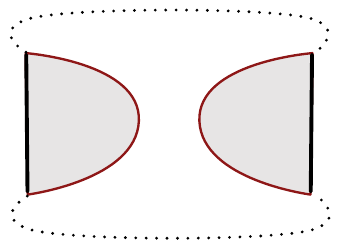}}}   
+
\vcenter{\hbox{\includegraphics{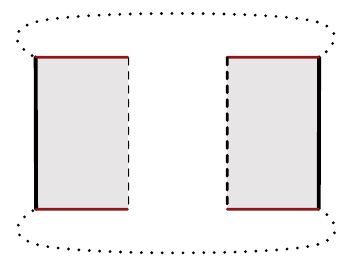}}}   \notag
   \\
   &= N_f e^{-2I_{BH}}+N_f^2e^{-2I_{cyl}}
\end{align}
Unlike the thin-shell case, there is no on-shell 2-boundary wormhole. However, there is another disconnected phase consisting of two copies of the cylinder. It also seems likely (based on an uncontrolled analytic continuation from $k < k_{max}$) that there is an off-shell contribution from the wormhole topology, which would add a second term $\sim N_f^2$ that we cannot calculate in the effective theory. On normalising the density matrix, the expression for purity is 
\begin{equation}\label{rhoss}
    \text{Tr}(\hat{\rho}_R^2)_{\rm saddles}=\frac{1}{N_f}+e^{-2(I_{cyl}-I_{BH})} \ .
\end{equation}
An off-shell wormhole, if it exists, could add a term $\sim e^{2I_{BH}}$, so we should bear this mind when considering whether to trust \eqref{rhoss}.
For a small number of flavours, the black hole phase dominates and the purity decreases with the increase in number of flavours. At $\log N_f=2(I_{cyl}-I_{BH})$, there is a phase transition and the cylinder phase begins to dominate and the purity takes a constant value independent of $N_f$. In 3d, the location of the phase transition point can be calculated analytically in terms of the parameters $\tau_0$ and $T$ using the results presented in Appendix \ref{app:bstatedetails} and Appendix \ref{app:cylinderdetails}:
\begin{equation}
   \log N_f=\frac{c\pi^2}{12\tau_0}-\frac{c}{3}(\tau_0+2\tanh^{-1}(T)) \ .
\end{equation}
In higher dimensions, the phase transition point can be determined numerically.

\subsubsection{Page curve for the thin-shell black hole}

We will now compute the entanglement entropy of the radiation,
\begin{equation}
  S_{vN}=-\text{Tr}(\hat \rho_R \log \hat \rho_R) \ .
\end{equation}
The von Neumann entropy is calculated using the replica trick,
\begin{equation}
  S_{vN}=-\partial_k\log\left(\frac{Z_k}{Z_1^k}\right)\bigg|_{k=1}
\end{equation}
where $Z_k$ is the gravitational partition function computing $\text{Tr}(\rho_R^k)$.  Let us first discuss the thin-shell black hole case. The geometries which have the correct boundary conditions to contribute to $\tr (\rho_R^k)$ for this case are
\begin{align}
    \text{Tr}(\rho_R^k)_{\rm saddles}
    &=
    \vcenter{\hbox{\includegraphics{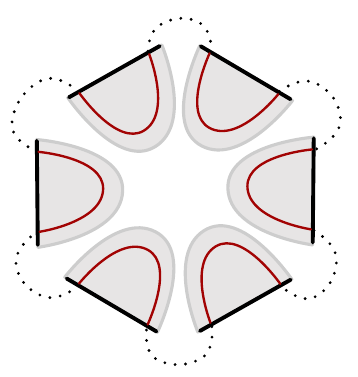}}}
    +\cdots
    +
     \vcenter{\hbox{\includegraphics{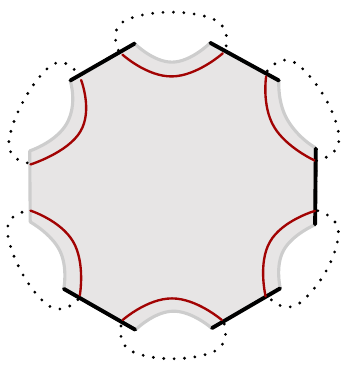}}}
    \\
    &= N_f e^{-kI_{BH}}+ \cdots + (N_f)^k e^{-I_k}
\end{align}
where the diagrams have $k=6$ and the dots are partially connected wormholes that are subleading away from the Page transition.
The disconnected phase has a contribution to the von Neumann entropy given by $S_{vN}^{\text{disc}}=\log N_f$ which shows that the radiation is in a maximally mixed state for small $N_f$. The contribution of the connected $\mathbb{Z}_k$ symmetric phase to the von Neumann entropy can be determined using the gravitational replica method of Lewkowycz and Maldacena \cite{Lewkowycz:2013nqa} which relates it to the entropy of the apparent horizon which is the fixed point of the $\mathbb{Z}_k$ replica symmetry,
\begin{equation}
   S_{vN}^{\text{conn}}=\partial_k(I_k-kI_1)\big |_{k=1}=S_0(\beta)
\end{equation}
where $S_0$ is one quarter the area of the apparent horizon at inverse temperature $\beta$ determined in terms of the parameters $\tau_0$ and $m$ using (\ref{tau0s}). Therefore, the entropy of radiation is 
\begin{equation} \label{ShellvN}
   S_{vN}=\text{min}\{\log N_f,S_0(\beta)\} \ .
\end{equation}
This result is analogous to the Page curve with the number of flavours playing the role of time. The saddles exchange dominance at the `Page time', $\log N_f^{\rm Page}=S_0(\beta)$ with the disconnected phase dominating at early times and the connected phase dominating at late times. This is consistent with the prediction of the island rule \cite{Penington:2019npb,Almheiri:2019psf} and we can thus view (\ref{ShellvN}) as a derivation of the island rule in this simple model for evaporation of a black hole with a time-symmetric apparent horizon formed from thin-shell collapse, as in \cite{Penington:2019kki,Almheiri:2019qdq}. For thin shells our toy model is nearly identical to the JT toy model described in \cite{Penington:2019kki} (except that we are limited to an on-shell analysis).

\subsubsection{Page curve for the $B$-state black hole}

%EDIT68
%The calculation of the entropy of radiation for the $B$-state black hole is more surprising.
For $B$-states, there are three gravitational phases contributing to the replica partition function,
\begin{align}
    \text{Tr}(\rho_R^k)_{\rm saddles} &= 
     \vcenter{\hbox{\includegraphics{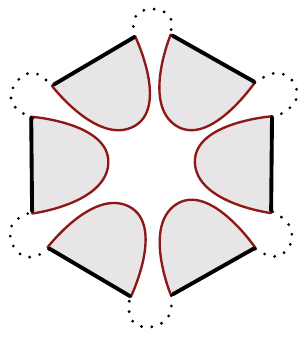}}}
     +
      \vcenter{\hbox{\includegraphics{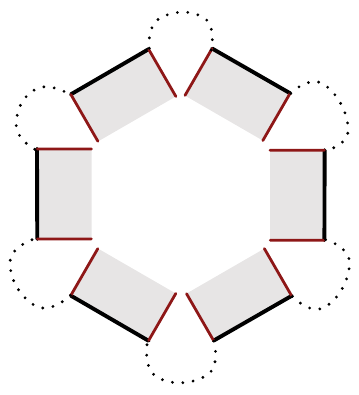}}}
      +
       \vcenter{\hbox{\includegraphics{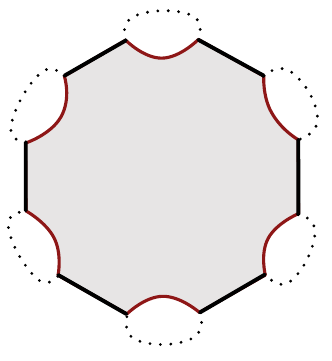}}}
       +\cdots
     \\
                                 &= N_f e^{-kI_{BH}}+(N_f)^k e^{-kI_{cyl}}+(N_f)^k e^{-I_k}+\cdots
\end{align}
where the dots are partially-connected wormholes and mixed phases. The middle term is the cylinder phase. The diagram shows $k=6$, but the last term --- the wormhole --- is only on-shell for $k < k_{max} \leq 2$. It is always on-shell for $k \to 1$ and therefore contributes to the von Neumann entropy. 
%EDIT68 next few paragraphs
The cylinder phase never dominates for $k \to 1$, so adding the three terms, we see that the entropy of radiation is given by
\begin{equation}
   S_{vN}=\text{min}\{\log N_f,S_0(\beta)\} \ .
\end{equation}
Again there is Page-like behavior with the transition at
   $\log N_f^{\text{Page}}=S_0(\beta)$.

\section{Discussion}

\subsubsection*{Summary}

We have considered pure-state black holes prepared by a Euclidean path integral that have a subdominant extremal surface at $t=0$, \textit{i.e.}, a time-symmetric apparent horizon. These black holes have two salient features in the Euclidean regime: 
\begin{itemize}
\item They are naturally assigned a coarse-grained entropy equal to one quarter the area of the extremal surface;
\item There is a sharp distinction, discernible to a local bulk observer, between the regions inside and outside the extremal surface.
\end{itemize}
What is the coarse-grained entropy in the dual CFT? To answer this question we followed a 3-step process:
\begin{enumerate}
\item Find the replica wormholes that come from branching around the extremal surface.
\item Calculate their gravitational action, ${I_k}$, and interpret it in the dual CFT as a replica partition function
\begin{align}\label{concrho}
e^{-I_k} = \tr \brho^k \ . 
\end{align}
\item  Read off from \eqref{concrho} the coarse-grained density matrix $\brho$, and the coarse-graining map $\C$ such that $\brho = \C(\rho)$. By design, this density matrix reproduces the apparent horizon entropy,
\begin{align}\label{areaentropyC}
S(\brho)  = \frac{\mbox{Area}(\gamma)}{4} \ .
\end{align} 
\end{enumerate}
This is like the derivation of the Ryu-Takayanagi formula by Lewkowycz and Maldacena \cite{Lewkowycz:2013nqa}, but in reverse. In that case, one starts with the left-hand side of the area-entropy relation --- it is the von Neumann entropy $S(\rho)$, where $\rho$ is a given CFT state --- and uses the AdS/CFT dictionary to calculate it in the bulk. This leads to the Ryu-Takayanagi formula \cite{Ryu:2006bv,Lewkowycz:2013nqa} and its fine-grained generalizations \cite{Hubeny:2007xt,Engelhardt:2014gca,Dong:2016hjy,Almheiri:2019psf,Penington:2019npb,Almheiri:2019qdq,Penington:2019kki}. In the present case, we started out knowing the right-hand side of \eqref{areaentropyC}, and our goal was to determine what $\brho$ appears on the left-hand side. So we followed the replica wormhole derivation in reverse. Along the way we showed that wormholes are related to CFT states with GHZ-like entanglement, and derived statistical properties of the CFT microstates dual to thin shells and EOW branes.

%EDIT67: Added next paragraph:
The way we derived the relationship between wormholes and GHZ-like states was somewhat indirect --- we calculated each side of the equation \eqref{zoverlap}, and found they are equal. There should be a more systematic derivation that starts from the standard AdS/CFT dictionary, applied to the CFT quantity $\langle \Psi_k|\Psi_k\rangle$. This would require a more detailed understanding of the gravitational path integral on a windmill; see figure \ref{fig:pinwheel}. This would also be interesting for the purpose of studying more elaborate types of $k$-party entanglement holographically, beyond the standard Ryu-Takayanagi formula which cannot be applied to this state (at least not directly).

The entire analysis has two significant limitations. First, the solutions were Euclidean. This may not be as severe a restriction as it sounds. Subdominant extremal surfaces in time-dependent, Lorentzian situations can probably be analyzed by analytically continuing the Euclidean wormholes to Lorentzian signature. Another interesting target for future study is the coarse-grained entropy of an apparent horizon that is not extremal. There is a gravitational procedure to turn any apparent horizon into an extremal surface \cite{Engelhardt:2017aux,Engelhardt:2018kcs} (see also \cite{Bousso:2020yxi,Engelhardt:2022qts}), so Euclidean methods might be applicable to this situation, as well.

The second limitation is that we have only studied black holes where the geometry is locally identical to an eternal black hole plus perturbations. The strategy sketched in steps (1), (2), and (3) and followed in this paper could be applied more generally but seems difficult to carry out in practice. For more general black holes it may be necessary to introduce an auxiliary code Hilbert space to define the coarse-graining procedure. For example, if we turn on a massless matter field (including gravitational excitations above the eternal black hole) then generally it cannot be assigned to either `inside' or `outside' the horizon, so our coarse-graining method does not apply.

This raises an important question:   Is holographic coarse graining a quantum channel? A quantum channel $\N_{A \to B}$ is a linear map from density matrices on $\H_A$ to density matrices on $\H_B$. The holographic coarse-graining map $\C$ defined in section \ref{s:replicaformalism} requires a decomposition of the state into the form $\rho = V \tilde{\rho} \tilde{V}^{\dagger}$, and without input from wormhole calculations on the gravity side, it is unclear how to define this decomposition uniquely. Therefore it is not clear whether $\C$ can be defined unambiguously on density matrices to formulate it as a quantum channel. Clearly it it acts like a quantum channel in some situations --- including all of the examples studied in this paper --- but the question is whether it is well defined on any density matrix as input. In appendix \ref{app:toychannel} we consider a simple toy model where there is a quantum channel $\C$ similar to the holographic coarse-graining map.

\bigskip

\noindent\textbf{Acknowledgments} \\
We thank Scott Collier, Sean Hartnoll, Alex Maloney, Dominik Neuenfeld, Pratik Rath, and Jorge Santos for discussions, as well as the organizers and participants of the workshops \textit{Low-dimensional Holography and Black Holes}, \textit{Qubits on the Horizon II}, and \textit{Physics Sessions Initiative 2022} for invigorating meetings, and Vijay Balasubramanian, Paul Ginsparg, Peter McMahon, Elliot Rosenberg and other members of the DOE QuantISED consortium for many conversations on related topics. This work is funded by NSF grant PHY-2014071. 

\appendix

\section{Details of EOW brane geometries}\label{app:bstatedetails}

In this appendix we provide additional details about the $B$-state geometries discussed in section \ref{ss:bstate}. The analysis of the brane equation of motion closely follows \cite{Fujita:2011fp,Cooper:2018cmb}. The action for gravity coupled to an EOW brane is given in \eqref{eowaction}.
The dynamical variables are the bulk metric $g_{\mu\nu}$ and the metric induced on the brane $h_{ij}$. Upon varying the action, we get the respective equations of motion,
\begin{equation} \label{eoweom}
    \begin{split}
        & R_{\mu\nu}-\frac{1}{2}g_{\mu \nu}R+\Lambda g_{\mu \nu}=0 \\
        & K_{ij}=Th_{ij}
    \end{split}
\end{equation}
The first equation is the Einstein equation for the bulk metric. The second equation can be thought of as a Neumann boundary condition which fixes the normal derivative of the bulk metric at the brane. We shall use this equation to determine the trajectory of the EOW brane through the bulk.

\subsection{The spherical boundary case}

\subsubsection{Brane trajectory}

We look for Euclidean AdS-Schwarzschild black hole solutions obtained by back-reaction of the brane. As described in section \ref{ss:bstate}, these are portions of the eternal black hole \eqref{ds2bh} cutoff at the brane trajectory, $\tau = u(r)$. (In this appendix, $u(r)$ means the EOW brane trajectory, called $u_B(r)$ in the main text.) 

For a spherically symmetric brane in the eternal black hole geometry, the Neumann condition is $K_{\Omega\Omega}=Th_{\Omega\Omega}$ where $\Omega$ represents any of the tranverse $S^{d-1}$ directions. This reduces to
\begin{equation} \label{branetrajectoryeqn}
   \frac{fu'}{\sqrt{fu'^2+\frac{1}{f}}}=\pm T r  .
\end{equation}
We thus have a first-order equation of motion for the brane trajectory,
\begin{equation}
    u'=\pm\frac{Tr}{f\sqrt{f-(Tr)^2}}
\end{equation}
The turning point for the brane where $u'\to \infty$ occurs at $r=r_0$ determined by the solution to $f(r_0)=T^2r_0^2$. We choose the endpoints of the brane to be at $\pm \tau_0$ and the turning point to be behind the horizon, so $u(r_0) =  \frac{\beta}{2}$.  Integrating (\ref{branetrajectoryeqn}) between the brane turning point and infinity, we have
\begin{equation} \label{braneendpt}
    \tau_0=\frac{\beta}{2}-\int_{r_0}^{\infty}\frac{dr}{f}\frac{Tr}{\sqrt{f-(Tr)^2}}
\end{equation}
This relation should be regarded as an implicit relation to determine the temperature of the black hole in terms of the brane tension $T$ and the brane end-point $\tau_0$. By numerically evaluating the integral $\int_{r_0}^{\infty}\frac{dr}{f}\frac{Tr}{\sqrt{f-(Tr)^2}}$, we can check that for any choice of the parameters, $\tau_0,T>0$, it is larger than $\frac{\beta}{4}$ which means the brane always goes behind the horizon.  The analysis simplifies considerably in 3d where the black holes in question are static BTZ black holes, with
\begin{equation}
    f(r)=r^2-r_H^2 , \qquad 
    \beta=\frac{2\pi}{r_H} \ .
\end{equation}
In this case, $\tau_0=\frac{\beta}{4}$ for any value of the brane tension, because
\begin{equation}
    \int_{r_0}^{\infty}\frac{dr}{f}\frac{Tr}{\sqrt{f-(Tr)^2}}=\frac{\pi}{2r_H}=\frac{\beta}{4} \ ,
\end{equation}
where $r_0=\frac{r_H}{\sqrt{1-T^2}}$.
This means that the temperature of the black hole formed by the back-reaction of the brane depends on the brane end-point, but it is independent of the tension. This is a feature of the brane dynamics in 3d and does not carry over to higher dimensions.

\subsection{On-shell action}

The action in \eqref{eowaction} requires counterterms to render it finite. There are counterterms associated to the brane that we will not find explicitly, but we will use a trick to calculate the regulated action.

After plugging in the brane equation of motion $K = Td$, the contribution to the on-shell action at the brane is
\begin{align}
I_{\rm brane} &=-\frac{ T}{8\pi G} \int_{\rm brane} d^d y \sqrt{h}  \ .
\end{align}
Write the full action as $I = I_{\rm grav} + I_{\rm brane}$, where $I_{\rm grav}$ is the usual gravitational action, including the standard counterterm and GHY term at the asymptotic boundary. We can split the gravitational action into two parts,
\begin{equation}
    I_{\text{grav}}=I_{\text{in}}+I_{\text{out}}
\end{equation}
where $I_{\text{in}}$ is the contribution to the bulk action from the region bounded by the brane and the angular slices $\tau=\pm \tau_0$, and $I_{\text{out}}$ is the contribution from the outside wedge. The Einstein equations set $R = -d(d+1)$.  Let us evaluate each of the three pieces in the total action,
\begin{equation}
    I=I_{\text{grav}}+I_{\text{brane}}=I_{\text{in}}+I_{\text{out}}+I_{\text{brane}}
\end{equation}
By definition, $I_{\text{out}}$ corresponds to a portion of the eternal black hole,
\begin{equation}
    I_{\text{out}}=\frac{2\tau_0}{\beta}I_0(\beta)
\end{equation}
where $I_0$ is the renormalised action for the eternal black hole. The bulk action for the interior region is
\begin{equation}
    \begin{split}
        I_{\text{in}}= & \frac{-1}{16\pi G}2\Omega_d\int_{\tau_0}^{\frac{\beta}{2}}d\tau \int_{r_H}^{u^{-1}(\tau)}dr r^{d-1} (-2d)\\
        = & \frac{\Omega_d}{4\pi G} \int_{r_0}^{\infty}dr |u'(r)|(r^d-r_H^d)
    \end{split}
\end{equation}
where $\Omega_d=\frac{2\pi^{\frac{d}{2}}}{\Gamma(\frac{d}{2})}$ is the area of the unit transverse sphere. The brane term evaluates to give
\begin{equation}
    \begin{split}
        I_{\text{brane}}=&-\frac{1}{8\pi G}2\Omega_d\int_{r_0}^{\infty}dr r^{d-1}\sqrt{\frac{1}{f}+f(u')^2}\,\,T\\
        =& -\frac{\Omega_d}{4\pi G}\int_{r_0}^{\infty}dr r^{d-2} f|u'|\\
        =& -\frac{\Omega_d}{4\pi G}\int_{r_0}^{\infty}dr |u'(r)| \left[(r^d-r_H^d)+(r^{d-2}-r_H^{d-2})\right]
    \end{split}
\end{equation}
where we used the brane equation of motion (\ref{branetrajectoryeqn}).
% in going from the first line to the second. 
Observe that the contribution from the left wedge, including both the bulk and brane terms, is
\begin{equation}\label{bstateIL}
\tilde{I}_L :=    I_{\text{in}}+I_{\text{brane}}=-\frac{\Omega_d}{4\pi G}\int_{r_0}^{\infty}dr |u'(r)|(r^{d-2}-r_H^{d-2})
\end{equation}
Both $I_{\text{in}}$ and $I_{\text{brane}}$ are divergent, so we should first regulate the divergence. We shall come back to the issue of renormalisation later on in this section.
The total action, before renormalising, is
\begin{align}
I &= \frac{2\tau_0}{\beta}I_0(\beta) + \tilde{I}_L(\beta) \ .
\end{align}
In 3D gravity, $d=2$, we see that $I_L = 0$; the brane term cancels with the gravitational contribution from the left wedge, so there is no need to renormalise. In 3D we also have $\tau_0 = \frac{\beta}{4}$, so the total action is
\begin{align}\label{d2action}
I^{(d=2)} &= \frac{2\tau_0}{\beta}I_{\rm BTZ}(\beta) = - \frac{c\pi^2}{24\tau_0} \ ,
\end{align}
independent of the brane tension. This calculation was also done in \cite{Fujita:2011fp,Cooper:2018cmb}.

\subsubsection*{Renormalisation}

$\tilde{I}_L$ defined in \eqref{bstateIL} is 
divergent for $d>2$ due to the behaviour of the integrand as $r\to \infty$. In terms of a large $r=R$ cutoff, $\tilde{I}_L \sim -\frac{\Omega_d}{4\pi G}\frac{T}{\sqrt{1-T^2}}R^{d-3}$ so it diverges logarithmically in $d=3$ and as a power law in $d>3$. This term has to be renormalised. To this end, we use the identity\footnote{
This equation also tells us how to regulate $\tilde{I}_L'(\beta)$. If $\tilde{I}_L'(\beta)$ is calculated directly by taking the derivative of \eqref{bstateIL}, it is finite, but its value depends on a prescription to calculate the derivative, since we are manipulating infinite quantities in the intermediate steps. After a long calculation, we find that the correct prescription to reproduce \eqref{miracleidentity} is to define the derivative of the infinite quantity in \eqref{bstateIL} by the identity  $\p_{\beta} \int_{r_0}^{\infty} H(r) \equiv \lim_{\epsilon \to 0} \left[ -(\p_\beta r_0) H(r_0+\epsilon) + \int_{r_0+\epsilon}^{\infty} \p_\beta H(r)\right]$, which is obvious for finite integrals but here it acts as a choice of regulator. The resulting expression for $\tilde{I}_L'(\beta)$ is finite and unambiguous, and we have checked analytically in $d=3$ and numerically in higher dimensions that it satisfies \eqref{miracleidentity}.
\label{regfootnote}}  
%\footnote{We derive this identity in the succeding discussion about fractional wormholes.},
\begin{equation} \label{miracleidentity}
    I_L '(\beta)=-2S_0(\beta)(1-\beta \partial_{\beta})\tau_0(\beta)
\end{equation}
derived in section \ref{s:bulkentropy}, 
where $S_0(\beta)=-(1-\beta\partial_{\beta})I_0(\beta)$ is the entropy of the eternal black hole at inverse temperature $\beta$, and $I_L$ is the renormalised version of $\tilde{I}_L$.  Using the result (\ref{inbrane}) for the planar boundary case derived in the next section as a boundary condition, we have $I_L(\beta)\to 0$ as $\beta \to 0$. Therefore, we can express the renormalised action as
\begin{equation}
    I_L(\beta)=-2\int_0^{\beta}dx S_0(x)(1-x\partial_x)\tau_0(x) \ .
\end{equation}
Thus, the renormalised on-shell action for spherical brane geometries is
\begin{equation}
     I =\frac{2\tau_0}{\beta}I_0(\beta)+I_L(\beta)
\end{equation}
with $\beta$ determined from (\ref{braneendpt}). This is the equation for the action used in the main text, see \eqref{singleI}.

\subsection{The planar boundary case}

We will now take the planar limit $r_H\to\infty$ of the results of the previous section. This corresponds to studying black brane solutions obtained by back-reaction of the EOW brane. In this limit, the analysis simplifies considerably and some of the results can also be expressed in closed form in $d>2$. It is convenient to work with coordinates where the dependence on $r_H$ of the black hole metric from the previous section,
\begin{equation}
ds^2=f(r)d\Tilde{\tau}^2+\frac{dr^2}{f(r)}+r^2d\Omega^2_{d-1}
\end{equation}
has been scaled out. To this end, we define rescaled coordinates, $(\rho,\tau,x)$ related to the black hole coordinates by,
\begin{equation}
    \begin{split}
        & \left(\cosh(\frac{d\rho}{2})\right)^{\frac{2}{d}}=\frac{r_H}{r}\\
        & \tau=\frac{\beta d}{4\pi}r_H\Tilde{\tau} \\
        & x=\frac{\beta d}{4\pi}r_H \phi
    \end{split}
\end{equation}
where $x$ and $\phi$ collectively represent the planar and spherical transverse directions respectively. The planar horizon is at $\rho=0$ in these coordinates. Now, $\tau\sim\tau+\beta$ so $\beta$ should be regarded as an inverse temperature parameter for the black brane geometry. In these coordinates, the metric for the black brane takes the form,
\begin{equation}
    ds^2=d\rho^2+\Tilde{g}^2(\rho)d\tau^2+\Tilde{h}^2(\rho)dx_{d-1}^2
\end{equation}
where the coefficient functions are
\begin{equation}
    \begin{split}
        & \Tilde{h}(\rho)=\frac{4\pi}{\beta d}\left(\cosh (\frac{d\rho}{2})\right)^{\frac{2}{d}}\\
        & \Tilde{g}(\rho)=\Tilde{h}(\rho)\tanh (\frac{d\rho}{2})
    \end{split}
\end{equation}

\subsubsection{The brane trajectory}

We parametrise the EOW brane trajectory by $\tau=u(\rho)$. The Neumann boundary condition gives the brane equation of motion,
\begin{equation} \label{braneintegral}
    u'=\pm\frac{1}{\Tilde{g}}\frac{A}{\sqrt{1-A^2}}
\end{equation}
where $A(\rho)=T\coth(\frac{d\rho}{2})$. The turning point of the brane trajectory occurs at, $A(\rho_0)=1$ and has a simple expression, $\rho_0=\frac{2}{d}\coth^{-1}(\frac{1}{T})$. 
%A solution to (\ref{braneintegral}) exists only if the brane tension is smaller than a critical value, $T<T_c=1$. 
On integrating (\ref{braneintegral}), we can express the temperature in terms the brane end point $\tau_0$ and brane tension $T$,
\begin{equation}
     \tau_0=\frac{\beta}{2}-\int_{\rho_0}^{\infty}\frac{d\rho}{\Tilde{g}}\frac{A}{\sqrt{1-A^2}}
\end{equation}
In what follows, we shall not need the explicit form of the integral. However, it is important to note that the dependence on $\beta$ factors out and we express the above result as
\begin{equation} \label{planartau0}
    \tau_0=\frac{\beta}{2}(1-\alpha(T))
\end{equation}
where
\begin{equation} \label{FT}
       \alpha(T)=\frac{1}{\pi}\int_{\rho_0}^{\infty}\frac{d\rho}{g}\frac{A}{\sqrt{1-A^2}}
\end{equation}
with $g(\rho)=\frac{\beta}{2\pi}\Tilde{g}(\rho)$, so $\alpha(T)$ has no temperature dependence. In the tensionless limit, $\alpha(T)\to \frac{1}{2}$ and numerically we can verify that for any non-zero value of the brane tension below the critical value, $\alpha(T)>\frac{1}{2}$.  In 3d, $\alpha(T)=\frac{1}{2}$.

\subsubsection{On-shell action}

We split the action into three terms,
\begin{equation}
   I=I_{\text{in}}+I_{\text{out}}+I_{\text{brane}}
\end{equation}
 with each term defined in a way similar to that in the spherical boundary case. We have the following expressions for each term,
\begin{equation}
 \begin{split}
       & I_{\text{out}}=\frac{2\tau_0}{\beta}I_0(\beta) \\
       & I_{\text{in}}=-I_{\text{brane}}=\frac{V_{\perp}}{8\pi G}\left(\frac{2}{d}\right)^d \int_{\rho_0}^{\rho_{\text{max}}}d\rho |u'(\rho)|(\cosh(d\rho)-1)
 \end{split}
\end{equation}
Here, $\rho_{\text{max}}$ is a large cutoff introduced to regulate the divergence and $V_{\perp}$ is the area of the tranverse $\mathbb{R}^{d-1}$. $I_0(\beta)$ is the renormalised action for the eternal black brane and by a simple scaling argument, it must take the form, $I_0(\beta)=-bV_{\perp}\beta^{1-d}$ where $b$ is a known positive constant.  Thus, we see that upon regulating the divergence,
\begin{equation} \label{inbrane}
I_L=   I_{\text{in}}+I_{\text{brane}}=0
\end{equation}
So, the renormalised on-shell action for this geometry is  
\begin{equation}
  I(\tau_0,T)=\frac{2\tau_0}{\beta}I_0(\beta)=-bV_{\perp}(2\tau_0)^{1-d}(\alpha(T))^d
\end{equation}

\section{Details of thin shell solutions}\label{app:thinshelldetails}

In this appendix we solve the equations for the Euclidean thin-shell black holes described in section \ref{ss:thinshell}; see figure \ref{fig:thinshell}. The method is similar to \cite{Keranen:2015fqa} but there are some sign differences coming from the fact that we assume the shell is behind the horizon and work in Euclidean signature.

Consider a spherically symmetric thin shell of pressureless perfect fluid (i.e. `dust') separating region $\mathcal{M}_-$ which is a portion of global AdS from region $\mathcal{M}_+$ which is a portion of the AdS-Schwarzschild black hole. The two sides $\mathcal{S}_\pm$ of the thin shell are at the boundaries of the two regions $\partial \mathcal{M}_\pm$ respectively. We choose coordinate systems covering the two regions,
\begin{equation} \label{2coord}
    \begin{split}
        & \mathcal{M}_+ : \quad x_+^\mu= (\tau_+,r_+,\Omega)\\
        & \mathcal{M}_- : \quad x_-^\mu=  (\tau_-,r_-,\Omega)
    \end{split}
\end{equation}
where the metrics take the form
\begin{equation}
    ds^2_{\pm}=f_{\pm}(r_\pm)d\tau_\pm^2+\frac{dr_\pm^2}{f_{\pm}(r_\pm)}+r_{\pm}^2d\Omega^2
\end{equation}
with $\tau_+\sim \tau_++\beta$ and $\tau_- \in \mathbb{R}$ where $\beta$ is the inverse temperature of the black hole. Since the geometry is spherically symmetric, we may choose the same set of angular coordinates denoted collectively by $\Omega$ for both regions. The functions $f_\pm$ are given by
\begin{equation}
    \begin{split}
       & f_+(r)=1+r^2-\frac{M}{r^{d-2}}\\
       & f_-(r)=1+r^2
    \end{split}
\end{equation}
where $M = r_H^{d-2}(1+r_H^2)$ is the ADM mass of the black hole. We will glue a region of the form $r_- < r_-^{\rm max}(\tau_-)$ to a region $r_+ < r_+^{\rm max}(\tau_+)$. The signs of these inequalities are appropriate to a shell created by an operator acting on the vacuum state (this selects  $r_- < r_-^{\rm max}(\tau_-)$ as the physical region) which is behind the horizon at $t=0$ (this selects $r_+ < r_+^{\rm max}(\tau_+)$); see figure \ref{fig:thinshell}. 
The equation of motion for the shell is derived from the Israel junction conditions. It is most convenient to formulate the junction conditions in terms of coordinates intrinsic to the shell (covering $\mathcal{S}_\pm$) which we choose to be,
\begin{equation}
    \mathcal{S}: \quad y^a=  (\ell, \Omega)
\end{equation}
where $\ell$ is the proper distance on the shell, as measured along a curve of fixed $\Omega$. We parametrise the trajectory of the shell by $(\tau_\pm(\ell),r_\pm(\ell))$ respectively in the two coordinate systems (\ref{2coord}). 
By the definition of $\ell$, we have,
\begin{equation} \label{dlsquared}
    f_{\pm}(r_\pm(\ell))\dot{\tau}^2_\pm+\frac{\dot{r}^2_\pm}{f_{\pm}(r_\pm(\ell))}=1
\end{equation}
where dots are derivatives with respect to $\ell$. 
The metrics induced on the two sides of the shell are
\begin{equation}
    h_{ab}(\mathcal{S}_\pm)=(e^\mu_a e^\nu_b g_{\mu\nu})_\pm
\end{equation}
where $(e^\mu_a)_\pm=\frac{\partial x^\mu_\pm}{\partial y^a}$ are the respective Jacobians. Explicitly, the induced metrics take the forma
\begin{equation} \label{indds2}
    ds^2(\mathcal{S}_\pm)=d\ell^2+r^2_\pm(\ell)d\Omega^2
\end{equation}
The first junction condition requires that the metric is continuous across the shell,
\begin{equation}
    h_{ab}(\mathcal{S}_+)=h_{ab}(\mathcal{S}_-)
\end{equation}
which using (\ref{indds2}) implies $ r_+(\ell)=r_-(\ell)$ which we shall call $r(\ell)$ henceforth. Using (\ref{dlsquared}), we have
\begin{equation} \label{taupm}
    \begin{split}
        & \dot{\tau}_+(\ell)=-\frac{1}{f_+(r)}\sqrt{f_+(r)-\dot{r}^2}\\
        & \dot{\tau}_-(\ell)=\frac{1}{f_-(r)}\sqrt{f_-(r)-\dot{r}^2}
    \end{split}
\end{equation}
We shall only consider the case where the turning point of the trajectory is on the $\tau_+=\frac{\beta}{2}$ slice in the black hole coordinates, so that the shell is behind the horizon at $t=0$. Furthermore, we choose to work with the branch of the trajectory where $\tau_+(\ell)<\frac{\beta}{2}$. In this branch, $\frac{d\tau_+}{d\ell}<0$ and $\frac{d\tau_-}{d\ell}>0$ which explains the signs in (\ref{taupm}).
We can obtain a differential equation for the shell trajectory by solving the second junction condition,
\begin{equation} \label{junc2}
    \Delta K_{ab}-\Delta K h_{ab}=-S_{ab}
\end{equation}
where $\Delta K_{ab}\equiv K_{ab}(\mathcal{S}_+)-K_{ab}(\mathcal{S}_-)$ with $K_{ab}(\mathcal{S}_\pm)$ being the extrinsic curvatures of $\mathcal{S}_\pm$ respectively. $S_{ab}$ is the surface stress tensor for the shell. Since we assume that the shell is made of a pressureless perfect fluid, we can parametrise the surface stress tensor as
\begin{equation}
    S_{ab}=-\sigma(\ell)U_a U_b
\end{equation}
with $U=\partial_{\ell}$ being the normalised velocity field ($U^aU_a=1$) for the shell and $\sigma(\ell)>0$ is a measure of the rest mass density of the shell. We can compute the extrinsic curvatures using,
\begin{equation}
    K_{ab}=e^{\mu}_a e^{\nu}_b \nabla_{\mu}n_{\nu}
\end{equation}
The unit normals to $\mathcal{S}_\pm$ are chosen to point away from the region $\mathcal{M}_-$. They can be computed using $n_\mu e^\mu_a=0$ and $n_\mu n^\mu=1$ and are given by
\begin{equation}
    \begin{split}
        & n_\mu^+dx^\mu_+=-\dot{r}d\tau_++\dot{\tau}_+dr_+ \\
        & n_\mu^-dx^\mu_-=-\dot{r}d\tau_-+\dot{\tau}_-dr_- \ .
    \end{split}
\end{equation}
The non-vanishing components of the extrinsic curvature can then be calculated to be
\begin{equation} \label{Extrinsiccurvature}
    \begin{split}
        & K_{\ell \ell}(\mathcal{S}_\pm)=\mp\frac{\dot{H}_\pm}{\dot{r}}\\
        & K_{\Omega \Omega}(\mathcal{S}_\pm)=\mp\frac{H_\pm}{r}h_{\Omega \Omega}
    \end{split}
\end{equation}
where $H_\pm(\ell)=\sqrt{f_\pm(r)-\dot{r}^2}$ and $\Omega$ collectively represents the angular directions. Contracting the junction condition, (\ref{junc2}) with $h^{ab}$, we have $\Delta K=- \frac{\sigma(\ell)}{d-1}$. The two independent equations of (\ref{junc2}) using (\ref{Extrinsiccurvature}) read,
\begin{equation}
    \begin{split}
        & \frac{H_++H_-}{r}=\frac{\sigma(\ell)}{d-1}\\
        & \frac{\dot{H}_++\dot{H}_-}{\dot{r}}=-\sigma(\ell)\frac{d-2}{d-1}
    \end{split}
\end{equation}
The two equations can be solved to give $\frac{H_++H_-}{r}=c r^{2-d}$ where $c$ is a constant. We shall choose the constant such that,
\begin{equation}
    \sigma(\ell)=(d-1)\frac{m}{r^{d-1}}
\end{equation}
so that $m$ can be interpreted as a rest mass for the shell. Thus, we have,
\begin{equation} \label{eomrdot}
    \sqrt{f_--\dot{r}^2}+ \sqrt{f_+-\dot{r}^2}=\frac{m}{r^{d-2}}
\end{equation}
The turning point ($\dot{r}=0$) for the shell trajectory denoted $r_0$ satisfies,
\begin{equation}
    r_0^{d-2}=\frac{m}{\sqrt{f_-(r_0)}+\sqrt{f_+(r_0)}}
\end{equation}
(\ref{eomrdot}) can be solved for $\dot{r}$ which is chosen to be non-negative to obtain
\begin{equation} \label{rdot}
    \dot{r}^2=1+r^2-\frac{r^{2(d-2)}}{4m^2}\left(\frac{m^2}{r^{2(d-2)}}+\frac{M}{r^{d-2}}\right)^2 \equiv B(r)
\end{equation}
Using (\ref{taupm}), we can write down a first order differential equation for the shell trajectory in the black hole coordinates,
\begin{equation}
    \frac{d\tau_+}{dr}=-\frac{1}{f_+(r)}\sqrt{\frac{f_+(r)}{B(r)}-1}
\end{equation}
In the notation of section \ref{s:wormholes}, $f_+=f$ and $\tau_+ = u_S$, so the solution is
\begin{align}\label{uSsolution}
u_S(r') &= \frac{\beta}{2} - \int_{r_0}^{r'}\frac{dr}{f(r)}\sqrt{\frac{f(r)}{B(r)}-1} \ .
\end{align}
The shell hits the boundary at $\tau_0 = u_S(\infty)$. 
%\begin{equation}
%    \tau_0=\frac{\beta}{2}-\int_{r_0}^{\infty}\frac{dr}{f_+(r)}\sqrt{\frac{f_+(r)}{B(r)}-1}
%\end{equation}
This should be viewed as an implicit equation to determine the ADM mass $M$ or equivalently the inverse temperature $\beta$ of the black hole in terms of the parameters $\tau_0,m$. 

\begin{figure}
\begin{center}
\begin{overpic}[width=3in,grid=false]{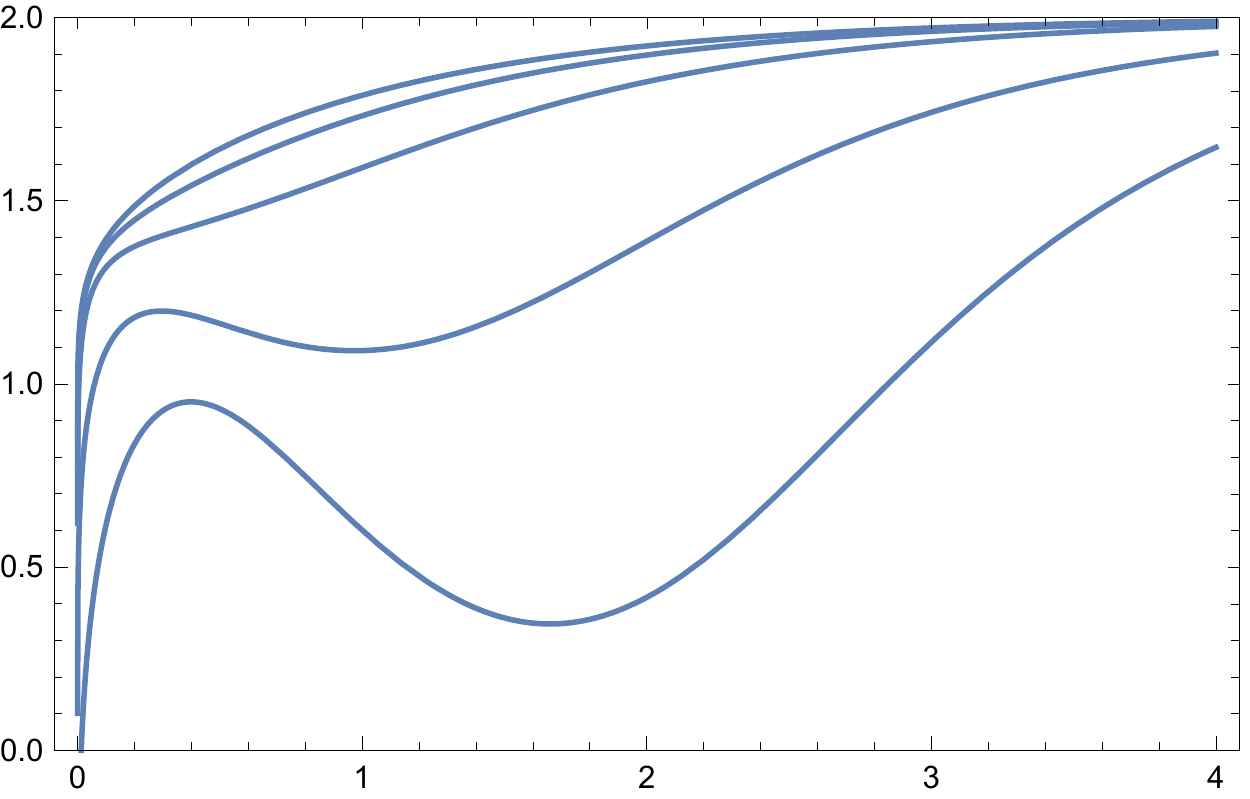}
\put (-10,30) {$\frac{4\tau_0}{\beta}$}
\put (50,-5) {$\lambda$}
\end{overpic}
\end{center}
\caption{Landing point $\tau_0$ of the thin shell, in units of $\frac{\beta}{4}$, as a function of $r_0 = r_H e^{\lambda}$. From bottom to top, $r_H = 0.1, 0.2, 0.5, 1, 10$.\label{fig:taushell}}
\end{figure}

For numerical purposes it is convenient to parameterize solutions by $r_H$ and $r_0$. By calculating the integral numerically for various parameters we find the following behavior. For any $r_H$, if we take $r_0 \gg r_H$, then the shell stays in the region $|\tau| > \frac{\beta}{4}$, and for very large $r_0$ the endpoint $\tau_0$ approaches $\frac{\beta}{2}$. Therefore, in this limit the allowed number of boundaries $k_{max}$ in the on-shell wormhole formed by taking a quotient becomes large. Figure \ref{fig:taushell} shows the behavior for other parameters.

\section{Details of the EOW brane cylinder phase}\label{app:cylinderdetails}

\subsection{Brane trajectory}

In this appendix, we analyse the disconnected phase for the EOW brane which consists of a pair of EOW branes embedded in vacuum global AdS, which has the metric
\begin{equation}
    ds^2=f(r)d\tau^2+\frac{dr^2}{f(r)}+r^2d\Omega^2_{d-1}
\end{equation}
where $\tau \in \mathbb{R}$ and $f(r)=1+r^2$. The EOW brane trajectory $\tau=u(r)$ obeys a first order differential equation similar to that for the BH case,
\begin{equation} \label{EOWtrajectory}
    \frac{d\tau}{dr}=\pm \frac{Tr}{f\sqrt{f-(Tr)^2}}
\end{equation}
with the $\frac{d\tau}{dr}<0$ branch corresponding to the brane in the $\tau>0$ region and the $\frac{d\tau}{dr}>0$ branch corresponds to the brane in the $\tau<0$ region. The choice of sign is dictated by the Neumann boundary condition, $K=Td$, which implies that the outward-pointing normal has positive expansion. The equation (\ref{EOWtrajectory}) can be integrated analytically to obtain the trajectory of the brane at $\tau>0$,
\begin{equation} \label{EOWgr}
    u(r)=\tau_0+ \tanh^{-1}\left(\frac{T}{\sqrt{T^2+(1-T^2)(1+r^2)^2}}\right)
\end{equation}
The turning point $\frac{d\tau}{dr}=0$ for the brane trajectory occurs at $r=0$ which using (\ref{EOWgr}) lands at
\begin{equation}
    \tau_{\text{max}}=\tau_0+ \tanh^{-1}(T)
\end{equation}
which can also be written as
\begin{equation}
    T=\tanh (\tau_{\text{max}}-\tau_0) \ .
\end{equation}

\subsection{On-shell action}

We split the action into three terms,
\begin{equation}
    I=I_{\text{in}}+I_{\text{out}}+I_{\text{brane}}
\end{equation}
where $I_{\text{out}}$ is the bulk action (EH+GHY+AdS counter-terms) evaluated in the cylindrical region between $\tau=\pm \tau_0$ and the AdS boundary. $I_{\text{in}}$ is the bulk action (EH term) evaluated in the dome region bounded by the brane and the $\tau=\tau_0$ slice, including its time-reversed partner. $I_{\text{brane}}$ is the brane action which includes the brane GHY term and the brane area term. Thus
\begin{equation}
    I_{\text{out}}=I_\text{AdS}(2\tau_0)
\end{equation}
where $I_\text{AdS}(2\tau_0)$ is the renormalised action for thermal AdS at $\beta=2\tau_0$. 
The bulk action for the interior region of both the domes added together is
\begin{equation}
    \begin{split}
        I_{\text{in}}= & \frac{-1}{16\pi G}2\Omega_d\int_{\tau_0}^{\tau_{\text{max}}}d\tau \int_{0}^{u^{-1}(\tau)}dr r^{d-1} (-2d)\\
        = & \frac{\Omega_d}{4\pi G} \int_{0}^{\infty}dr |u'(r)|r^d
    \end{split}
\end{equation}
The brane term evaluates to give,
\begin{equation}
    \begin{split}
        I_{\text{brane}}=&-\frac{1}{8\pi G}2\Omega_d\int_{0}^{\infty}dr r^{d-1}\sqrt{\frac{1}{f}+f(u')^2}\,\,T\\
        =& -\frac{\Omega_d}{4\pi G}\int_{0}^{\infty}dr r^{d-2} f|u'|\\
        =& -\frac{\Omega_d}{4\pi G}\int_{0}^{\infty}dr |u'(r)| \left(r^{d-2}+r^d\right)
    \end{split}
\end{equation}
Observe that,
\begin{equation}
    I_{\text{in}}+I_{\text{brane}}=-\frac{\Omega_d}{4\pi G}\int_{0}^{\infty}dr |u'(r)|r^{d-2}
\end{equation}
So the total action is given by
\begin{equation} \label{cylaction}
    I(\tau_0,T)= I_{\text{AdS}}(2\tau_0)-\frac{\Omega_d}{4\pi G}\int_{0}^{\infty}dr |u'(r)|r^{d-2}
\end{equation}
The second term is divergent for $d>2$ due to the behaviour of the integrand as $r\to \infty$, so the action has to be renormalised for $d>2$. For $d=2$, the total action can be evaluated analytically and is given by
\begin{equation}
\begin{split}
    I(\tau_0, T)=& I_{\text{AdS}}(2\tau_0)-\frac{1}{2G} \tanh^{-1} (T) \\
    =& -\frac{c}{6}\left(\tau_0+2\tanh^{-1} (T)\right)
    \end{split}
\end{equation}

\subsubsection*{Renormalisation for $d>2$}

As in appendix \ref{app:bstatedetails}, we will renormalise the action by a trick that does not require finding the explicit counterterms. We have already renormalised the action of the black hole phase, so the idea is to renormalise the cylinder phase by defining the action relative to the black hole. 
Let us define the divergent term in (\ref{cylaction}) regulated using a large cutoff at $r=\hat{R}$,
\begin{equation} \label{IGtilde}
  \Tilde{I}_G(T):=-\frac{\Omega_d}{4\pi G}\int_{0}^{\hat{R}}dr |u'(r)|r^{d-2}
\end{equation}
In terms of the cutoff, $ \Tilde{I}_G(T)\sim -\frac{\Omega_d}{4\pi G}\frac{T}{\sqrt{1-T^2}}\hat{R}^{d-3}$ at leading order. We define a renormalised version of the term (\ref{IGtilde}) by
\begin{equation} \label{IG}
  I_G := \Tilde{I}_G-\Tilde{I}_L+I_L
\end{equation}
To show that $I_G$ is finite, we need to show that $\Tilde{I}_G-\Tilde{I}_L$ is finite. However, since $\Tilde{I}_G$ and $\Tilde{I}_L$ are evaluated using different cutoffs $\hat{R}$ and $R$ respectively, we need to first find a relation between the two cutoffs. To this end, we relate the cutoffs to a common cutoff at $z=\epsilon$ in the Fefferman-Graham coordinates using
\begin{equation}
  \frac{dz}{z}=-\frac{dr}{\sqrt{f(r)}}
\end{equation}
Upon integrating this equation at large $r$ using the condition that $r= \frac{1}{z}$ for large $r$ to fix the integration constant, we get
\begin{equation} \label{FG}
   \begin{split}
      & \hat{R}=\frac{1}{\epsilon}-\frac{\epsilon}{4} \\
      & R=\frac{1}{\epsilon}-\frac{\epsilon}{4}+\frac{r_H^{d-2}(1+r_H^2)}{2d}\epsilon^{d-1}+O(\epsilon^{d})
    \end{split}
\end{equation}
The difference between the integrands in $\Tilde{I}_G$ and $\Tilde{I}_L$ is $O(\frac{1}{r^2})$ for large $r$, so the possible divergence in $ \Tilde{I}_G-\Tilde{I}_L$ comes from
\begin{equation} \label{GLdiff}
    \Tilde{I}_G-\Tilde{I}_L \supset -\frac{\Omega_d}{4\pi G}\int_{R}^{\hat{R}}|u'(r)|r^{d-2}
\end{equation}
The integrand in this term is $O(r^{d-4})$ for large $r$. So, using (\ref{FG}), we can check that the term in (\ref{GLdiff}) is $O(\frac{1}{R^3})$ so vanishes in the limit $R\to \infty$. Thus, we have shown that $\Tilde{I}_G-\Tilde{I}_L $ is finite which means that $I_G$ in (\ref{IG}) is also finite. However, to check that the renormalisation scheme in (\ref{IG}) is consistent, we need to show that $I_G$ is independent of $r_H$. This is not obvious since the RHS of (\ref{IG}) involves expressions which depend on $r_H$. Treating the cutoff, $\hat{R}=\hat{R}(R,r_H)$, consider
\begin{equation} \label{rHder}
    \partial_{r_H} I_G = \partial_{r_H}\Tilde{I}_G-\partial_{r_H}\Tilde{I}_L+\partial_{r_H}I_L
\end{equation}
with the partial derivatives taken with $T$ held fixed. A prescription to evaluate $\partial_{r_H}\Tilde{I}_L$ was discussed in footnote (\ref{regfootnote}) following which it was observed that $\partial_{r_H}I_L=\partial_{r_H}\Tilde{I}_L$. Now, consider the remaining term in (\ref{rHder}),
\begin{equation}
    \partial_{r_H}\Tilde{I}_G= -\frac{\Omega_d}{4\pi G}\frac{\partial \hat{R}}{\partial r_H}(|u'(r)|r^{d-2})\big|_{r=\hat{R}}
\end{equation}
Using (\ref{FG}) to evaluate $\frac{\partial \hat{R}}{\partial r_H}$, we see that $\partial_{r_H}\Tilde{I}_G=0$ as the cutoff, $R\to \infty$. Thus, we have also shown that $\partial_{r_H}I_G=0$ so that $I_G=I_G(T)$ and is independent of $\tau_0$. Therefore the renormalised action for the cylinder phase is 
\begin{equation}
   I_{cyl}(\tau_0,T)=I_{AdS}(2\tau_0)+I_G(T) \ .
\end{equation}
The action of thermal AdS can also be expressed in terms of the Casimir energy as $I_{AdS}(\beta) = \beta E_{vac}$. 

\section{Coarse graining with a quantum channel}\label{app:toychannel}

In this appendix we define a simple toy model for coarse graining by a quantum channel that is similar, but not identical, to the holographic coarse graining map $\C$ defined in section \ref{s:replicaformalism}. The goal is to construct a channel that strips off an isometric `dressing' operator, then completely dephases. 

Consider a finite-dimensional quantum system with Hilbert space $\H = \H_A \oplus \H_B$, equipped with an isometry
\begin{align}
V: \H_A \to \H_B \ .
\end{align}
For any state $|a\rangle \in \H_A$, we refer to $V|a\rangle \in \H_B$ as a `dressed' state. 
In the analogy to large-$N$ CFT, we view the subspace $\H_A$ as the states around energy $E$ and the subspace $\H_B$ as the states around energy $E'$, with $E' > E$. The isometry $V$ corresponds to dressing a heavy state by single-trace operators, or in bulk language, dressing a black hole by adding matter outside the apparent horizon.

Define the projectors
\begin{align}
P_V &= VV^\dagger , \quad \Pvc = \id_B - P_V \ ,
\end{align}
which satisfy $P^2=P$ and
\begin{align}\label{PVzero}
\Pvc V = V^\dagger \Pvc = 0 \ .
\end{align}
Denote the completely dephasing channels on $\H_A$ and $\H_B$ in the energy basis by $\D_A$ and $\D_B$. In terms of projectors onto energy eigenstates,
\begin{align}
\D_A(\rho) &=\sum_{p\in A} P_p \rho P_p , \quad \D_B(\rho) = \sum_{n \in B} P_n \rho P_n
\end{align}
where the notation $p\in A$ indicates a sum over an orthonormal basis of energy eigenstates in $\H_A$, and similarly for $n \in B$.

In this toy model we define a coarse-graining map $\C: \L(\H_B) \to \L(\H_B)$ by
\begin{align}
C(\rho) &=  V \D_A(V^\dagger \rho V)V^\dagger + \D_B (\Pvc \rho \Pvc) \ .
\end{align}
Then we have 
{\itshape
\begin{enumerate}
\item $\C$ is a quantum channel.
\item The von Neumann entropy is non-decreasing under $\C$, $S(\C(\rho)) \geq S(\rho)$. 
\end{enumerate}
}
\noindent Statement $(1)$ follows from the Choi-Kraus representation
\begin{align}
C(\rho) = \sum_{p \in A} VP_p V^\dagger \rho V P_\rho V^\dagger + \sum_{n \in B} P_n \Pvc \rho \Pvc P_n \ ,
\end{align}
which is easily shown to satisfy $\sum_i A_i^\dagger A_i = \id_B$. 
Statement $(2)$ follows from an argument almost identical to \eqref{entropyIncrease} using \eqref{PVzero}.
We could also extend $\C$ to act on the full system, $\C': \L(H) \to \L(H)$, by acting with the diagonal projection on $\L(\H_A)$, without changing the following discussion in any essential way.

\bigskip

To understand what $\C$ does, let's apply it to a pure state. Any pure state $|\psi\rangle \in \H_B$ can be expressed as
\begin{align}
|\psi\rangle = \sum_{p \in A} Va_p |p\rangle + \sum_{n \in B} b_n \Pvc|n\rangle \ .
\end{align}
The coarse-graining map acts on $\rho_\psi = |\psi\rangle\langle \psi|$ as
\begin{align}
\C(\rho_\psi) &= \sum_{p\in A} |a_p|^2 V |p\rangle \langle p| V^\dagger
+ \sum_{n \in B} |b_n'|^2 |n\rangle \langle n|
\end{align}
with $b_n' = \sum_{m \in B} \langle n| \Pvc|m\rangle b_m$. 

We see that $\C$ acts on a dressed state $|\psi\rangle = V |\tilde{\psi}\rangle$ by completely dephasing $\H_A$, and it acts on an undressed state $|\psi\rangle = \Pvc |\psi\rangle$ by completely dephasing $\H_B$. If we think of $V$ as dressing a state by infrared degrees of freedom, then this matches the intuitive notion of coarse graining --- the coarse-grained density matrix $\brho = \C(\rho)$ in this toy model retains the quantum correlations among the `IR' degrees of freedom created by $V$ while discarding the off-diagonal matrix elements connecting the undressed, `UV' microstates. This is similar to the holographic coarse-graining map. In fact, it is identical if we restrict to the class of states studied in this paper; however, we do not know whether it is possible to decompose the full Hilbert space of a large-$N$ CFT in a way similar to the toy model. The bulk suggests that it should be possible.

\renewcommand{\baselinestretch}{1}\small
\bibliographystyle{ourbst}
\bibliography{biblio1}

\providecommand{\href}[2]{#2}\begingroup\raggedright\begin{thebibliography}{10}

\bibitem{Maldacena:2004rf}
J.~M. Maldacena and L.~Maoz, {{Wormholes in AdS}},
  \href{http://dx.doi.org/10.1088/1126-6708/2004/02/053}{JHEP {\bf 02}, 053,
  2004},
  [\href{http://arxiv.org/abs/arXiv:hep-th/0401024}{{arXiv:hep-th/0401024}}].

\bibitem{Arkani-Hamed:2007cpn}
N.~Arkani-Hamed, J.~Orgera and J.~Polchinski, {{Euclidean wormholes in string
  theory}}, \href{http://dx.doi.org/10.1088/1126-6708/2007/12/018}{JHEP {\bf
  12}, 018, 2007},
  [\href{http://arxiv.org/abs/arXiv:0705.2768}{{arXiv:0705.2768 [hep-th]}}].

\bibitem{Cotler:2016fpe}
J.~S. Cotler, G.~Gur-Ari, M.~Hanada, J.~Polchinski, P.~Saad, S.~H. Shenker,
  D.~Stanford, A.~Streicher and M.~Tezuka, {{Black Holes and Random Matrices}},
  \href{http://dx.doi.org/10.1007/JHEP05(2017)118}{JHEP {\bf 05}, 118, 2017},
  [\href{http://arxiv.org/abs/arXiv:1611.04650}{{arXiv:1611.04650 [hep-th]}}].
  [Erratum: JHEP 09, 002 (2018)].

\bibitem{Saad:2019lba}
P.~Saad, S.~H. Shenker and D.~Stanford, {{JT gravity as a matrix integral}},
  2019, [\href{http://arxiv.org/abs/arXiv:1903.11115}{{arXiv:1903.11115
  [hep-th]}}].

\bibitem{Penington:2019kki}
G.~Penington, S.~H. Shenker, D.~Stanford and Z.~Yang, {{Replica wormholes and
  the black hole interior}},  2019,
  [\href{http://arxiv.org/abs/arXiv:1911.11977}{{arXiv:1911.11977 [hep-th]}}].

\bibitem{Afkhami-Jeddi:2020ezh}
N.~Afkhami-Jeddi, H.~Cohn, T.~Hartman and A.~Tajdini, {{Free partition
  functions and an averaged holographic duality}},
  \href{http://dx.doi.org/10.1007/JHEP01(2021)130}{JHEP {\bf 01}, 130, 2021},
  [\href{http://arxiv.org/abs/arXiv:2006.04839}{{arXiv:2006.04839 [hep-th]}}].

\bibitem{Maloney:2020nni}
A.~Maloney and E.~Witten, {{Averaging over Narain moduli space}},
  \href{http://dx.doi.org/10.1007/JHEP10(2020)187}{JHEP {\bf 10}, 187, 2020},
  [\href{http://arxiv.org/abs/arXiv:2006.04855}{{arXiv:2006.04855 [hep-th]}}].

\bibitem{Chandra:2022bqq}
J.~Chandra, S.~Collier, T.~Hartman and A.~Maloney, {{Semiclassical 3D gravity
  as an average of large-c CFTs}},  2022,
  [\href{http://arxiv.org/abs/arXiv:2203.06511}{{arXiv:2203.06511 [hep-th]}}].

\bibitem{Almheiri:2019qdq}
A.~Almheiri, T.~Hartman, J.~Maldacena, E.~Shaghoulian and A.~Tajdini, {{Replica
  Wormholes and the Entropy of Hawking Radiation}},
  \href{http://dx.doi.org/10.1007/JHEP05(2020)013}{JHEP {\bf 05}, 013, 2020},
  [\href{http://arxiv.org/abs/arXiv:1911.12333}{{arXiv:1911.12333 [hep-th]}}].

\bibitem{greenberger1989going}
D.~M. Greenberger, M.~A. Horne and A.~Zeilinger, {Going beyond bell?s theorem},
   in \emph{Bell?s theorem, quantum theory and conceptions of the universe},
  pp.~69--72.
\newblock Springer, 1989.

\bibitem{Hayden:2011ag}
P.~Hayden, M.~Headrick and A.~Maloney, {{Holographic Mutual Information is
  Monogamous}}, \href{http://dx.doi.org/10.1103/PhysRevD.87.046003}{Phys. Rev.
  D {\bf 87}, 046003, 2013},
  [\href{http://arxiv.org/abs/arXiv:1107.2940}{{arXiv:1107.2940 [hep-th]}}].

\bibitem{Balasubramanian:2014hda}
V.~Balasubramanian, P.~Hayden, A.~Maloney, D.~Marolf and S.~F. Ross,
  {{Multiboundary Wormholes and Holographic Entanglement}},
  \href{http://dx.doi.org/10.1088/0264-9381/31/18/185015}{Class. Quant. Grav.
  {\bf 31}, 185015, 2014},
  [\href{http://arxiv.org/abs/arXiv:1406.2663}{{arXiv:1406.2663 [hep-th]}}].

\bibitem{Susskind:2014yaa}
L.~Susskind, {{ER=EPR, GHZ, and the consistency of quantum measurements}},
  \href{http://dx.doi.org/10.1002/prop.201500094}{Fortsch. Phys. {\bf 64},
  72--83, 2016}, [\href{http://arxiv.org/abs/arXiv:1412.8483}{{arXiv:1412.8483
  [hep-th]}}].

\bibitem{Bao:2015bfa}
N.~Bao, S.~Nezami, H.~Ooguri, B.~Stoica, J.~Sully and M.~Walter, {{The
  Holographic Entropy Cone}},
  \href{http://dx.doi.org/10.1007/JHEP09(2015)130}{JHEP {\bf 09}, 130, 2015},
  [\href{http://arxiv.org/abs/arXiv:1505.07839}{{arXiv:1505.07839 [hep-th]}}].

\bibitem{Stanford:2020wkf}
D.~Stanford, {{More quantum noise from wormholes}},  2020,
  [\href{http://arxiv.org/abs/arXiv:2008.08570}{{arXiv:2008.08570 [hep-th]}}].

\bibitem{Gross:1982cv}
D.~J. Gross, M.~J. Perry and L.~G. Yaffe, {{Instability of Flat Space at Finite
  Temperature}}, \href{http://dx.doi.org/10.1103/PhysRevD.25.330}{Phys. Rev. D
  {\bf 25}, 330--355, 1982}.

\bibitem{Marolf:2021kjc}
D.~Marolf and J.~E. Santos, {{AdS Euclidean wormholes}},
  \href{http://dx.doi.org/10.1088/1361-6382/ac2cb7}{Class. Quant. Grav. {\bf
  38}, 224002, 2021},
  [\href{http://arxiv.org/abs/arXiv:2101.08875}{{arXiv:2101.08875 [hep-th]}}].

\bibitem{Cooper:2018cmb}
S.~Cooper, M.~Rozali, B.~Swingle, M.~Van~Raamsdonk, C.~Waddell and D.~Wakeham,
  {{Black hole microstate cosmology}},
  \href{http://dx.doi.org/10.1007/JHEP07(2019)065}{JHEP {\bf 07}, 065, 2019},
  [\href{http://arxiv.org/abs/arXiv:1810.10601}{{arXiv:1810.10601 [hep-th]}}].

\bibitem{Keranen:2015fqa}
V.~Keranen, H.~Nishimura, S.~Stricker, O.~Taanila and A.~Vuorinen,
  {{Gravitational collapse of thin shells: Time evolution of the holographic
  entanglement entropy}}, \href{http://dx.doi.org/10.1007/JHEP06(2015)126}{JHEP
  {\bf 06}, 126, 2015},
  [\href{http://arxiv.org/abs/arXiv:1502.01277}{{arXiv:1502.01277 [hep-th]}}].

\bibitem{Anous:2016kss}
T.~Anous, T.~Hartman, A.~Rovai and J.~Sonner, {{Black Hole Collapse in the 1/c
  Expansion}}, \href{http://dx.doi.org/10.1007/JHEP07(2016)123}{JHEP {\bf 07},
  123, 2016}, [\href{http://arxiv.org/abs/arXiv:1603.04856}{{arXiv:1603.04856
  [hep-th]}}].

\bibitem{Ryu:2006bv}
S.~Ryu and T.~Takayanagi, {{Holographic derivation of entanglement entropy from
  AdS/CFT}}, \href{http://dx.doi.org/10.1103/PhysRevLett.96.181602}{Phys. Rev.
  Lett. {\bf 96}, 181602, 2006},
  [\href{http://arxiv.org/abs/arXiv:hep-th/0603001}{{arXiv:hep-th/0603001}}].

\bibitem{Hubeny:2007xt}
V.~E. Hubeny, M.~Rangamani and T.~Takayanagi, {{A Covariant holographic
  entanglement entropy proposal}},
  \href{http://dx.doi.org/10.1088/1126-6708/2007/07/062}{JHEP {\bf 07}, 062,
  2007}, [\href{http://arxiv.org/abs/arXiv:0705.0016}{{arXiv:0705.0016
  [hep-th]}}].

\bibitem{Engelhardt:2017aux}
N.~Engelhardt and A.~C. Wall, {{Decoding the Apparent Horizon: Coarse-Grained
  Holographic Entropy}},
  \href{http://dx.doi.org/10.1103/PhysRevLett.121.211301}{Phys. Rev. Lett. {\bf
  121}, 211301, 2018},
  [\href{http://arxiv.org/abs/arXiv:1706.02038}{{arXiv:1706.02038 [hep-th]}}].

\bibitem{Swingle:2009bg}
B.~Swingle, {{Entanglement Renormalization and Holography}},
  \href{http://dx.doi.org/10.1103/PhysRevD.86.065007}{Phys. Rev. D {\bf 86},
  065007, 2012}, [\href{http://arxiv.org/abs/arXiv:0905.1317}{{arXiv:0905.1317
  [cond-mat.str-el]}}].

\bibitem{Yang:2015uoa}
Z.~Yang, P.~Hayden and X.-L. Qi, {{Bidirectional holographic codes and sub-AdS
  locality}}, \href{http://dx.doi.org/10.1007/JHEP01(2016)175}{JHEP {\bf 01},
  175, 2016}, [\href{http://arxiv.org/abs/arXiv:1510.03784}{{arXiv:1510.03784
  [hep-th]}}].

\bibitem{Hayden:2016cfa}
P.~Hayden, S.~Nezami, X.-L. Qi, N.~Thomas, M.~Walter and Z.~Yang, {{Holographic
  duality from random tensor networks}},
  \href{http://dx.doi.org/10.1007/JHEP11(2016)009}{JHEP {\bf 11}, 009, 2016},
  [\href{http://arxiv.org/abs/arXiv:1601.01694}{{arXiv:1601.01694 [hep-th]}}].

\bibitem{Almheiri:2014lwa}
A.~Almheiri, X.~Dong and D.~Harlow, {{Bulk Locality and Quantum Error
  Correction in AdS/CFT}},
  \href{http://dx.doi.org/10.1007/JHEP04(2015)163}{JHEP {\bf 04}, 163, 2015},
  [\href{http://arxiv.org/abs/arXiv:1411.7041}{{arXiv:1411.7041 [hep-th]}}].

\bibitem{Pastawski:2015qua}
F.~Pastawski, B.~Yoshida, D.~Harlow and J.~Preskill, {{Holographic quantum
  error-correcting codes: Toy models for the bulk/boundary correspondence}},
  \href{http://dx.doi.org/10.1007/JHEP06(2015)149}{JHEP {\bf 06}, 149, 2015},
  [\href{http://arxiv.org/abs/arXiv:1503.06237}{{arXiv:1503.06237 [hep-th]}}].

\bibitem{note:toappear}
To appear.

\bibitem{Lewkowycz:2013nqa}
A.~Lewkowycz and J.~Maldacena, {{Generalized gravitational entropy}},
  \href{http://dx.doi.org/10.1007/JHEP08(2013)090}{JHEP {\bf 08}, 090, 2013},
  [\href{http://arxiv.org/abs/arXiv:1304.4926}{{arXiv:1304.4926 [hep-th]}}].

\bibitem{Barankov:2008qq}
R.~Barankov and A.~Polkovnikov, {{Microscopic diagonal entropy and its
  connection to basic thermodynamic relations}},
  \href{http://dx.doi.org/10.1016/j.aop.2010.08.004}{Annals Phys. {\bf 326},
  486--499, 2011},
  [\href{http://arxiv.org/abs/arXiv:0806.2862}{{arXiv:0806.2862
  [cond-mat.stat-mech]}}].

\bibitem{DAlessio:2015qtq}
L.~D'Alessio, Y.~Kafri, A.~Polkovnikov and M.~Rigol, {{From quantum chaos and
  eigenstate thermalization to statistical mechanics and thermodynamics}},
  \href{http://dx.doi.org/10.1080/00018732.2016.1198134}{Adv. Phys. {\bf 65},
  239--362, 2016},
  [\href{http://arxiv.org/abs/arXiv:1509.06411}{{arXiv:1509.06411
  [cond-mat.stat-mech]}}].

\bibitem{Roy:2015pga}
S.~R. Roy and D.~Sarkar, {{Hologram of a pure state black hole}},
  \href{http://dx.doi.org/10.1103/PhysRevD.92.126003}{Phys. Rev. D {\bf 92},
  126003, 2015},
  [\href{http://arxiv.org/abs/arXiv:1505.03895}{{arXiv:1505.03895 [hep-th]}}].

\bibitem{Sonner:2017hxc}
J.~Sonner and M.~Vielma, {{Eigenstate thermalization in the Sachdev-Ye-Kitaev
  model}}, \href{http://dx.doi.org/10.1007/JHEP11(2017)149}{JHEP {\bf 11}, 149,
  2017}, [\href{http://arxiv.org/abs/arXiv:1707.08013}{{arXiv:1707.08013
  [hep-th]}}].

\bibitem{Hunter-Jones:2017raw}
N.~Hunter-Jones, J.~Liu and Y.~Zhou, {{On thermalization in the SYK and
  supersymmetric SYK models}},
  \href{http://dx.doi.org/10.1007/JHEP02(2018)142}{JHEP {\bf 02}, 142, 2018},
  [\href{http://arxiv.org/abs/arXiv:1710.03012}{{arXiv:1710.03012 [hep-th]}}].

\bibitem{Pollack:2020gfa}
J.~Pollack, M.~Rozali, J.~Sully and D.~Wakeham, {{Eigenstate Thermalization and
  Disorder Averaging in Gravity}},
  \href{http://dx.doi.org/10.1103/PhysRevLett.125.021601}{Phys. Rev. Lett. {\bf
  125}, 021601, 2020},
  [\href{http://arxiv.org/abs/arXiv:2002.02971}{{arXiv:2002.02971 [hep-th]}}].

\bibitem{Marolf:2020vsi}
D.~Marolf, S.~Wang and Z.~Wang, {{Probing phase transitions of holographic
  entanglement entropy with fixed area states}},
  \href{http://dx.doi.org/10.1007/JHEP12(2020)084}{JHEP {\bf 12}, 084, 2020},
  [\href{http://arxiv.org/abs/arXiv:2006.10089}{{arXiv:2006.10089 [hep-th]}}].

\bibitem{Altland:2021rqn}
A.~Altland, D.~Bagrets, P.~Nayak, J.~Sonner and M.~Vielma, {{From operator
  statistics to wormholes}},
  \href{http://dx.doi.org/10.1103/PhysRevResearch.3.033259}{Phys. Rev. Res.
  {\bf 3}, 033259, 2021},
  [\href{http://arxiv.org/abs/arXiv:2105.12129}{{arXiv:2105.12129 [hep-th]}}].

\bibitem{Saad:2021uzi}
P.~Saad, S.~Shenker and S.~Yao, {{Comments on wormholes and factorization}},
  2021, [\href{http://arxiv.org/abs/arXiv:2107.13130}{{arXiv:2107.13130
  [hep-th]}}].

\bibitem{Freivogel:2021ivu}
B.~Freivogel, D.~Nikolakopoulou and A.~F. Rotundo, {{Wormholes from Averaging
  over States}},  2021,
  [\href{http://arxiv.org/abs/arXiv:2105.12771}{{arXiv:2105.12771 [hep-th]}}].

\bibitem{Cotler:2022rud}
J.~Cotler and K.~Jensen, {{A precision test of averaging in AdS/CFT}},  2022,
  [\href{http://arxiv.org/abs/arXiv:2205.12968}{{arXiv:2205.12968 [hep-th]}}].

\bibitem{Engelhardt:2018kcs}
N.~Engelhardt and A.~C. Wall, {{Coarse Graining Holographic Black Holes}},
  \href{http://dx.doi.org/10.1007/JHEP05(2019)160}{JHEP {\bf 05}, 160, 2019},
  [\href{http://arxiv.org/abs/arXiv:1806.01281}{{arXiv:1806.01281 [hep-th]}}].

\bibitem{Engelhardt:2021mue}
N.~Engelhardt, G.~Penington and A.~Shahbazi-Moghaddam, {{A world without
  pythons would be so simple}},
  \href{http://dx.doi.org/10.1088/1361-6382/ac2de5}{Class. Quant. Grav. {\bf
  38}, 234001, 2021},
  [\href{http://arxiv.org/abs/arXiv:2102.07774}{{arXiv:2102.07774 [hep-th]}}].

\bibitem{Brown:2019rox}
A.~R. Brown, H.~Gharibyan, G.~Penington and L.~Susskind, {{The
  Python\textquoteright{}s Lunch: geometric obstructions to decoding Hawking
  radiation}}, \href{http://dx.doi.org/10.1007/JHEP08(2020)121}{JHEP {\bf 08},
  121, 2020}, [\href{http://arxiv.org/abs/arXiv:1912.00228}{{arXiv:1912.00228
  [hep-th]}}].

\bibitem{Engelhardt:2021qjs}
N.~Engelhardt, G.~Penington and A.~Shahbazi-Moghaddam, {{Finding pythons in
  unexpected places}}, \href{http://dx.doi.org/10.1088/1361-6382/ac3e75}{Class.
  Quant. Grav. {\bf 39}, 094002, 2022},
  [\href{http://arxiv.org/abs/arXiv:2105.09316}{{arXiv:2105.09316 [hep-th]}}].

\bibitem{Renner:2021qbe}
R.~Renner and J.~Wang, {{The black hole information puzzle and the quantum de
  Finetti theorem}},  2021,
  [\href{http://arxiv.org/abs/arXiv:2110.14653}{{arXiv:2110.14653 [hep-th]}}].

\bibitem{Almheiri:2021jwq}
A.~Almheiri and H.~W. Lin, {{The Entanglement Wedge of Unknown Couplings}},
  2021, [\href{http://arxiv.org/abs/arXiv:2111.06298}{{arXiv:2111.06298
  [hep-th]}}].

\bibitem{Qi:2021oni}
X.-L. Qi, Z.~Shangnan and Z.~Yang, {{Holevo information and ensemble theory of
  gravity}}, \href{http://dx.doi.org/10.1007/JHEP02(2022)056}{JHEP {\bf 02},
  056, 2022}, [\href{http://arxiv.org/abs/arXiv:2111.05355}{{arXiv:2111.05355
  [hep-th]}}].

\bibitem{Karch:2000ct}
A.~Karch and L.~Randall, {{Locally localized gravity}},
  \href{http://dx.doi.org/10.1088/1126-6708/2001/05/008}{JHEP {\bf 05}, 008,
  2001},
  [\href{http://arxiv.org/abs/arXiv:hep-th/0011156}{{arXiv:hep-th/0011156}}].

\bibitem{Takayanagi:2011zk}
T.~Takayanagi, {{Holographic Dual of BCFT}},
  \href{http://dx.doi.org/10.1103/PhysRevLett.107.101602}{Phys. Rev. Lett. {\bf
  107}, 101602, 2011},
  [\href{http://arxiv.org/abs/arXiv:1105.5165}{{arXiv:1105.5165 [hep-th]}}].

\bibitem{Fujita:2011fp}
M.~Fujita, T.~Takayanagi and E.~Tonni, {{Aspects of AdS/BCFT}},
  \href{http://dx.doi.org/10.1007/JHEP11(2011)043}{JHEP {\bf 11}, 043, 2011},
  [\href{http://arxiv.org/abs/arXiv:1108.5152}{{arXiv:1108.5152 [hep-th]}}].

\bibitem{Hartman:2013qma}
T.~Hartman and J.~Maldacena, {{Time Evolution of Entanglement Entropy from
  Black Hole Interiors}}, \href{http://dx.doi.org/10.1007/JHEP05(2013)014}{JHEP
  {\bf 05}, 014, 2013},
  [\href{http://arxiv.org/abs/arXiv:1303.1080}{{arXiv:1303.1080 [hep-th]}}].

\bibitem{Kourkoulou:2017zaj}
I.~Kourkoulou and J.~Maldacena, {{Pure states in the SYK model and
  nearly-$AdS_2$ gravity}},  2017,
  [\href{http://arxiv.org/abs/arXiv:1707.02325}{{arXiv:1707.02325 [hep-th]}}].

\bibitem{Chen:2020uac}
H.~Z. Chen, R.~C. Myers, D.~Neuenfeld, I.~A. Reyes and J.~Sandor, {{Quantum
  Extremal Islands Made Easy, Part I: Entanglement on the Brane}},
  \href{http://dx.doi.org/10.1007/JHEP10(2020)166}{JHEP {\bf 10}, 166, 2020},
  [\href{http://arxiv.org/abs/arXiv:2006.04851}{{arXiv:2006.04851 [hep-th]}}].

\bibitem{Chen:2020hmv}
H.~Z. Chen, R.~C. Myers, D.~Neuenfeld, I.~A. Reyes and J.~Sandor, {{Quantum
  Extremal Islands Made Easy, Part II: Black Holes on the Brane}},
  \href{http://dx.doi.org/10.1007/JHEP12(2020)025}{JHEP {\bf 12}, 025, 2020},
  [\href{http://arxiv.org/abs/arXiv:2010.00018}{{arXiv:2010.00018 [hep-th]}}].

\bibitem{Miyaji:2021ktr}
M.~Miyaji, T.~Takayanagi and T.~Ugajin, {{Spectrum of End of the World Branes
  in Holographic BCFTs}}, \href{http://dx.doi.org/10.1007/JHEP06(2021)023}{JHEP
  {\bf 06}, 023, 2021},
  [\href{http://arxiv.org/abs/arXiv:2103.06893}{{arXiv:2103.06893 [hep-th]}}].

\bibitem{Akal:2021foz}
I.~Akal, Y.~Kusuki, N.~Shiba, T.~Takayanagi and Z.~Wei, {{Holographic moving
  mirrors}}, \href{http://dx.doi.org/10.1088/1361-6382/ac2c1b}{Class. Quant.
  Grav. {\bf 38}, 224001, 2021},
  [\href{http://arxiv.org/abs/arXiv:2106.11179}{{arXiv:2106.11179 [hep-th]}}].

\bibitem{Suzuki:2022yru}
Y.-k. Suzuki and S.~Terashima, {{On the Dynamics in the AdS/BCFT
  Correspondence}},  2022,
  [\href{http://arxiv.org/abs/arXiv:2205.10600}{{arXiv:2205.10600 [hep-th]}}].

\bibitem{Izumi:2022opi}
K.~Izumi, T.~Shiromizu, K.~Suzuki, T.~Takayanagi and N.~Tanahashi, {{Brane
  Dynamics of Holographic BCFTs}},  2022,
  [\href{http://arxiv.org/abs/arXiv:2205.15500}{{arXiv:2205.15500 [hep-th]}}].

\bibitem{Rozali:2019day}
M.~Rozali, J.~Sully, M.~Van~Raamsdonk, C.~Waddell and D.~Wakeham, {{Information
  radiation in BCFT models of black holes}},
  \href{http://dx.doi.org/10.1007/JHEP05(2020)004}{JHEP {\bf 05}, 004, 2020},
  [\href{http://arxiv.org/abs/arXiv:1910.12836}{{arXiv:1910.12836 [hep-th]}}].

\bibitem{deutsch1991quantum}
J.~M. Deutsch, {Quantum statistical mechanics in a closed system}, {Physical
  Review A {\bf 43}, 2046, 1991}.

\bibitem{Srednicki:1994mfb}
M.~Srednicki, {{Chaos and Quantum Thermalization}},  1994,
  [\href{http://arxiv.org/abs/arXiv:cond-mat/9403051}{{arXiv:cond-mat/9403051}}].

\bibitem{Akers:2021fut}
C.~Akers, C.~Akers and G.~Penington, {{Quantum minimal surfaces from quantum
  error correction}},
  \href{http://dx.doi.org/10.21468/SciPostPhys.12.5.157}{SciPost Phys. {\bf
  12}, 157, 2022},
  [\href{http://arxiv.org/abs/arXiv:2109.14618}{{arXiv:2109.14618 [hep-th]}}].

\bibitem{Belin:2021ryy}
A.~Belin, J.~de~Boer and D.~Liska, {{Non-Gaussianities in the Statistical
  Distribution of Heavy OPE Coefficients and Wormholes}},  2021,
  [\href{http://arxiv.org/abs/arXiv:2110.14649}{{arXiv:2110.14649 [hep-th]}}].

\bibitem{Penington:2019npb}
G.~Penington, {{Entanglement Wedge Reconstruction and the Information
  Paradox}}, \href{http://dx.doi.org/10.1007/JHEP09(2020)002}{JHEP {\bf 09},
  002, 2020}, [\href{http://arxiv.org/abs/arXiv:1905.08255}{{arXiv:1905.08255
  [hep-th]}}].

\bibitem{Almheiri:2019psf}
A.~Almheiri, N.~Engelhardt, D.~Marolf and H.~Maxfield, {{The entropy of bulk
  quantum fields and the entanglement wedge of an evaporating black hole}},
  \href{http://dx.doi.org/10.1007/JHEP12(2019)063}{JHEP {\bf 12}, 063, 2019},
  [\href{http://arxiv.org/abs/arXiv:1905.08762}{{arXiv:1905.08762 [hep-th]}}].

\bibitem{Akers:2020pmf}
C.~Akers and G.~Penington, {{Leading order corrections to the quantum extremal
  surface prescription}}, \href{http://dx.doi.org/10.1007/JHEP04(2021)062}{JHEP
  {\bf 04}, 062, 2021},
  [\href{http://arxiv.org/abs/arXiv:2008.03319}{{arXiv:2008.03319 [hep-th]}}].

\bibitem{Chen:2020tes}
Y.~Chen, V.~Gorbenko and J.~Maldacena, {{Bra-ket wormholes in gravitationally
  prepared states}}, \href{http://dx.doi.org/10.1007/JHEP02(2021)009}{JHEP {\bf
  02}, 009, 2021},
  [\href{http://arxiv.org/abs/arXiv:2007.16091}{{arXiv:2007.16091 [hep-th]}}].

\bibitem{Engelhardt:2014gca}
N.~Engelhardt and A.~C. Wall, {{Quantum Extremal Surfaces: Holographic
  Entanglement Entropy beyond the Classical Regime}},
  \href{http://dx.doi.org/10.1007/JHEP01(2015)073}{JHEP {\bf 01}, 073, 2015},
  [\href{http://arxiv.org/abs/arXiv:1408.3203}{{arXiv:1408.3203 [hep-th]}}].

\bibitem{Dong:2016hjy}
X.~Dong, A.~Lewkowycz and M.~Rangamani, {{Deriving covariant holographic
  entanglement}}, \href{http://dx.doi.org/10.1007/JHEP11(2016)028}{JHEP {\bf
  11}, 028, 2016},
  [\href{http://arxiv.org/abs/arXiv:1607.07506}{{arXiv:1607.07506 [hep-th]}}].

\bibitem{Bousso:2020yxi}
R.~Bousso, V.~Chandrasekaran, P.~Rath and A.~Shahbazi-Moghaddam, {{Gravity dual
  of Connes cocycle flow}},
  \href{http://dx.doi.org/10.1103/PhysRevD.102.066008}{Phys. Rev. D {\bf 102},
  066008, 2020},
  [\href{http://arxiv.org/abs/arXiv:2007.00230}{{arXiv:2007.00230 [hep-th]}}].

\bibitem{Engelhardt:2022qts}
N.~Engelhardt and r.~Folkestad, {{Canonical purification of evaporating black
  holes}}, \href{http://dx.doi.org/10.1103/PhysRevD.105.086010}{Phys. Rev. D
  {\bf 105}, 086010, 2022},
  [\href{http://arxiv.org/abs/arXiv:2201.08395}{{arXiv:2201.08395 [hep-th]}}].

\end{thebibliography}\endgroup
\end{document}